\documentclass[preprint,onecolumn,nofootinbib]{revtex4}
\pdfoutput=1
\usepackage[colorlinks=true,linkcolor=blue,urlcolor=blue,filecolor=black,citecolor=red,pdfstartview=FitV,pdftitle={},pdfsubject={},pdfkeywords={},pdfpagemode=None,bookmarksopen=true]{hyperref}
\usepackage{graphicx}
\usepackage{amsmath}
\usepackage{amsfonts}
\usepackage{amssymb}
\usepackage{color}%
\usepackage{dcolumn}
\usepackage{CJK}
\newcommand{\f}{\begin{equation}}
\newcommand{\ff}{\end{equation}}
\newcommand{\fa}{\begin{eqnarray}}
\newcommand{\ffa}{\end{eqnarray}}

\begin{document}
\title{EM duality and Quasi-normal modes from higher derivatives with homogeneous disorder}
\author{Guoyang Fu $^{1}$}
\thanks{FuguoyangEDU@163.com}
\author{Jian-Pin Wu $^{2}$}
\thanks{jianpinwu@yzu.edu.cn}
\affiliation{$^1$ Department of Physics, School of Mathematics and Physics, Bohai University, Jinzhou 121013, China \ \\
$^2$ Center for Gravitation and Cosmology, College of Physical Science and Technology, Yangzhou University, Yangzhou 225009, China\ \\
}
\begin{abstract}

We study the electromagnetic (EM) duality from $6$ derivative theory with homogeneous disorder.
We find that with the change of the sign of the coupling parameter $\gamma_1$ of the $6$ derivative theory,
the particle-vortex duality with homogeneous disorder holds more well than that without homogeneous disorder.
The properties of quasinormal modes (QNMs) of this system are also explored.
Similarly, for large $\gamma_1$, with the change of the sign of $\gamma_1$, the correspondence between the low frequency poles of the original theory
and the ones of its dual theory, which is strongly violated in the absence of homogeneous disorder, holds well.
Some new branch cuts of QNMs emerge for some system parameters.

\end{abstract}
\maketitle
\section{Introduction}

Holographic quantum critical (QC) dynamics at zero density has been deeply explored \cite{Myers:2010pk,Ritz:2008kh,WitczakKrempa:2012gn,WitczakKrempa:2013ht,Witczak-Krempa:2013nua,Witczak-Krempa:2013aea,Katz:2014rla,Sachdev:2011wg,Hartnoll:2016apf,Bai:2013tfa,Myers:2016wsu,Lucas:2017dqa}.
It provides a powerful tool to study strongly coupled systems without quasi-particles descriptions.
By studying the transports, in particular the optical conductivity, from a probe Maxwell field coupled to the Weyl tensor $C_{\mu\nu\rho\sigma}$
on top of the $4$ dimensional Schwarzschild-AdS (SS-AdS) black brane background\footnote{Since the Maxwell field is the $1$ derivative term and the Weyl tensor
is the $2$ derivative term, the coupling term as $C_{\mu\nu\rho\sigma}F^{\mu\nu}F^{\rho\sigma}$ is the $4$ derivative term. Therefore, we also refer to this theory as $4$ derivative theory.} \cite{Myers:2010pk,Sachdev:2011wg,Hartnoll:2016apf,Ritz:2008kh,WitczakKrempa:2012gn,
WitczakKrempa:2013ht,Witczak-Krempa:2013nua,Witczak-Krempa:2013aea,Katz:2014rla},
they observe a non-trivial frequency dependent conductivity attributing to the introduction of Weyl tensor,
which breaks the electromagnetic (EM) duality.
It exhibits a peak, which resembles the particle response and we refer as the Damle-Sachdev (DS) peak\footnote{Since the peak results from the particle-hole symmetry at zero charge density
but not the breaking of translation symmetry, we refer as DS peak instead of Drude peak.} \cite{Damle:1997rxu},
or a dip, which is similar with the behavior of the vortex response.
Such behavior is analogous to the one in the superfluid-insulator quantum critical point (QCP) \cite{Myers:2010pk,Sachdev:2011wg,Hartnoll:2016apf}.
In addition, the DC conductivity has a bound, which can not arrive at zero in the allowed region of coupling parameter.
When higher derivative (HD) terms are introduced, an arbitrarily sharp DS peak can be observed at low frequency in the optical conductivity
depending on the coupling parameters,
and the bound of conductivity in $4$ derivative theory is violated such that a zero DC conductivity can be obtained at specific parameter \cite{Witczak-Krempa:2013aea}.
In particular, its behavior is quite closely analogous to that of the $O(N)$ superfluid-insulator model in the limit of large-$N$ \cite{Damle:1997rxu}.
Another progress is the construction of neutral scalar hair black brane by coupling Weyl tensor with neutral scalar field,
which provides a framework to describe QC dynamics and one away from QCP \cite{Myers:2016wsu,Lucas:2017dqa,Wu:2018xjy}.

Also, we can introduce a mechanism of microscopic translational symmetry broken, for example \cite{Donos:2012js,Donos:2013eha,Vegh:2013sk,Andrade:2013gsa,Donos:2014uba,Grozdanov:2015qia,Blake:2013bqa,Ling:2015epa,Ling:2015exa,Kuang:2017cgt,Kuang:2018ymh},
which leads to the momentum dissipation of the dual field theory and is also referred as homogeneous disorder,
into the holographic QC system \cite{Myers:2010pk,Ritz:2008kh,WitczakKrempa:2012gn,WitczakKrempa:2013ht,Witczak-Krempa:2013nua,Witczak-Krempa:2013aea,Katz:2014rla,Sachdev:2011wg,Hartnoll:2016apf,Bai:2013tfa,Myers:2016wsu,Lucas:2017dqa},
such that we can study the effects from homogeneous disorder on these systems \cite{Wu:2016jjd,Fu:2017oqa,Wu:2018pig,Wu:2018vlj}.
We find that for $4$ derivative theory studied in \cite{Wu:2016jjd}, the homogeneous disorder drives the incoherent metallic state with DS peak
into the one with a dip for $\gamma>0$ and an oppositive scenario is found for $\gamma<0$.
But for the $6$ derivative theory explored in \cite{Fu:2017oqa},
the homogeneous disorder cannot make the peak (gap) transform into its contrary.
Another interesting phenomena is that for $4$ derivative theory there is a specific value of $\hat{\alpha}=2/\sqrt{3}$,
for which the particle-vortex duality, corresponding to EM duality in bulk, almost exactly holds with the change of the sign of $\gamma$ \cite{Wu:2016jjd}.

On the other hand, to study the holographic QC dynamics, a good method we can used is the quasinormal modes (QNMs)
of a gravitational theory on the bulk AdS spacetime.
Recently, the structure of the QNMs of $4/6$ derivative theory over SS-AdS has been explored \cite{WitczakKrempa:2012gn,Witczak-Krempa:2013aea}.
In particular, the particle-vortex duality has also been discussed.
Further, when the homogeneous disorder in $4$ derivative theory is introduced, more rich pole structures are exhibited \cite{Wu:2018vlj}.
In this paper, we shall study the EM duality and the QNMs of $6$ derivative theory with homogeneous disorder.
Our paper is organized as what follows.
In Section \ref{sec-setup}, we present a brief introduction on the $6$ derivative theory with homogeneous disorder.
EM duality is discussed in Section \ref{sec-em-duality}.
In Section \ref{sec-qnm}, we study the properties of the QNMs.
A brief discussion is presented in Section \ref{sec-con}.
In Appendix \ref{sec-stability}, we also give a brief discussion on the instabilities of gauge mode by the QNMs.

\section{Holographic setup}\label{sec-setup}

We consider the following neutral Einstein-axions (EA) theory \cite{Andrade:2013gsa},
\fa
\label{ac-ax}
S_0=\int d^4x\sqrt{-g}\Big(R+\frac{6}{L^2}-\frac{1}{2}\sum_{I=x,y}(\partial \phi_I)^2\Big)
\,.
\ffa
In the above action, we have introduced
a pair of spatial linear dependent axionic fields, $\phi_I=\alpha x_I$ with $I=x,y$ and $\alpha$ being a constant,
which are responsible for dissipating the momentum of the dual boundary field.
$L$ is the radius of the AdS spacetimes.

From the EA action (\ref{ac-ax}), we have
a neutral black brane solution \cite{Andrade:2013gsa}
\fa
\label{bl-br}
ds^2=\frac{L^2}{u^2}\Big(-f(u)dt^2+\frac{1}{f(u)}du^2+dx^2+dy^2\Big)\,,
\ffa
where
\fa
\label{fu}
f(u)=(1-u)p(u)\,,~~~~~~~
p(u)=\frac{\sqrt{1+6\hat{\alpha}^2}-2\hat{\alpha}^2-1}{\hat{\alpha}^2}u^2+u+1\,.
\ffa
The AdS boundary is at $u=0$ and the horizon locates at $u=1$.
$\hat{\alpha}\equiv\alpha/4\pi T$
with the Hawking temperature $T=p(1)/4\pi$.
Since the translational symmetry breaks, the momentum dissipates.
But the geometry is homogeneous and so we refer to this mechanism as homogeneous disorder
and $\hat{\alpha}$ denotes the strength of disorder.
Therefore, the background describes a specific thermal excited state with homogeneous disorder.

Over this background, we study the following HD action for gauge field coupling with Weyl tensor
\fa
\label{ac-SA}
S_A=\int d^4x\sqrt{-g}\Big(-\frac{1}{8g_F^2}F_{\mu\nu}X^{\mu\nu\rho\sigma}F_{\rho\sigma}\Big)\,,
\ffa
where the tensor $X$ is an infinite family of HD terms
\cite{Witczak-Krempa:2013aea}
\fa
X_{\mu\nu}^{\ \ \rho\sigma}&=&
I_{\mu\nu}^{\ \ \rho\sigma}-8\gamma_{1,1}L^2 C_{\mu\nu}^{\ \ \rho\sigma}
-4L^4\gamma_{2,1}C^2I_{\mu\nu}^{\ \ \rho\sigma}
-8L^4\gamma_{2,2}C_{\mu\nu}^{\ \ \alpha\beta}C_{\alpha\beta}^{\ \ \rho\sigma}
\nonumber
\\
&&
-4L^6\gamma_{3,1}C^3I_{\mu\nu}^{\ \ \rho\sigma}
-8L^6\gamma_{3,2}C^2C_{\mu\nu}^{\ \ \rho\sigma}
-8L^6\gamma_{3,3}C_{\mu\nu}^{\ \ \alpha_1\beta_1}C_{\alpha_1\beta_1}^{\ \ \ \alpha_2\beta_2}C_{\alpha_2\beta_2}^{\ \ \ \rho\sigma}
+\ldots
\,.
\label{X-tensor}
\ffa
Note that $I_{\mu\nu}^{\ \ \rho\sigma}=\delta_{\mu}^{\ \rho}\delta_{\nu}^{\ \sigma}-\delta_{\mu}^{\ \sigma}\delta_{\nu}^{\ \rho}$
is an identity matrix and $C^n=C_{\mu\nu}^{\ \ \alpha_1\beta_1}C_{\alpha_1\beta_1}^{\ \ \ \alpha_2\beta_2}\ldots C_{\alpha_{n-1}\beta_{n-1}}^{\ \ \ \mu\nu}$.
In the above equations, the factor of $L$ is introduced such that the coupling parameters $g_F$ and $\gamma_{i,j}$ are dimensionless.
But for later convenience, we work in units where $L=1$ in what follows and set $g_F=1$ in the numerical calculation.
When $X_{\mu\nu}^{\ \ \rho\sigma}=I_{\mu\nu}^{\ \ \rho\sigma}$, the theory reduces to the standard Maxwell theory.
For convenience, we denote $\gamma_{1,1}=\gamma$ and $\gamma_{2,i}=\gamma_i (i=1,2)$.
In this paper, our main focus is the $6$ derivative terms, i.e., $\gamma_1$ and $\gamma_2$ terms.
At the same time, since the effect of both $\gamma_1$ and $\gamma_2$ terms is similar,
we only turn on $\gamma_1$ term through this paper.
Note that when other parameters are turned off, $\gamma_1$ is confined in the region $\gamma_1\leq 1/48$
over SS-AdS black brane background\footnote{In \cite{Fu:2017oqa}, they also study the instabilities of the gauge mode and the causality in CFT,
in addition to the constraint from $\texttt{Re}\sigma(\omega)\geq 0$,
they confirm that the constraint $\gamma_1\leq 1/48$ also holds over EA-AdS geometry.} \cite{Witczak-Krempa:2013aea}.
And then, from the action (\ref{ac-SA}), we can write down the equation of motion (EOM) as,
\fa
\nabla_{\nu}(X^{\mu\nu\rho\sigma}F_{\rho\sigma})=0\,.
\label{eom-Max}
\ffa

As \cite{Myers:2010pk}, the dual EM theory of \eqref{ac-SA} can be constructed, which is
\fa
\label{ac-SB}
S_B=\int d^4x\sqrt{-g}\Big(-\frac{1}{8\hat{g}_F}G_{\mu\nu}\widehat{X}^{\mu\nu\rho\sigma}G_{\rho\sigma}\Big)\,.
\ffa
$G_{\mu\nu}\equiv\partial_{\mu}B_{\nu}-\partial_{\nu}B_{\mu}$ is the dual field strength
and $\hat{g}_F$ is the coupling constant, which relates $g_F$ as  $\hat{g}_F^2\equiv 1/g_F^2$.
$\widehat{X}$ is the dual tensor, which satisfies
\fa
\widehat{X}_{\mu\nu}^{\ \ \rho\sigma}=-\frac{1}{4}\varepsilon_{\mu\nu}^{\ \ \alpha\beta}(X^{-1})_{\alpha\beta}^{\ \ \gamma\lambda}\varepsilon_{\gamma\lambda}^{\ \ \rho\sigma}\,,
~~~~~~
\frac{1}{2}(X^{-1})_{\mu\nu}^{\ \ \rho\sigma}X_{\rho\sigma}^{\ \ \alpha\beta}\equiv I_{\mu\nu}^{\ \ \alpha\beta}\,.
\label{X-hat}
\ffa
And then, the EOM of the dual theory (\ref{ac-SB}) can be wrote down as
\fa
\nabla_{\nu}(\widehat{X}^{\mu\nu\rho\sigma}G_{\rho\sigma})=0\,.
\label{eom-Max-B}
\ffa

For the four dimensional standard Maxwell theory,
$\widehat{X}_{\mu\nu}^{\ \ \rho\sigma}=I_{\mu\nu}^{\ \ \rho\sigma}$.
At this moment, both the theories (\ref{ac-SA}) and (\ref{ac-SB})
are identical and so the standard Maxwell theory is self-dual.
Once the HD terms are introduced, such self-duality is broke.
However, if $\gamma_1$ is small, we have
\fa
&&
(X^{-1})_{\mu\nu}^{\ \ \rho\sigma}=I_{\mu\nu}^{\ \ \rho\sigma}+4\gamma_1C^2 I_{\mu\nu}^{\ \ \rho\sigma}+\mathcal{O}(\gamma_1^2)\,,
\label{Xin}
\\
&&
\widehat{X}_{\mu\nu}^{\ \ \rho\sigma}=(X^{-1})_{\mu\nu}^{\ \ \rho\sigma}+\mathcal{O}(\gamma_1^2)\,.
\ffa
It indicates that for small $\gamma_1$, there is an approximate duality between both theories (\ref{ac-SA}) and (\ref{ac-SB})
with the change of the sign of $\gamma_1$.

\section{Conductivity and EM duality}\label{sec-em-duality}

Turning on the perturbation
$
A_{y}(t,u)\sim e^{-i\omega t}A_{y}(u)
$,
we can write down the EOM of gauge field \cite{Myers:2010pk,Wu:2016jjd,Fu:2017oqa}
\fa
A''_y
+\Big(\frac{f'}{f}+\frac{X'_6}{X_6}\Big)A'_y
+\frac{\mathfrak{p}^2\hat{\omega}^2}{f^2}\frac{X_2}{X_6}A_y
=0\,,
\label{Ma-Ay}
\ffa
where $X_i$, $i=1,\ldots,6$, are the components of $X_{A}^{\ B}$ defined as
$
X_{A}^{\ B}=\{X_1(u),X_2(u),X_3(u),X_4(u),X_5(u),X_6(u)\}\,,
$
with
$
A,B\in\{tx,ty,tu,xy,xu,yu\}\,.
$
Due to the isotropy of background,
$X_1(u)=X_2(u)$ and $X_5(u)=X_6(u)$.
In addition, we have defined the dimensionless frequency in the above equation
$
\hat{\omega}\equiv\frac{\omega}{4\pi T}=\frac{\omega}{\mathfrak{p}}\,,
$
with
$
\mathfrak{p}\equiv p(1)=4\pi T
$.
Letting $X_i\rightarrow\widehat{X}_i=1/X_i$, we can obtain the dual EOM.
Solving the EOM (\ref{Ma-Ay}) or the dual EOM with ingoing boundary condition at the horizon,
we can read off the conductivity in terms of
\fa
\sigma(\hat{\omega})=\frac{\partial_uA_y(u,\hat{\omega})}{i\hat{\omega}\mathfrak{p} A_y(u,\hat{\omega})}\,.
\label{con-def}
\ffa

It has been illustrated in the last section that
for very small $\gamma_1$, there is an approximate duality between both the original EM theory and its dual theory with the change of the sign of $\gamma_1$
as that for $4$ derivative theory \cite{Myers:2010pk,Wu:2016jjd}.
At the same time, there is an inverse relation for the conductivity of the original theory and its dual theory\footnote{This relation has been proved in \cite{Myers:2010pk,WitczakKrempa:2012gn}.
Also it has been derived for a specific class of CFTs in \cite{Herzog:2007ij}.}\cite{Myers:2010pk,WitczakKrempa:2012gn}
\fa
\sigma_{\ast}(\hat{\omega};\hat{\alpha},\gamma_1)=\frac{1}{\sigma(\hat{\omega};\hat{\alpha},\gamma_1)}\,,
\label{sigma-B-A}
\ffa
where $\sigma_{\ast}$ is the conductivity of the dual theory.
Therefore, for small $\gamma_1$, we conclude that \cite{Wu:2016jjd}
\fa
\sigma_{\ast}(\hat{\omega};\hat{\alpha},\gamma_1)\approx\sigma(\hat{\omega};\hat{\alpha},-\gamma_1)
\,,~~~~~~~|\gamma_1|\ll1\,.
\label{sigma-v1}
\ffa
It indicates that the dual optical conductivity
is approximately equal to its original one with the change of the sign of $\gamma_1$.
For $4$ derivative theory, it has been studied and found that the conductivity of the
dual EM theory is not precisely equal to that of its original theory with the change of the sign of $\gamma$ \cite{Myers:2010pk}.
But when the homogeneous disorder is introduced, we find that
this relation holds more well for a specific value of $\hat{\alpha}=2/\sqrt{3}$ than other value of $\hat{\alpha}$.
Here, we shall study the EM duality from $6$ derivative theory.
\begin{figure}
\center{
\includegraphics[scale=0.58]{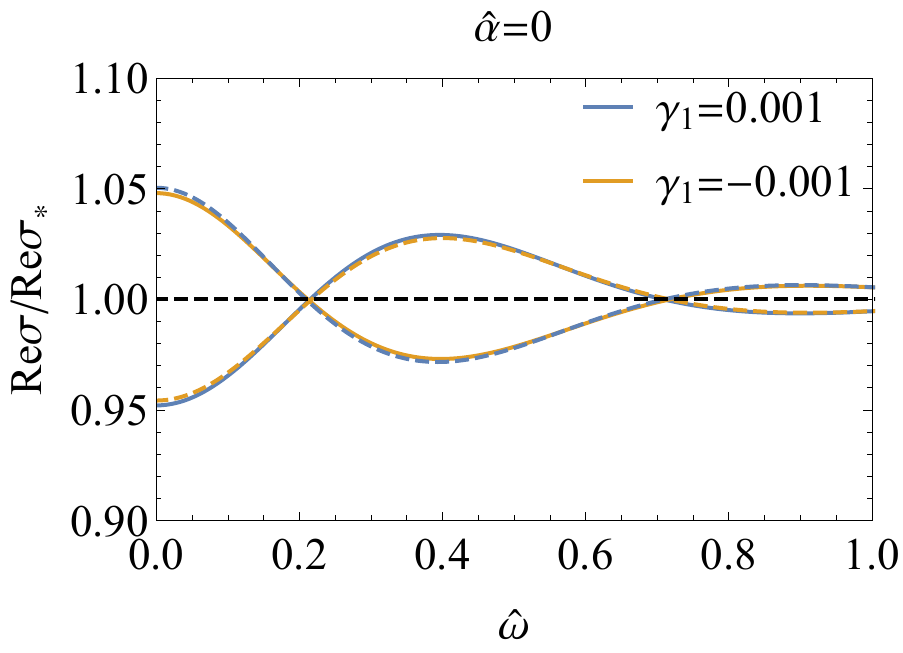}\ \hspace{0.1cm}
\includegraphics[scale=0.55]{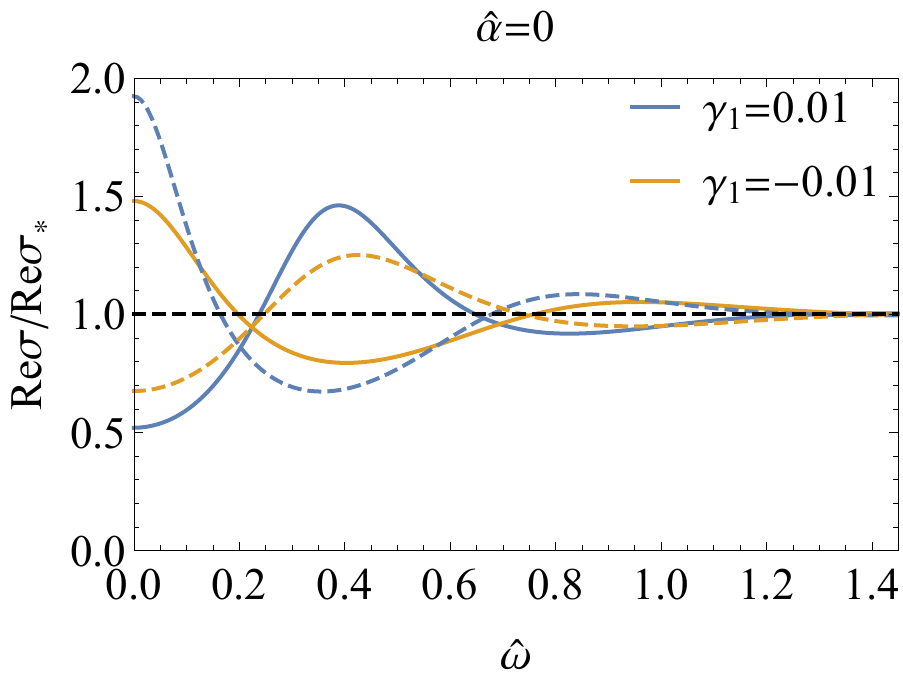}\ \hspace{0.1cm}
\includegraphics[scale=0.53]{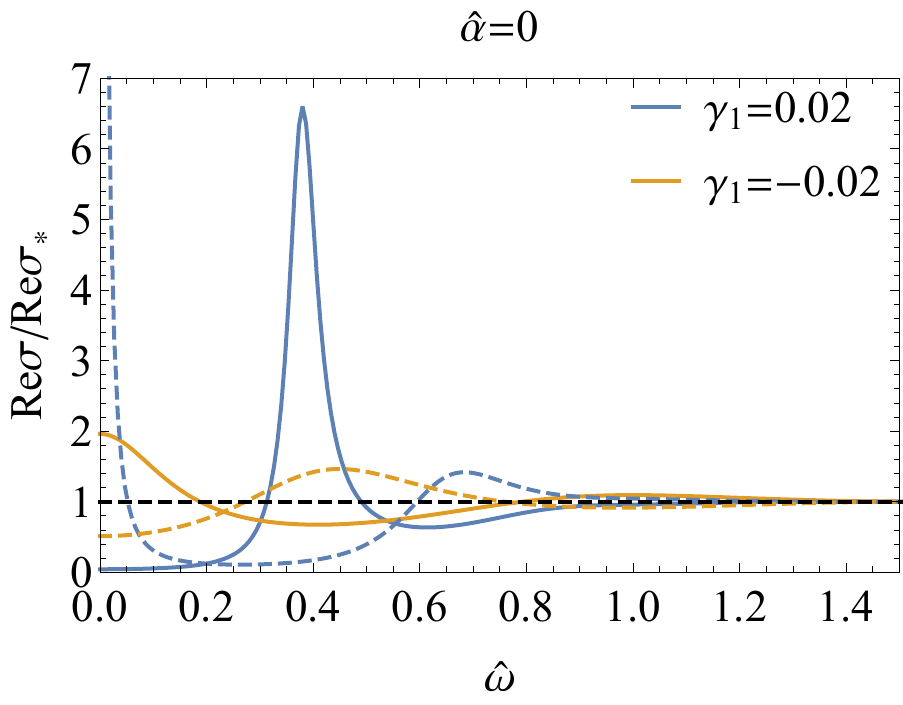}\ \\
\includegraphics[scale=0.58]{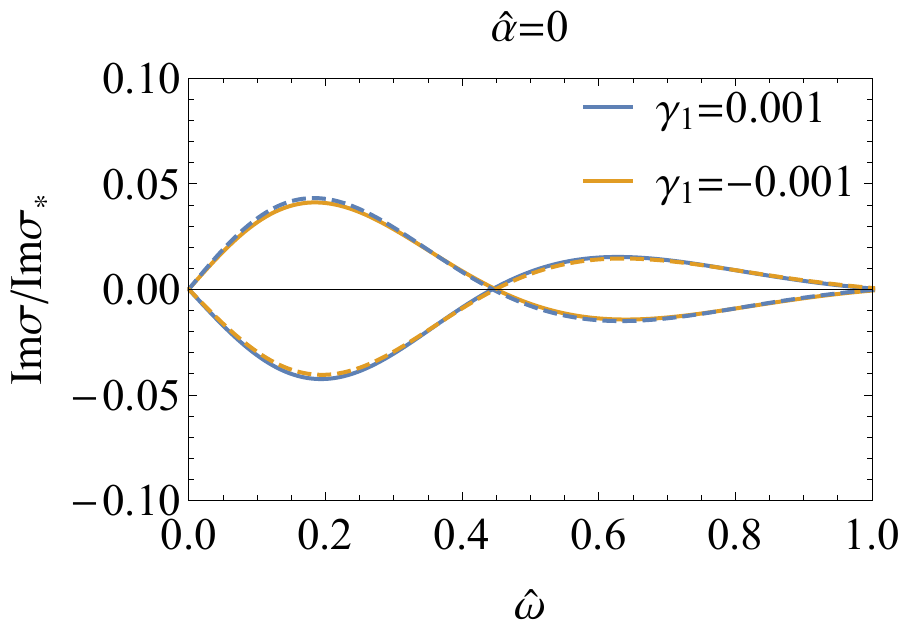}\ \hspace{0.1cm}
\includegraphics[scale=0.55]{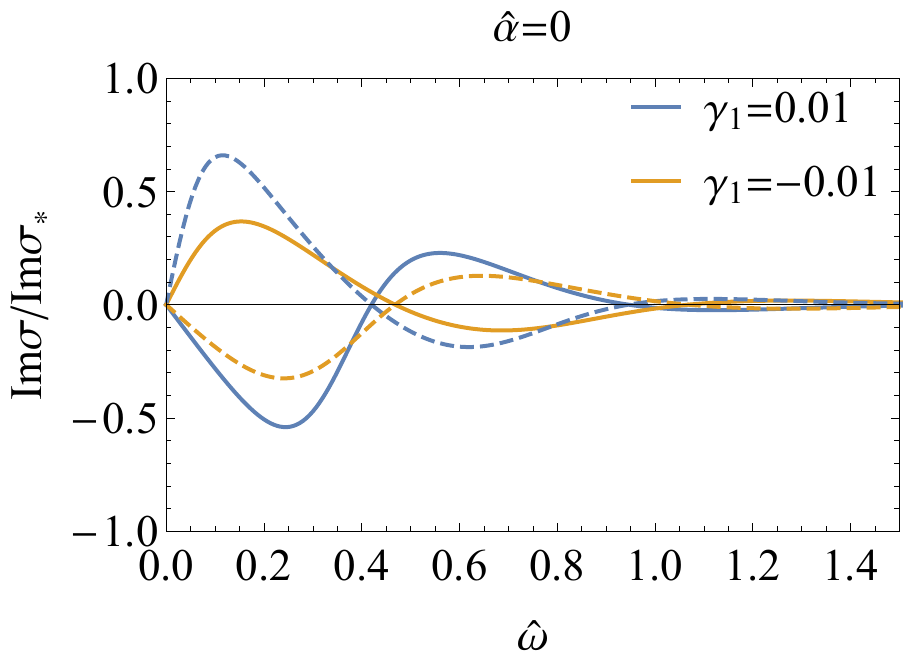}\ \hspace{0.1cm}
\includegraphics[scale=0.53]{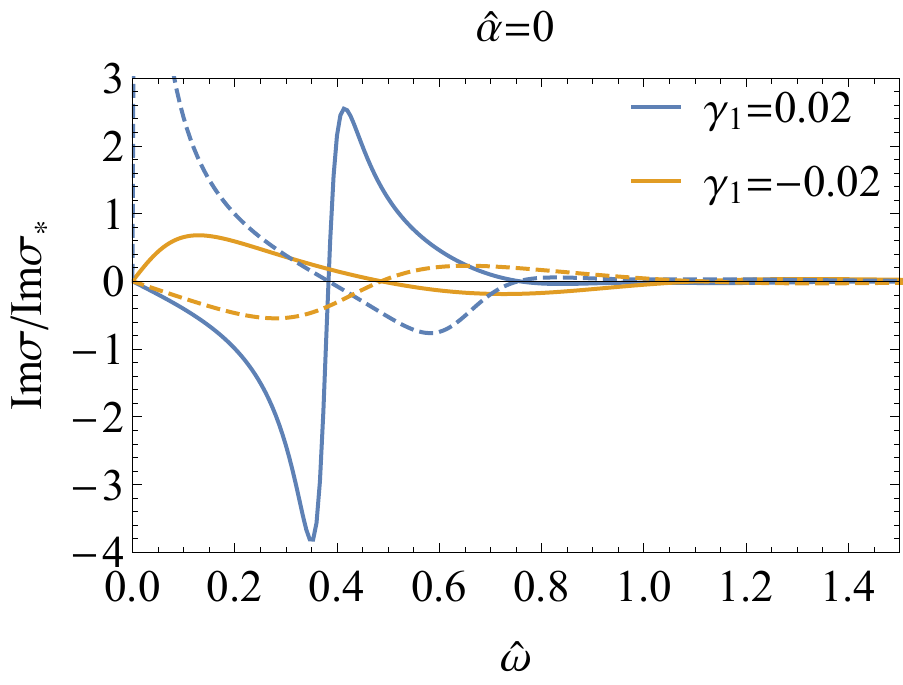}\ \\
}
\caption{\label{fig_em_con_alpha_0}The real part (the plot above) and the imaginary part (the plot below) of the optical conductivity as the function of
$\hat{\omega}$ for  $\hat{\alpha}=0$ and various values of $\gamma_1$.
The solid curves are the conductivity $\sigma$ of the original EM theory (\ref{ac-SA}) and
the dashed curves display the conductivity $\sigma_{\ast}$ of the EM dual theory (\ref{ac-SB}).
}
\end{figure}

FIG.\ref{fig_em_con_alpha_0} exhibits the optical conductivity as the function of
$\hat{\omega}$ for $\hat{\alpha}=0$ and various values of $\gamma_1$.
We find that for small $|\gamma_1|$ ($|\gamma_1|=0.001$), with the change of the sign of $\gamma_1$,
the relation \eqref{sigma-v1} holds very well.
With the increase of $|\gamma_1|$, the relation \eqref{sigma-v1} violates.
In particular for $\gamma_1=0.02$, the real part of the conductivity at low frequency is a pseudogap-like behavior.
Correspondingly, it dual conductivity $\texttt{Re}\sigma_{\ast}$ at low frequency becomes sharper, which is Drude-like.
But for $\gamma_1=-0.02$, the real parts of the conductivity of the original theory and its dual one at low frequency are only
peak and dip, respectively. At this moment, the relation \eqref{sigma-v1} is strongly violated.
\begin{figure}
\center{
\includegraphics[scale=0.51]{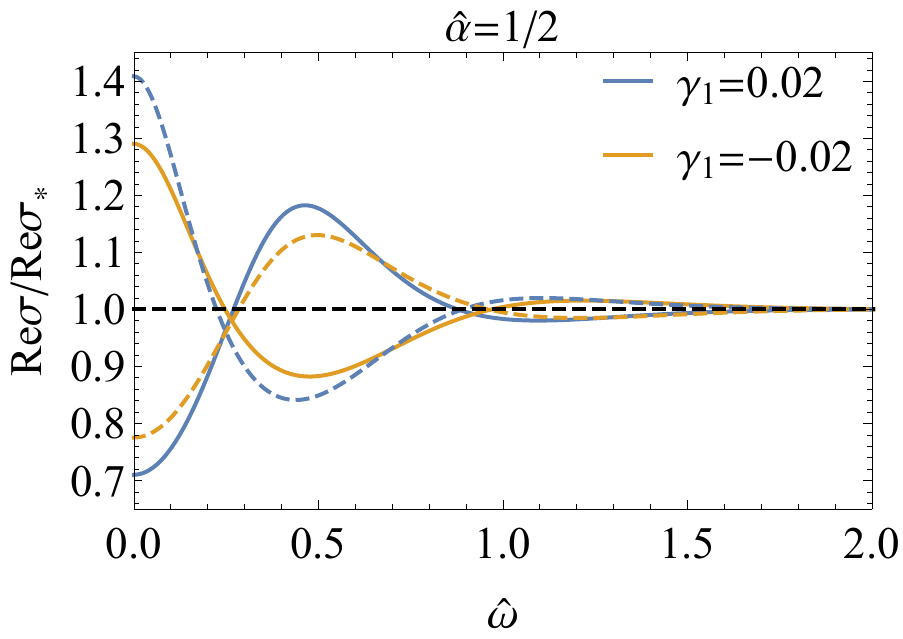}\ \hspace{0.4cm}
\includegraphics[scale=0.52]{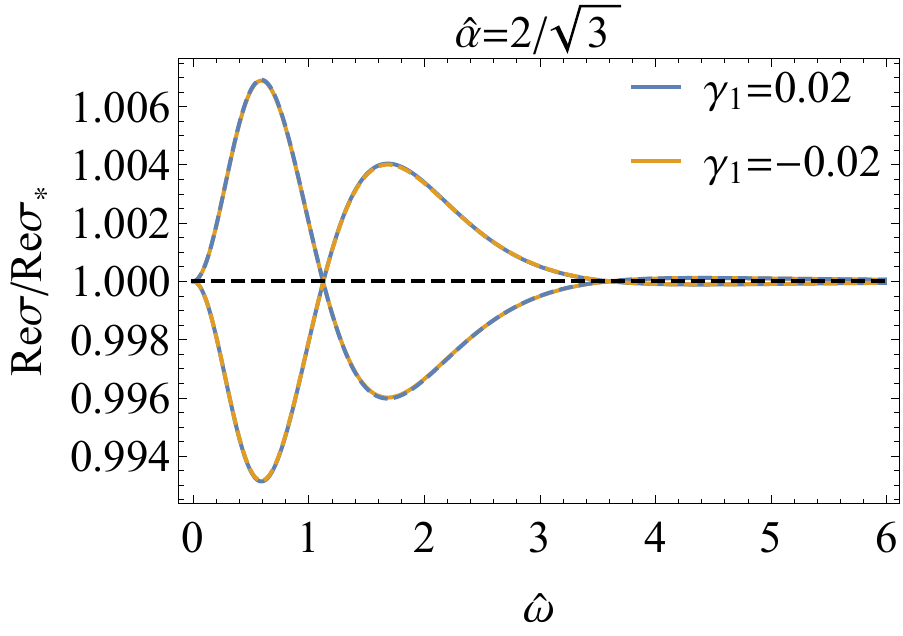}\ \hspace{0.4cm}
\includegraphics[scale=0.5]{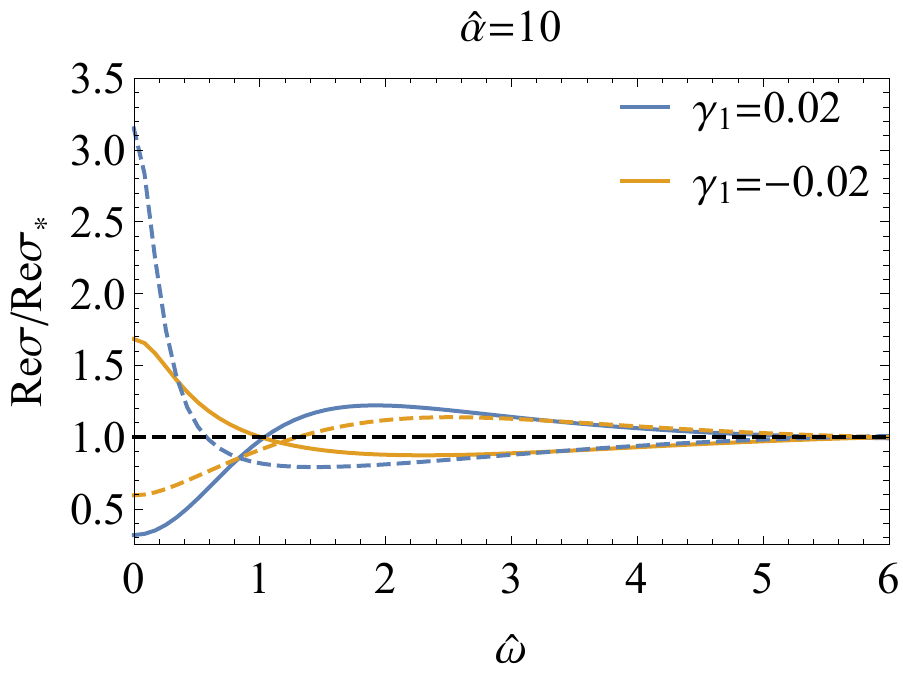}\ \\
\includegraphics[scale=0.51]{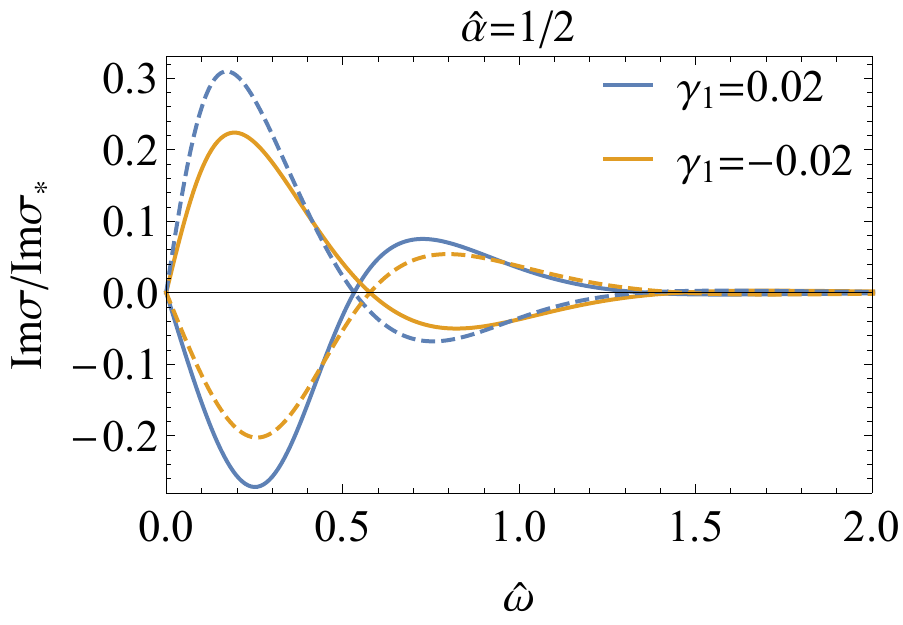}\ \hspace{0.4cm}
\includegraphics[scale=0.52]{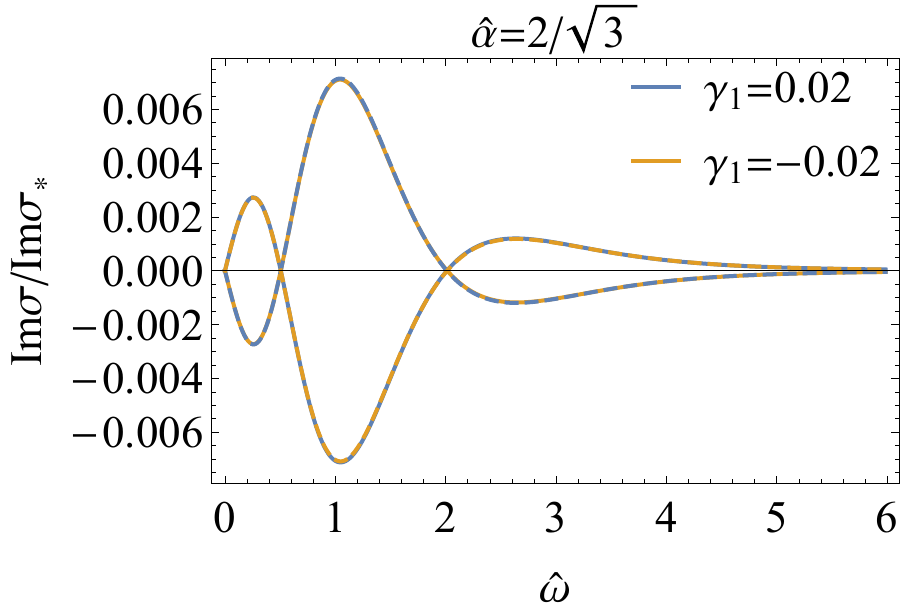}\ \hspace{0.4cm}
\includegraphics[scale=0.5]{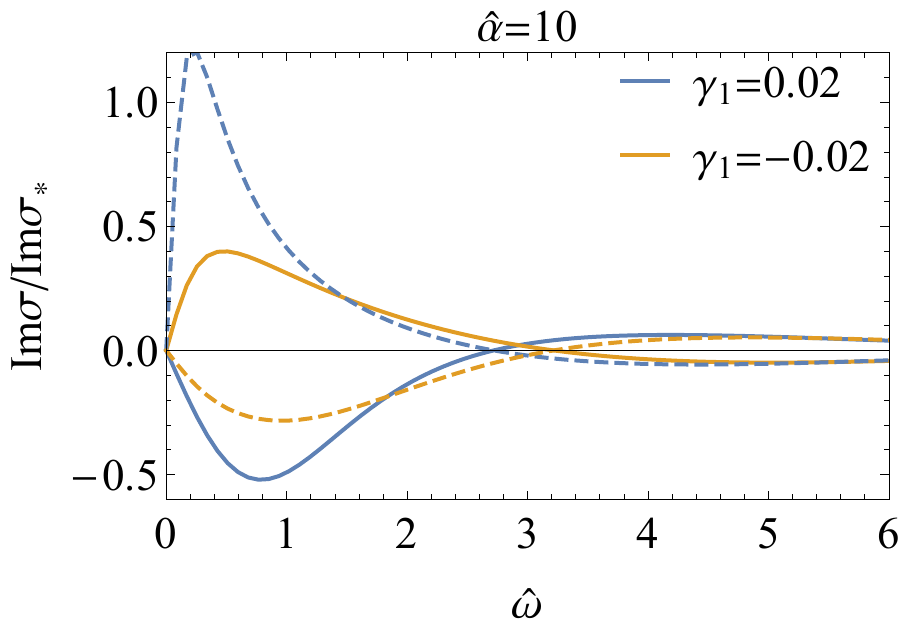}\ \\
}
\caption{\label{fig_em_con_diff_alpha}The real part (the plot above) and the imaginary part (the plot below) of the optical conductivity as the function of
$\hat{\omega}$ for $|\gamma_1|=0.02$ and various values of $\hat{\alpha}$.
The solid curves are the conductivity $\sigma$ of the original EM theory (\ref{ac-SA}) and
the dashed curves display the conductivity $\sigma_{\ast}$ of the EM dual theory (\ref{ac-SB}).}
\end{figure}

Next, we explore the EM duality for $6$ derivative theory when the homogeneous disorder is introduced.
At the first sight, we find that for the specific value of $\hat{\alpha}=2/\sqrt{3}$,
the optical conductivity of the original theory is almost (approximate but not exact)
the same as that of the dual one when the sign of $\gamma_1$ changes,
i.e., the relation \eqref{sigma-v1} holds very well. It is similar to that from $4$ derivative theory \cite{Wu:2016jjd}.
For other value of $\hat{\alpha}$, the relation \eqref{sigma-v1} also approximately holds.
It seems that the homogeneous disorder make the pseudogap-like of the low frequency conductivity of the original theory becomes small dip,
while suppresses the sharp peak of the EM dual theory.

\begin{figure}
\center{
\includegraphics[scale=0.65]{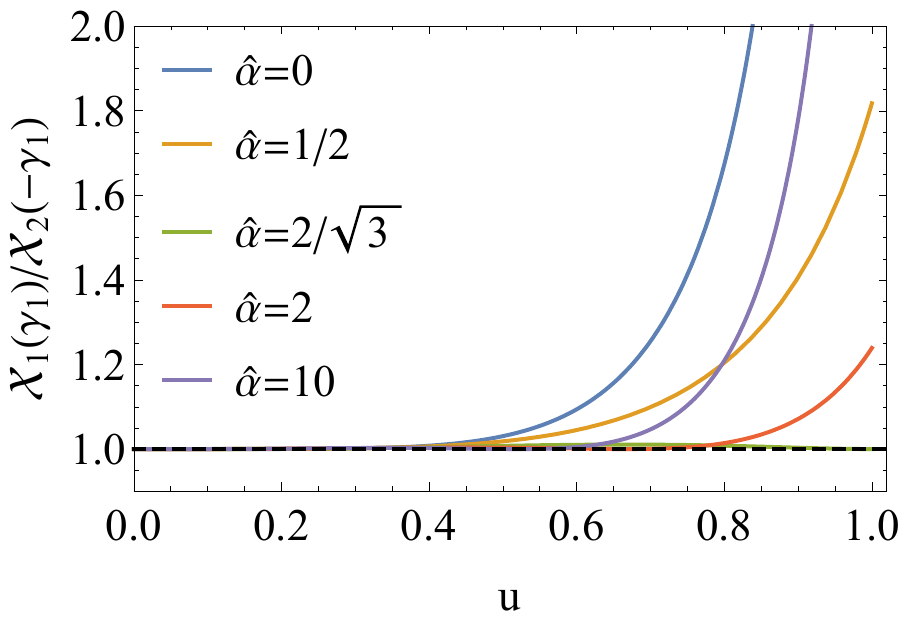}\ \\
\caption{\label{fig-x1vsx2-hd} $\mathcal{X}_1(\gamma_1)/\mathcal{X}_2(-\gamma_1)$ as the function of $u$ with different $\hat{\alpha}$.
Here we set $\gamma_1=0.02$.}}
\end{figure}

It is hard to obtain an analytical understanding on this phenomenon.
But we can follow the method of \cite{Wu:2016jjd} and attempt to taste kind of analytical similarity between the EOM
\eqref{Ma-Ay} and its dual one, which can be explicitly wrote down
\fa
A''_y
+\Big(\frac{f'}{f}-\frac{X'_6}{X_6}\Big)A'_y
+\frac{\mathfrak{p}^2}{f^2}\frac{X_6}{X_2}\hat{\omega}^2A_y
=0\,.
\label{Max-Ay-d-v2}
\ffa
Since $\frac{X_2}{X_6}=\frac{X_6}{X_2}$ for any $\hat{\alpha}$, the difference between Eq.\eqref{Ma-Ay} and \eqref{Max-Ay-d-v2} is in the coefficients of $A'_y$, which are
\fa
&&
\mathcal{X}_1(\gamma_1)\equiv\frac{X'_6}{X_6}=\frac{-\frac{16}{3} \gamma _1 u^3 f''(u)^2-\frac{8}{3} \gamma _1 u^4 f^{(3)}(u) f''(u)}{1-\frac{4}{3}
   \gamma _1 u^4 f''(u)^2}\,,\label{mathX1}\\
&&
\mathcal{X}_2(\gamma_1)\equiv-\frac{X'_6}{X_6}=-\frac{-\frac{16}{3} \gamma _1 u^3 f''(u)^2-\frac{8}{3} \gamma _1 u^4 f^{(3)}(u) f''(u)}{1-\frac{4}{3}
   \gamma _1 u^4 f''(u)^2}\,.\label{mathX2}
\ffa
And then, we explicitly plot the ration $\frac{\mathcal{X}_1}{\mathcal{X}_2}\Big|_{\gamma_1=0.02}$ as the function of $u$ with different $\hat{\alpha}$ in FIG.\ref{fig-x1vsx2-hd}.
For $\hat{\alpha}=0$, the value of $\frac{\mathcal{X}_1}{\mathcal{X}_2}$ is maximum departure from $1$ near the horizon,
which indicates a most deviation between \eqref{Ma-Ay} and \eqref{Max-Ay-d-v2}.
Once the homogeneous disorder is introduced, this deviation become small.
Specially, for $\hat{\alpha}=2/\sqrt{3}$, this value approaches to $1$ near the horizon, which implies a most similar between \eqref{Ma-Ay} and \eqref{Max-Ay-d-v2}.
On the other hand, it is well known that the low frequency behavior of the conductivity is mainly controlled by the near horizon geometry.
Therefore, to some extent, the above comparison in the original EOM and its dual one helps us understand the phenomenon shown in FIG.\ref{fig_em_con_diff_alpha}.

\section{The properties of QNMs}\label{sec-qnm}

By definition, QNMs are the poles of the Green's function and are directly related to the conductivity.
Summing an increasing number of QNMs and the corresponding residues one should obtain
a better and better approximation of the conductivity itself.
In this section, we shall study the properties of QNMs of the transverse gauge mode along $y$ direction, i.e., $A_y$, from $6$ derivative term with homogeneous disorder.
In particular, we shall also study the particle-vortex duality in the complex frequency panel in the holographic CFT.

It is high efficient to use the pseudospectral methods outlined in \cite{Jansen:2017oag} to solve the QNM equations
and obtain the QNMs. To this end, we shall work in the advanced Eddington-Finkelstein coordinate, which is
\fa
ds^2=\frac{1}{u^2}\Big(-f(u)dt^2-2dtdu+dx^2+dy^2\Big)\,.
\label{dsAEF}
\ffa
Correspondingly, the EOM (\ref{Ma-Ay}) can be changed as
\fa
A_y' \left(\frac{f'}{f}+\frac{2 i \mathfrak{p} \hat{\omega} }{f}+\frac{X_6'}{X_6}\right)+\frac{i \mathfrak{p} \hat{\omega}
   A_y X_2'}{f X_6}+A_y''
   =0\,.
   \label{Ma-Ay-r}
\ffa
For the dual theory, the corresponding EOM can be obtained by letting $A_{\mu}\rightarrow B_{\mu}$ and $X_i\rightarrow \widehat{X}_i=1/X_i$.
Imposing the ingoing boundary at the horizon, we numerically solve Eq. \eqref{Ma-Ay-r} to obtain QNMs.
\begin{figure}
\center{
\includegraphics[scale=0.24]{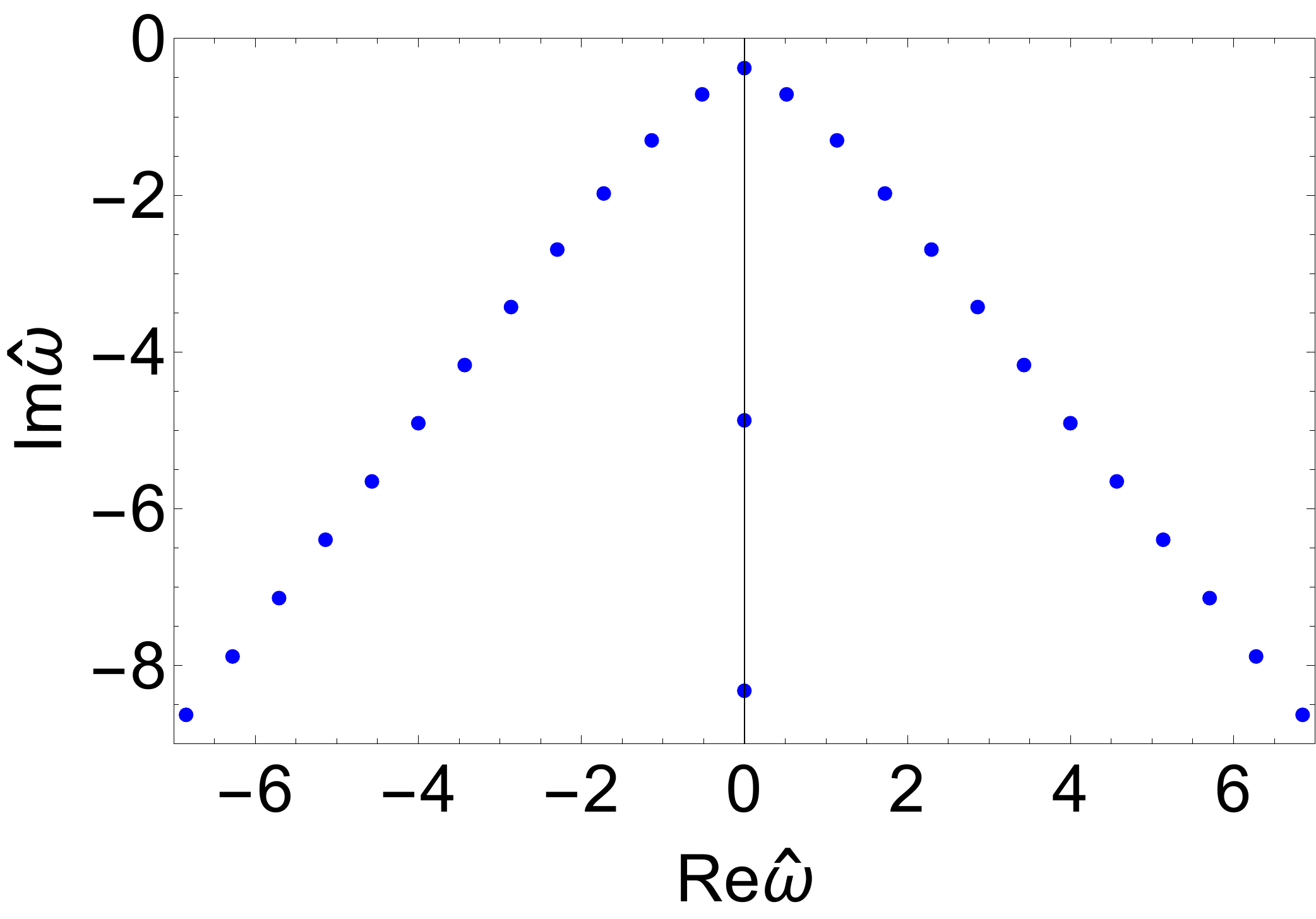}\ \hspace{0.1cm}
\includegraphics[scale=0.24]{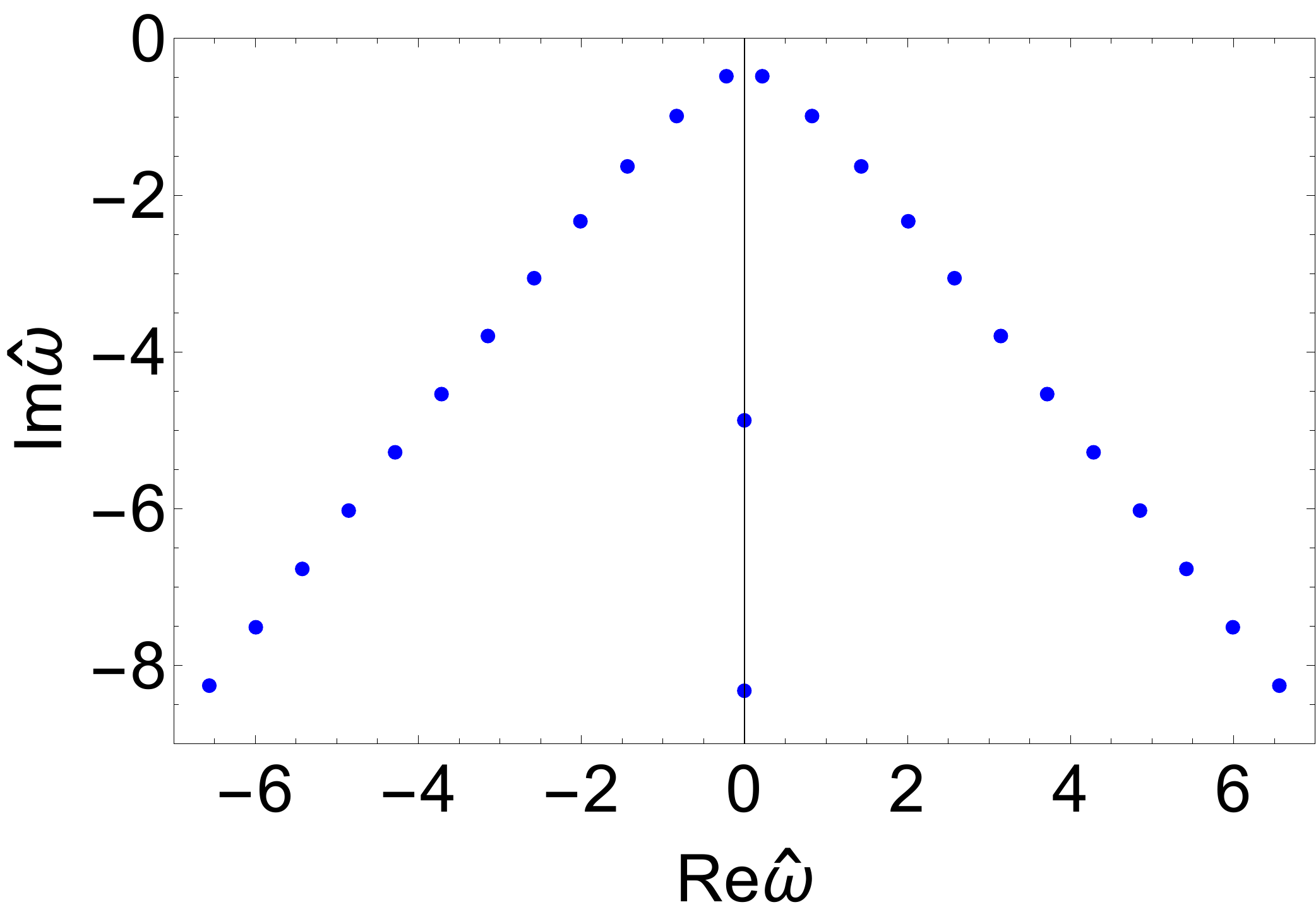}\ \\
\includegraphics[scale=0.24]{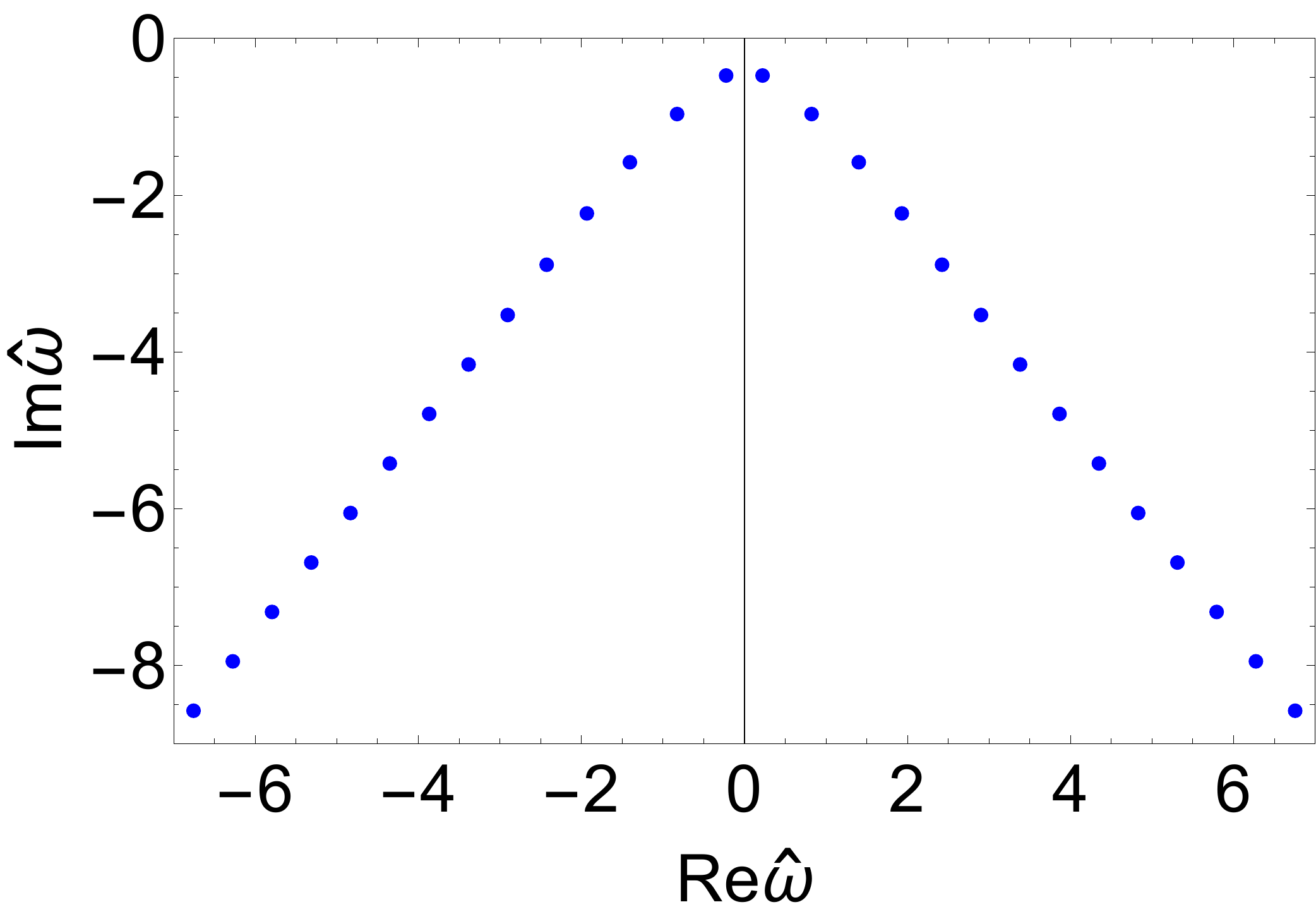}\ \hspace{0.1cm}
\includegraphics[scale=0.24]{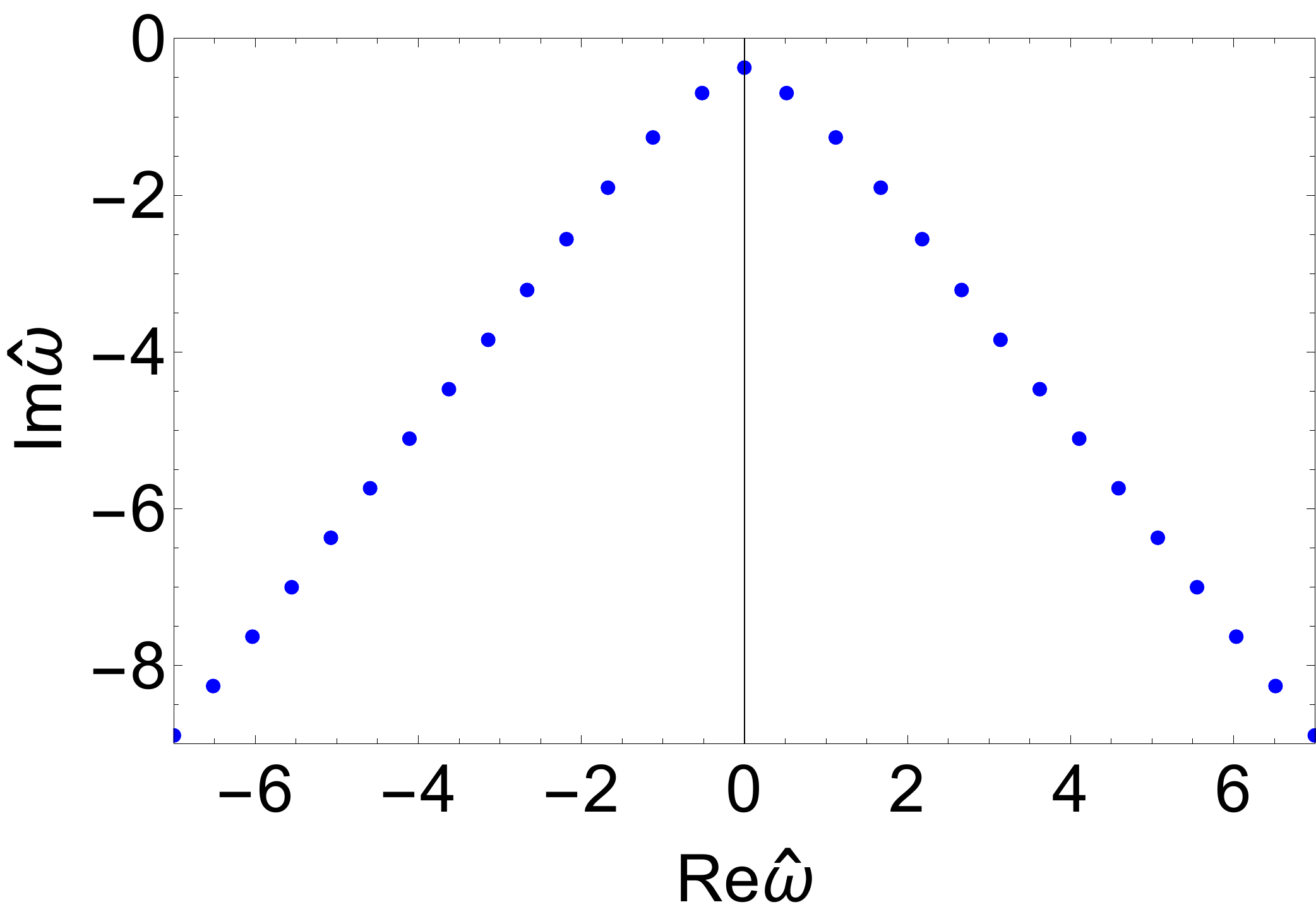}\ \\}
\caption{\label{fig_beta1_0p001} QNMs (blue spots) of the gauge mode for $|\gamma_1|=0.001$ (the panels above are for $\gamma_1=-0.001$
and the ones below for $\gamma_1=-0.001$) and $\hat{\alpha}=0$.
The left panels are the QNMs of the gauge mode and the right ones are that of the dual gauge mode.}
\end{figure}
\begin{figure}
\center{
\includegraphics[scale=0.24]{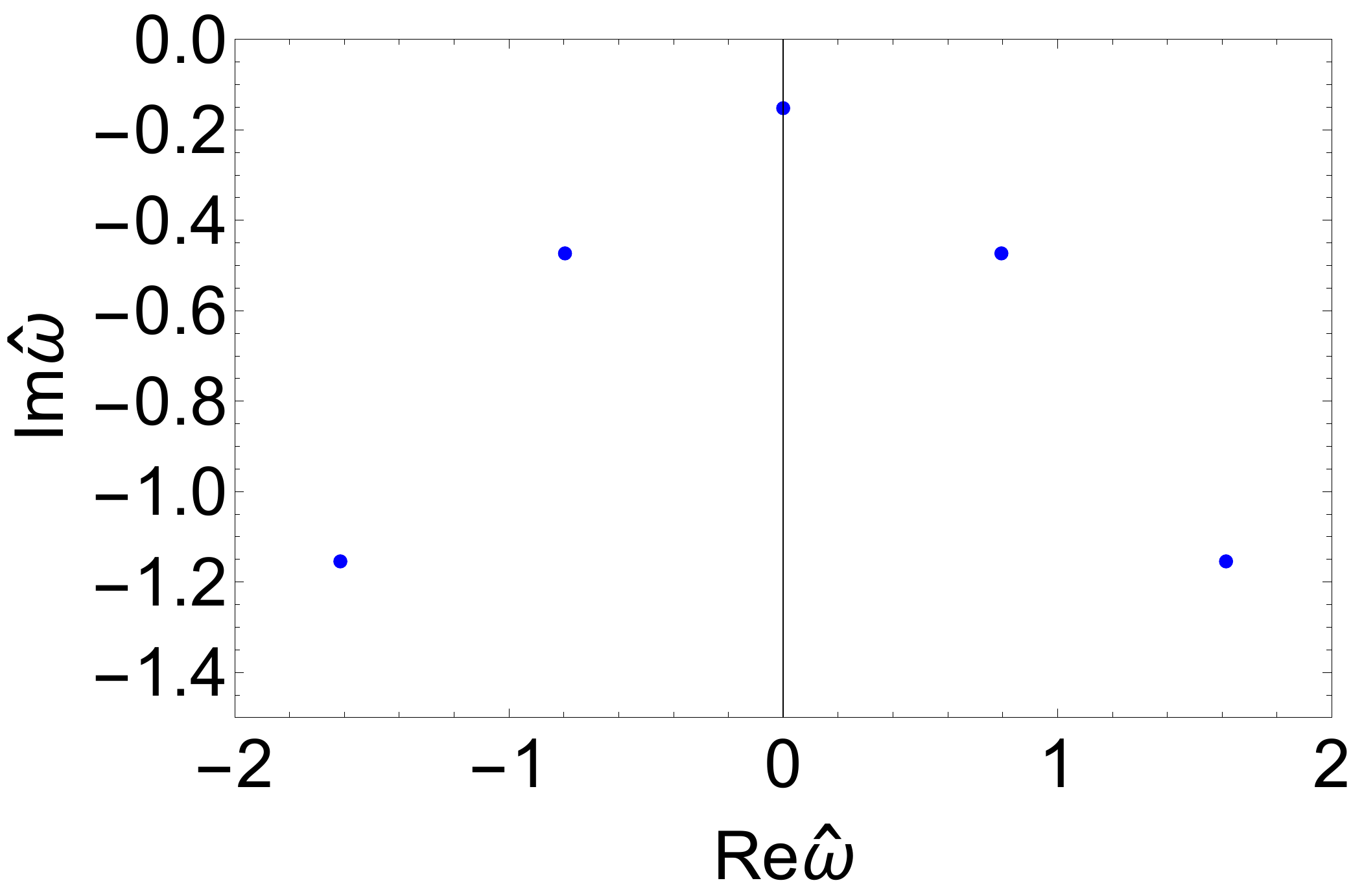}\ \hspace{0.1cm}
\includegraphics[scale=0.24]{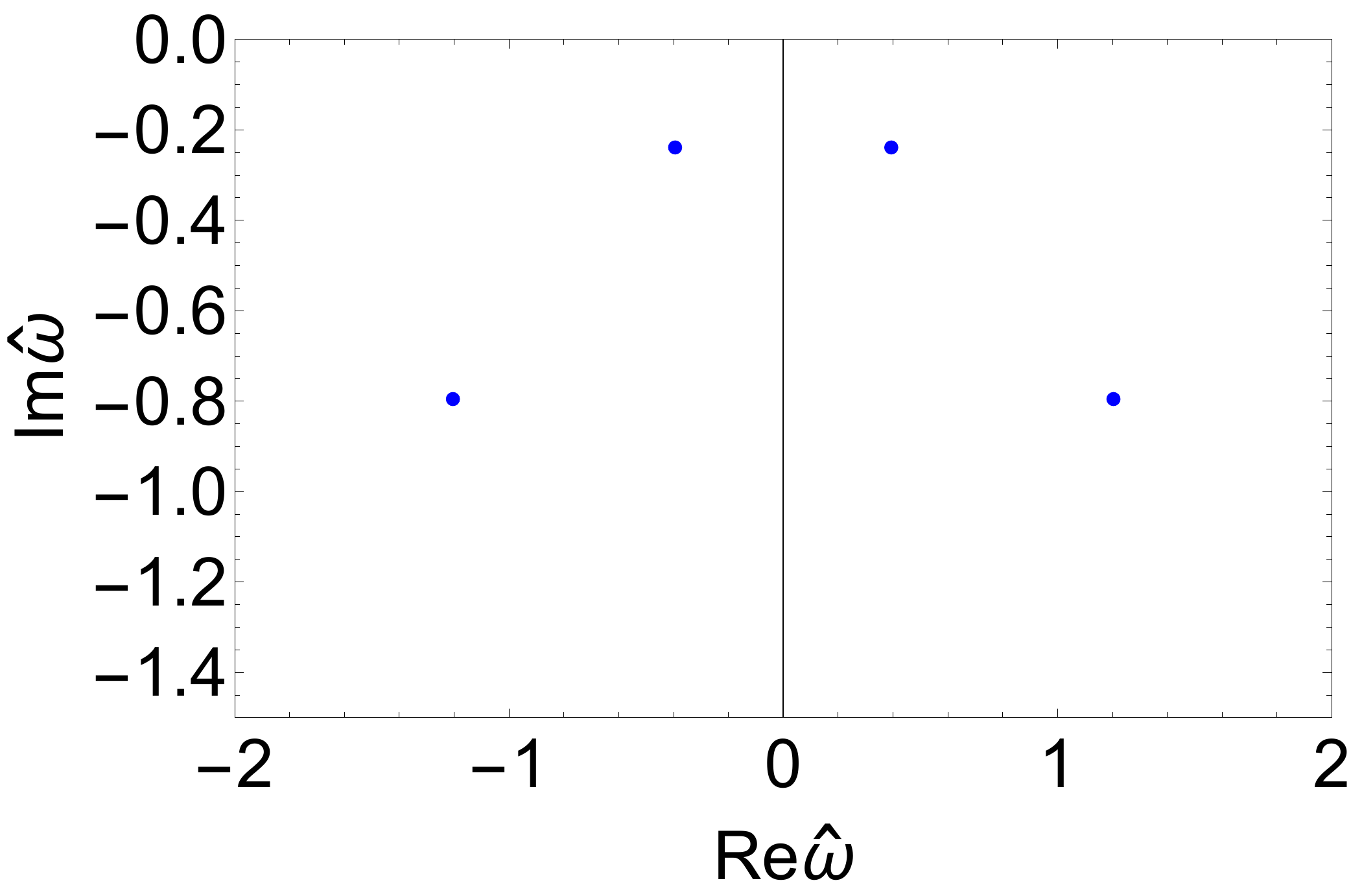}\ \\
\includegraphics[scale=0.24]{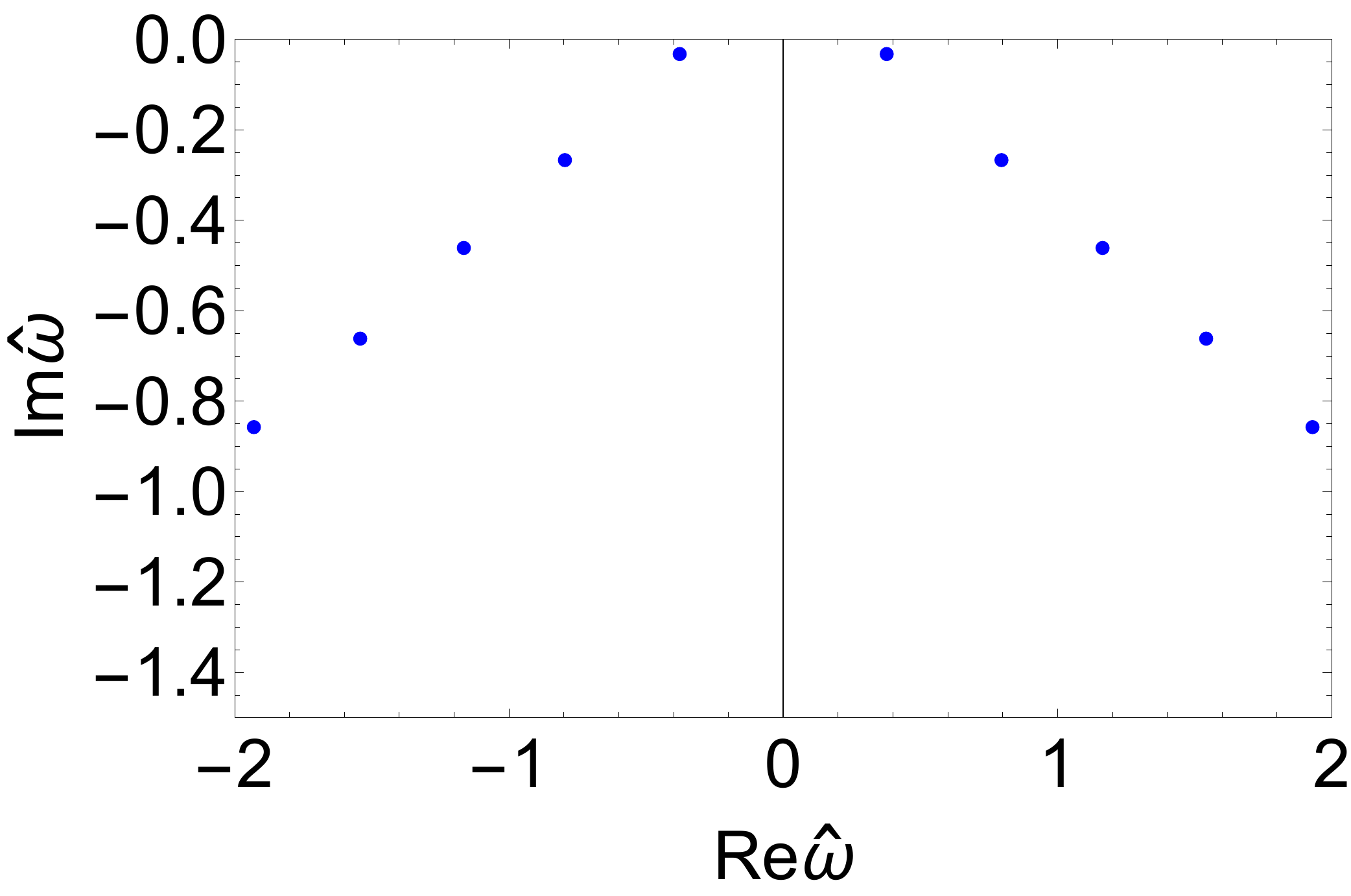}\ \hspace{0.1cm}
\includegraphics[scale=0.24]{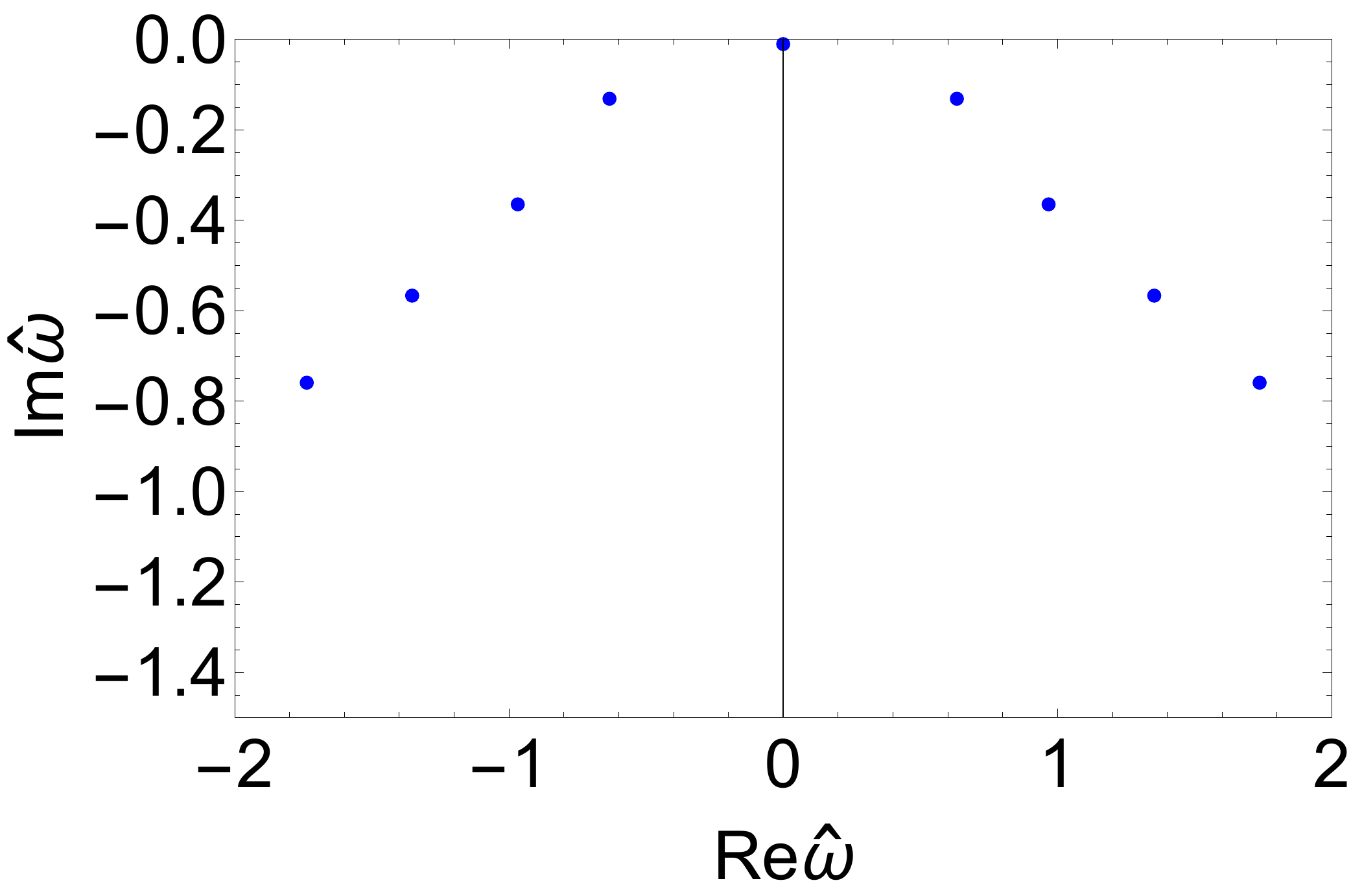}\ \\}
\caption{\label{fig_beta1_0p02} QNMs (blue spots) of the gauge mode for $|\gamma_1|=0.02$ (the panels above are for $\gamma_1=-0.02$
and the ones below for $\gamma_1=-0.02$) and $\hat{\alpha}=0$.
The left panels are the QNMs of the gauge mode and the right ones are that of the dual gauge mode.}
\end{figure}
\begin{figure}
\center{
\includegraphics[scale=0.24]{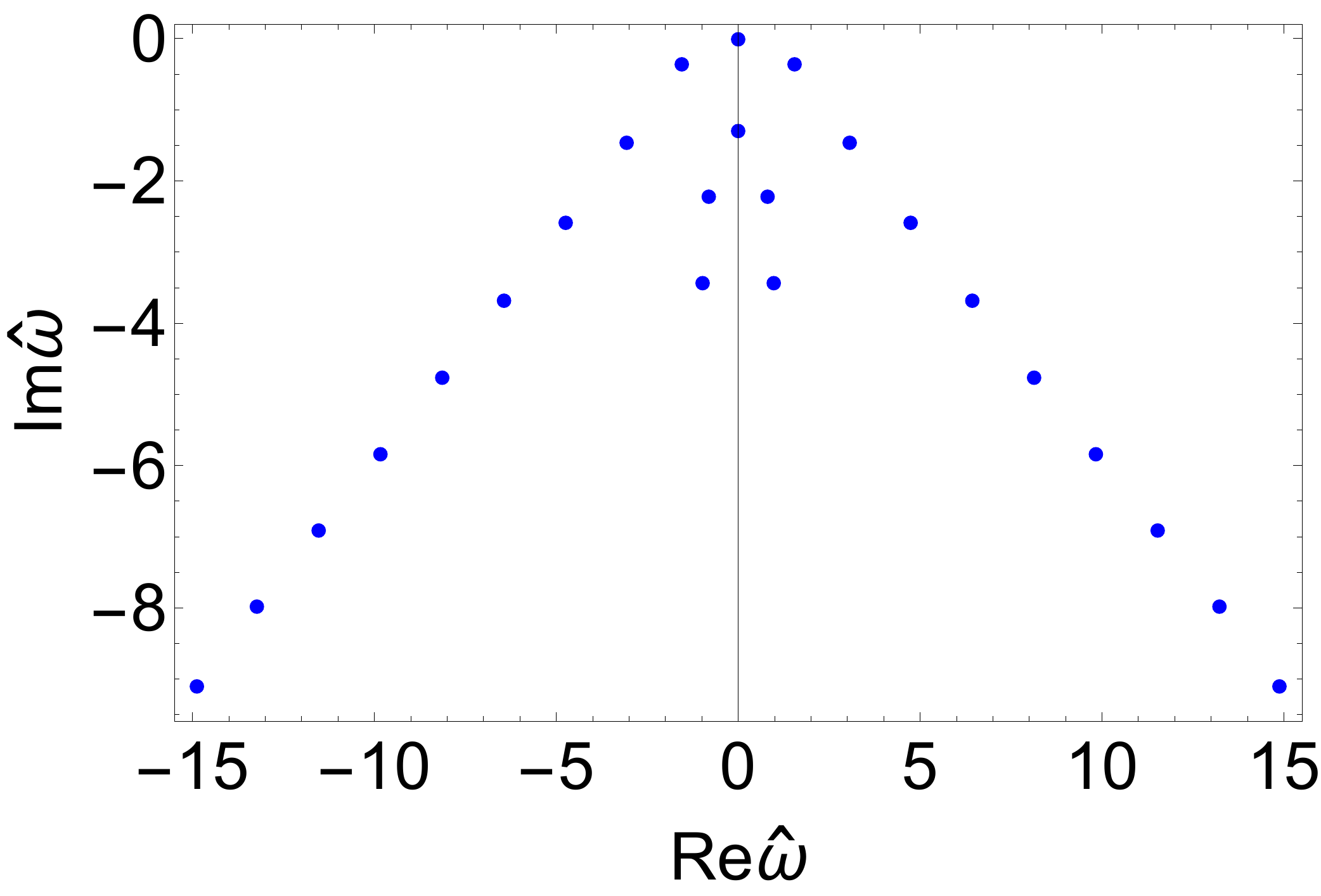}\ \hspace{0.1cm}
\includegraphics[scale=0.23]{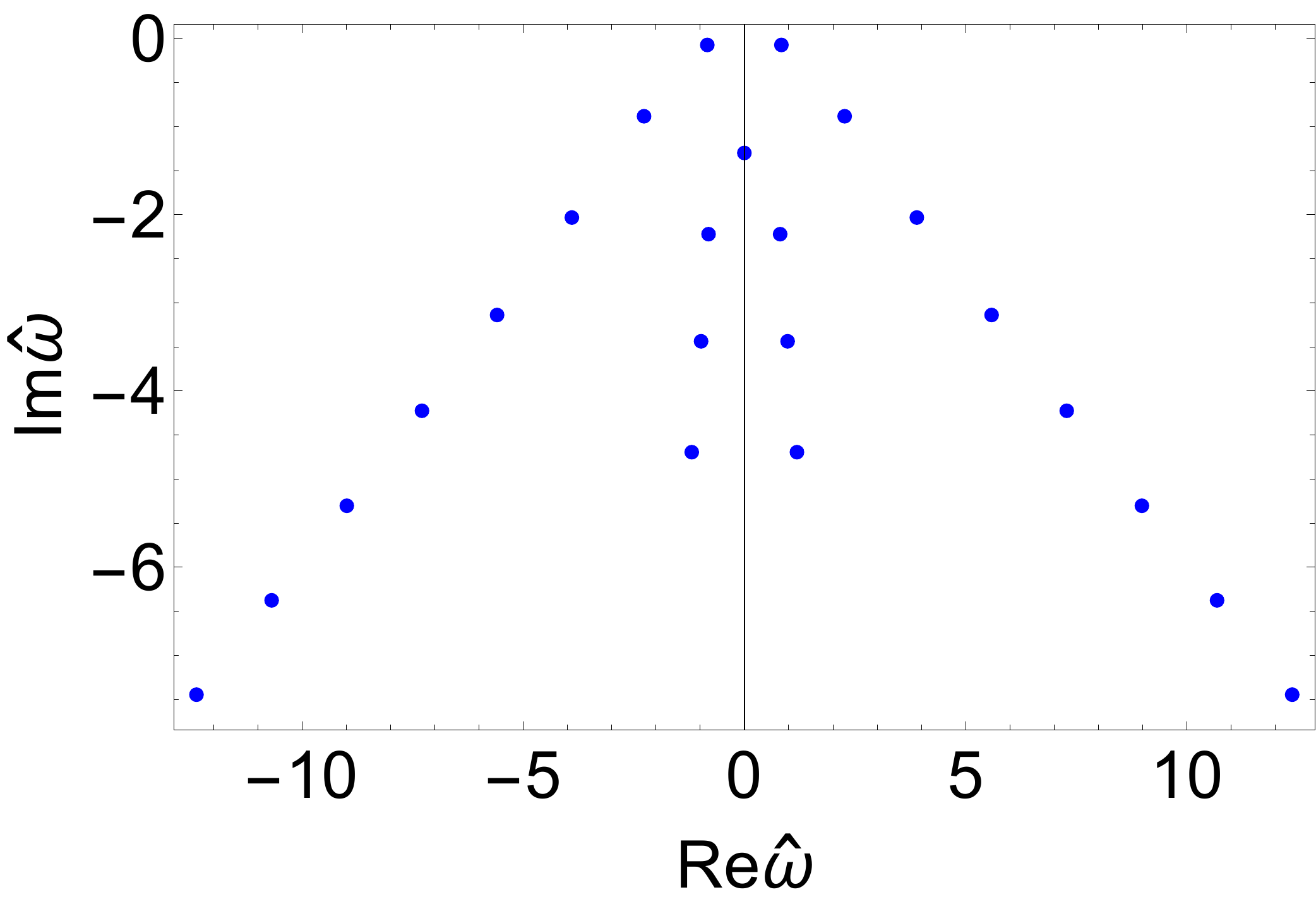}\ \\}
\caption{\label{fig_beta1_n1} QNMs (blue spots) of the gauge mode for $\gamma_1=-1$ and $\hat{\alpha}=0$.
The left panels are the QNMs of the gauge mode and the right ones are that of the dual gauge mode.}
\end{figure}

We first study the case without homogeneous disorder.
FIG.\ref{fig_beta1_0p001} and \ref{fig_beta1_0p02} show QNMs of the gauge mode with $\hat{\alpha}=0$ for $|\gamma_1|=0.001$ and $0.02$, respectively.
The left panels are the QNMs of the gauge mode and the right ones are that of the dual gauge mode.
We find that for small $|\gamma_1|$, for example, $|\gamma_1|=0.001$ in FIG.\ref{fig_beta1_0p001},
the qualitative correspondence between the poles of $\texttt{Re}\sigma(\hat{\omega};\gamma_1)$
and the ones of $\texttt{Re}\sigma_{\ast}(\hat{\omega};-\gamma_1)$ holds well at low frequency.
But for the large frequency region, there are some modes emerging at the imaginary frequency axis for negative $\gamma_1$,
which results in the violation of the particle-vortex duality with the change of the sign of $\gamma_1$.
But with the increase of $\gamma_1$, this correspondence begins to violate in complex frequency panel
even in the low frequency region (see FIG.\ref{fig_beta1_0p02} for $|\gamma_1|=0.02$), as that in real frequency axis.
We also plot the QNMs for $\gamma_1=-1$ in FIG.\ref{fig_beta1_n1}.
We see that another branch cuts, which are closer to the imaginary frequency axis, emerges in the complex frequency panel.
\begin{figure}
\center{
\includegraphics[scale=0.24]{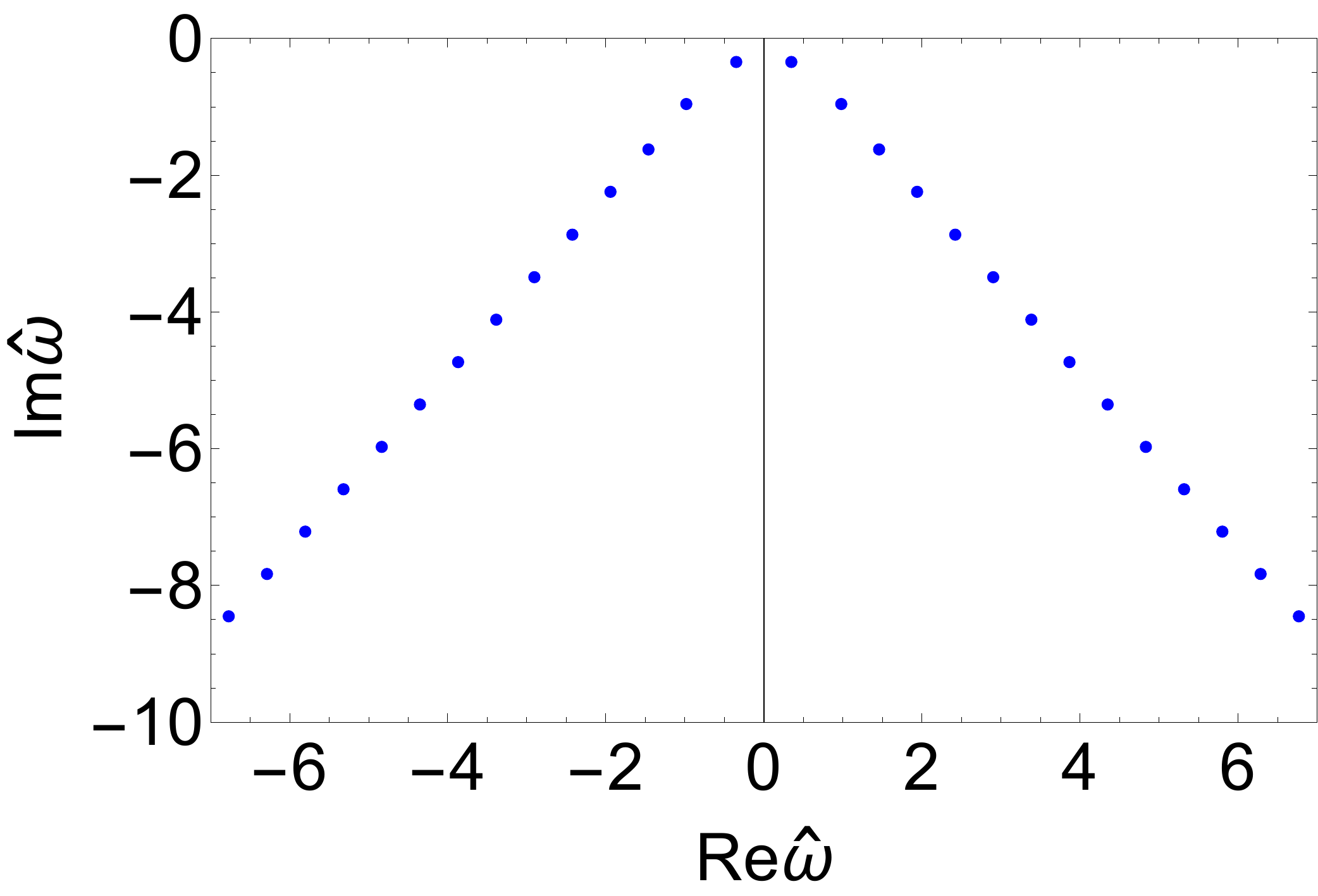}\ \hspace{0.1cm}
\includegraphics[scale=0.24]{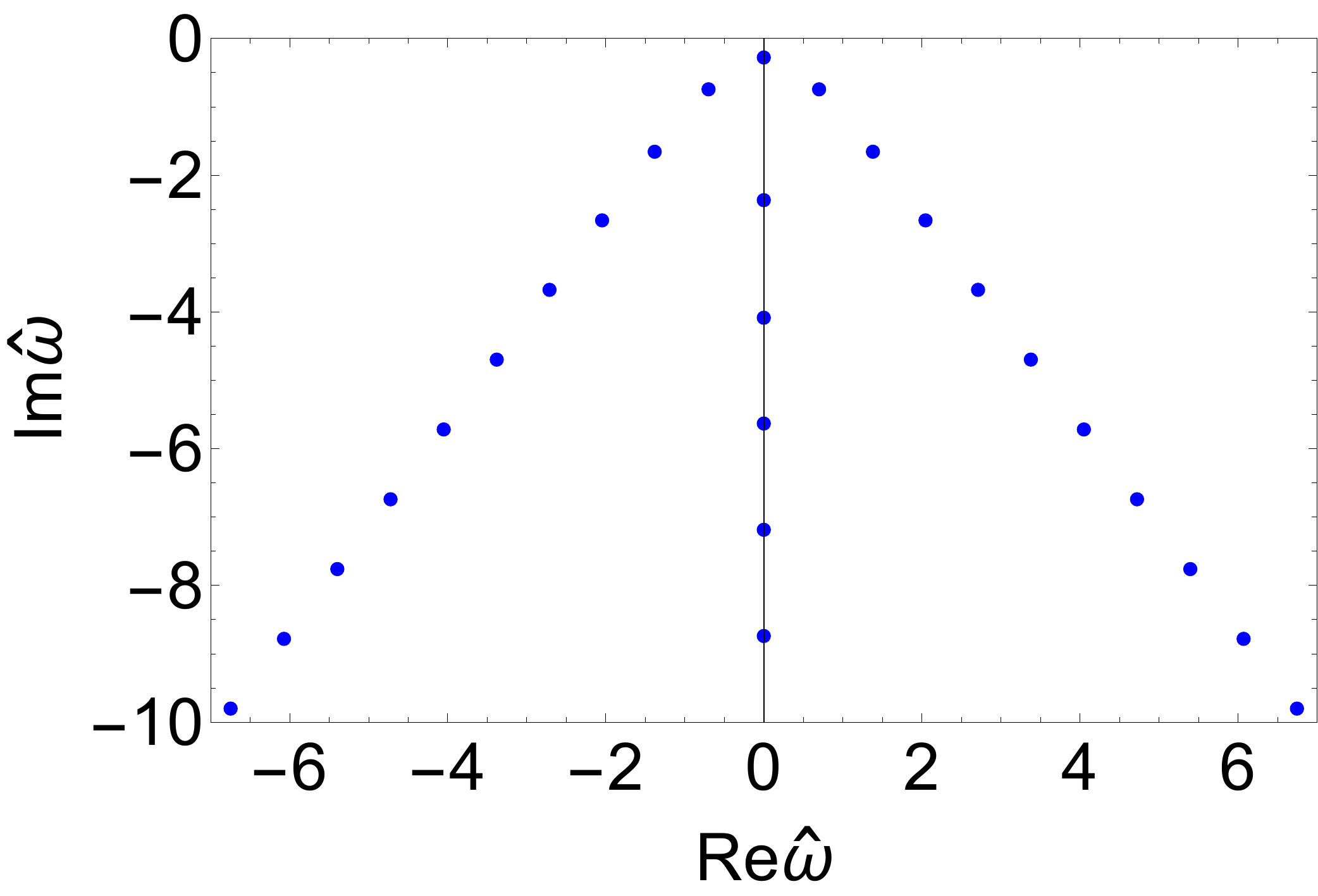}\ \hspace{0.1cm}
\includegraphics[scale=0.24]{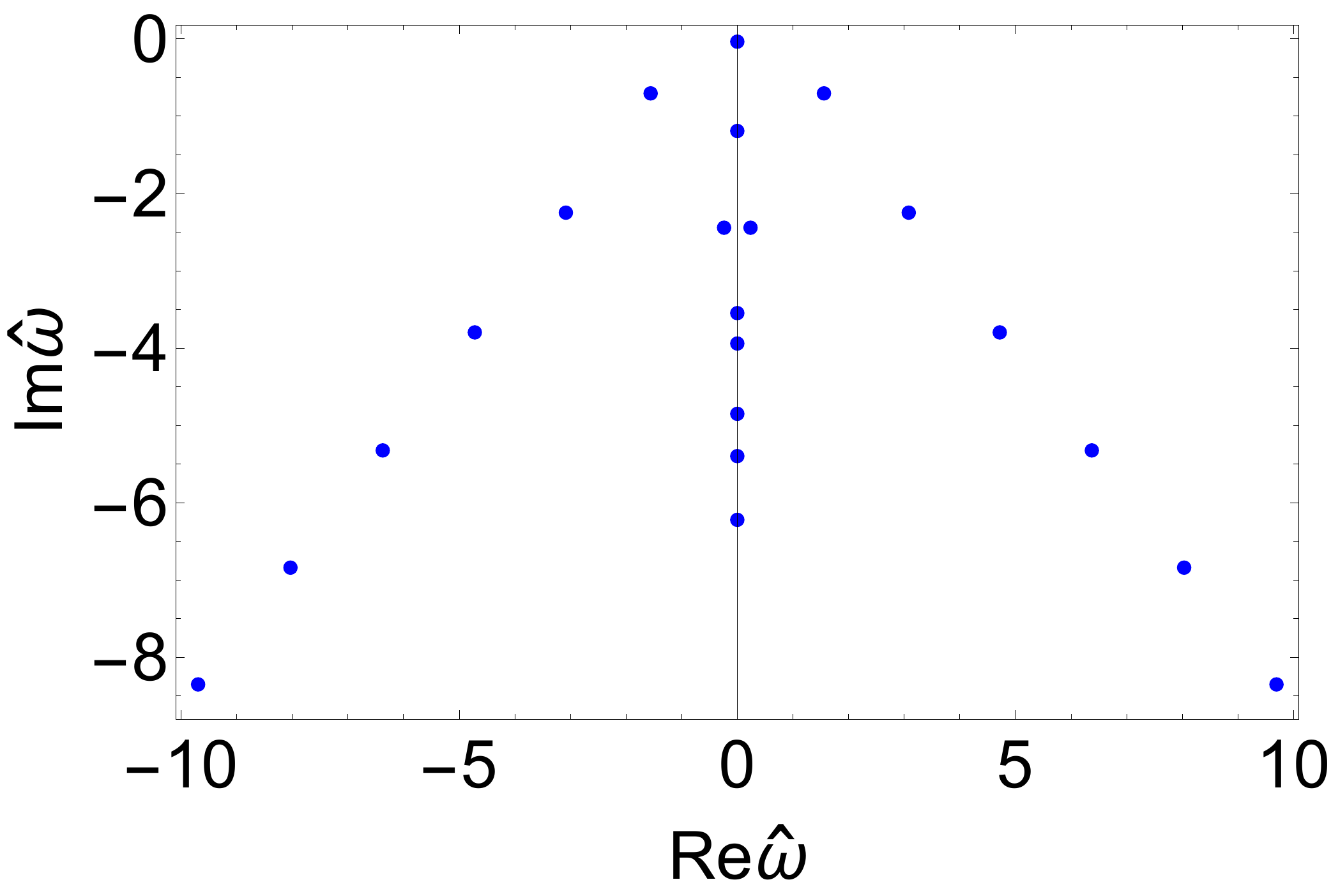}\ \\
\includegraphics[scale=0.24]{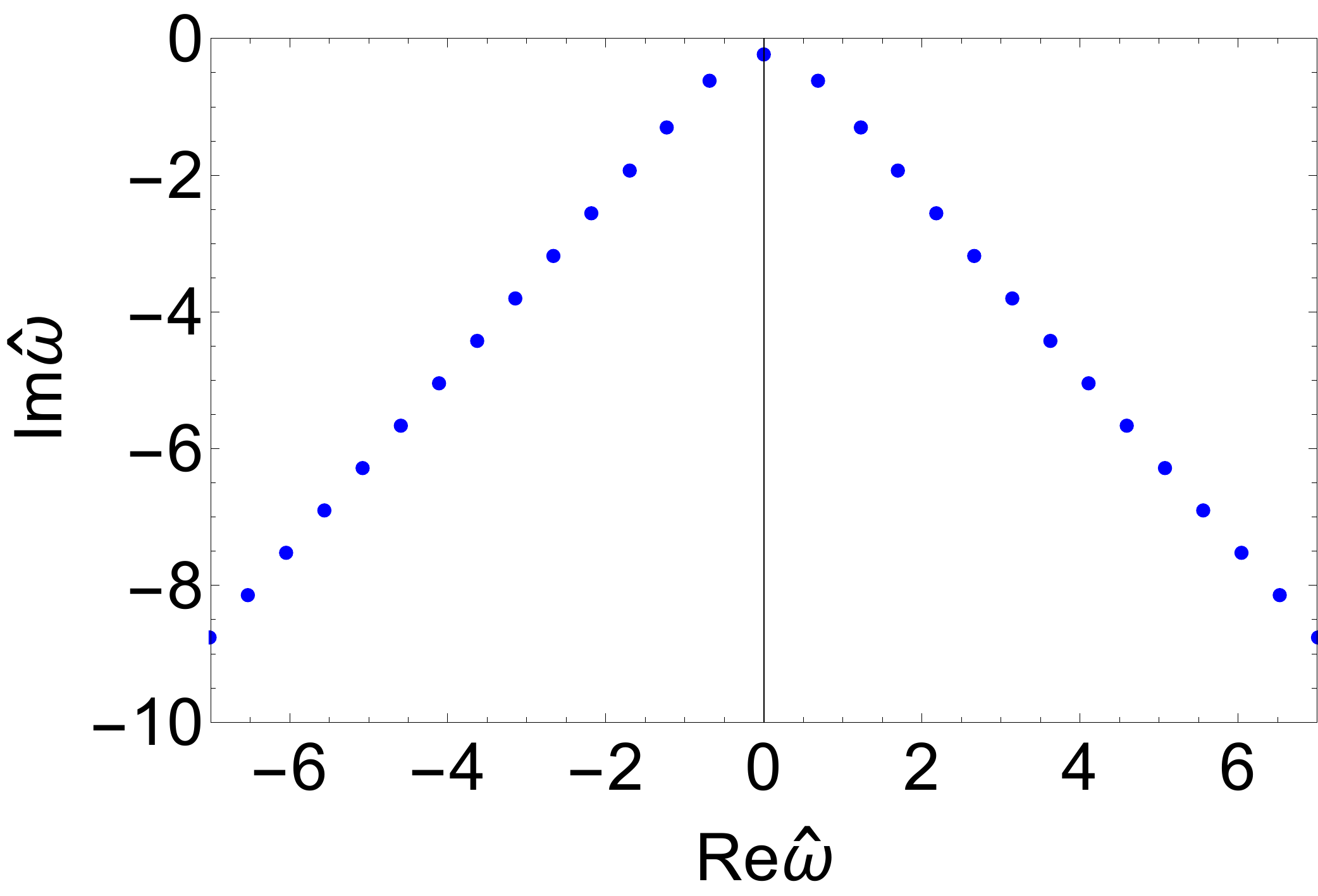}\ \hspace{0.1cm}
\includegraphics[scale=0.24]{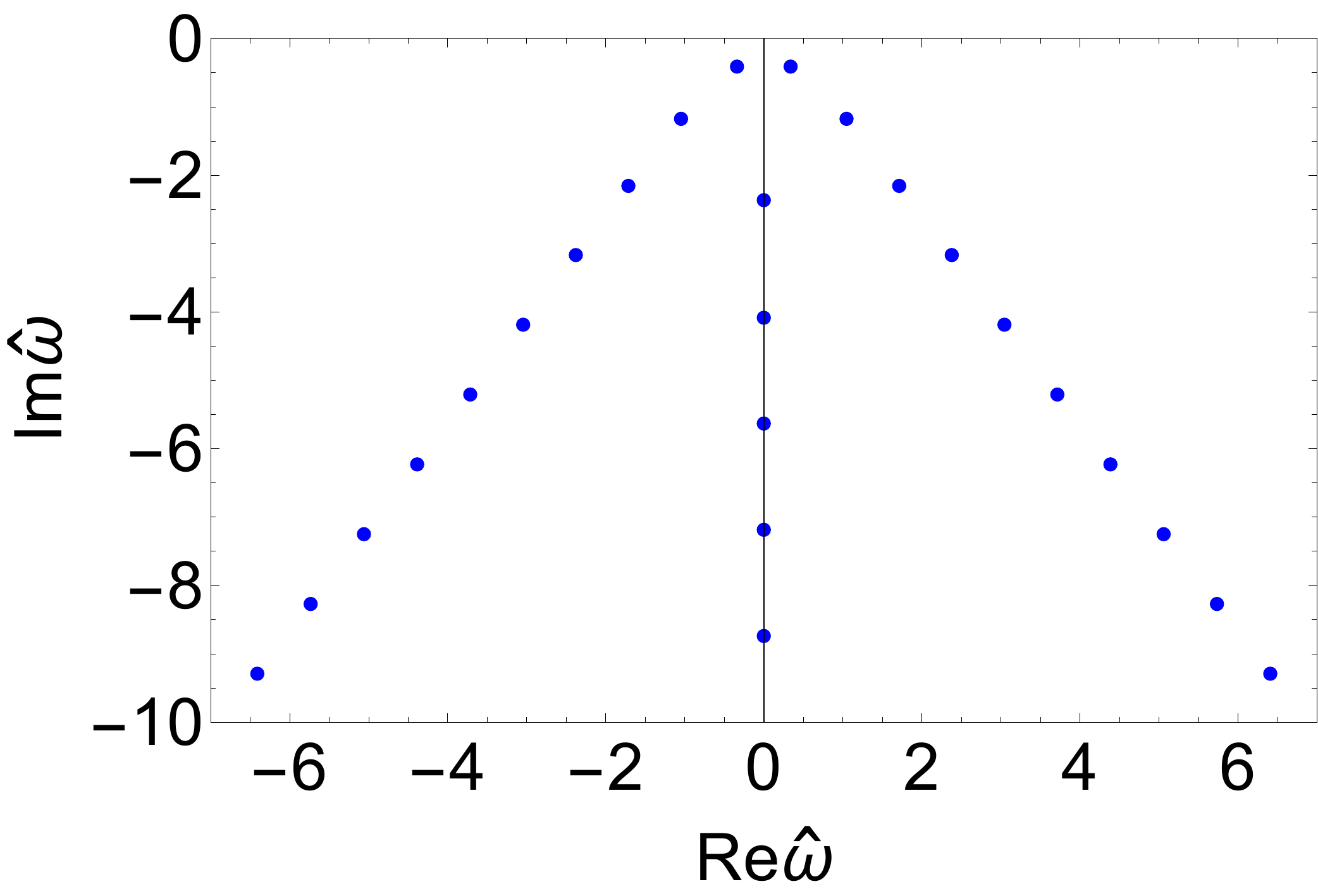}\ \hspace{0.1cm}
\includegraphics[scale=0.24]{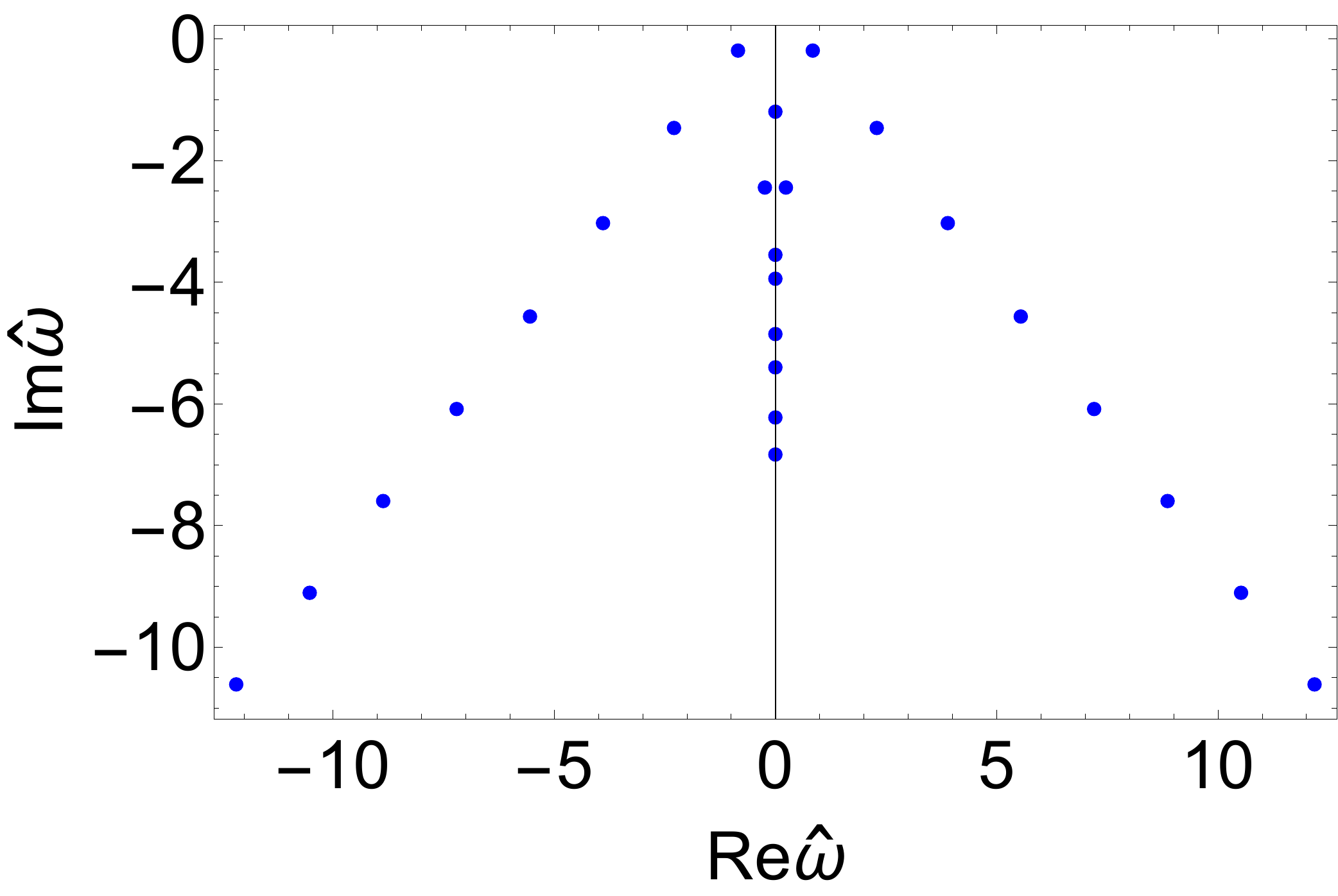}\ \\}
\caption{\label{fig_alpha_0p5} QNMs (blue spots) of the gauge mode for $\hat{\alpha}=1/2$.
The first, second and third columns are for $\gamma_1=0.02$, $-0.02$ and $-1$, respectively.
The panels above are for the gauge mode, while the ones below are the dual gauge mode.}
\end{figure}
\begin{figure}
\center{
\includegraphics[scale=0.24]{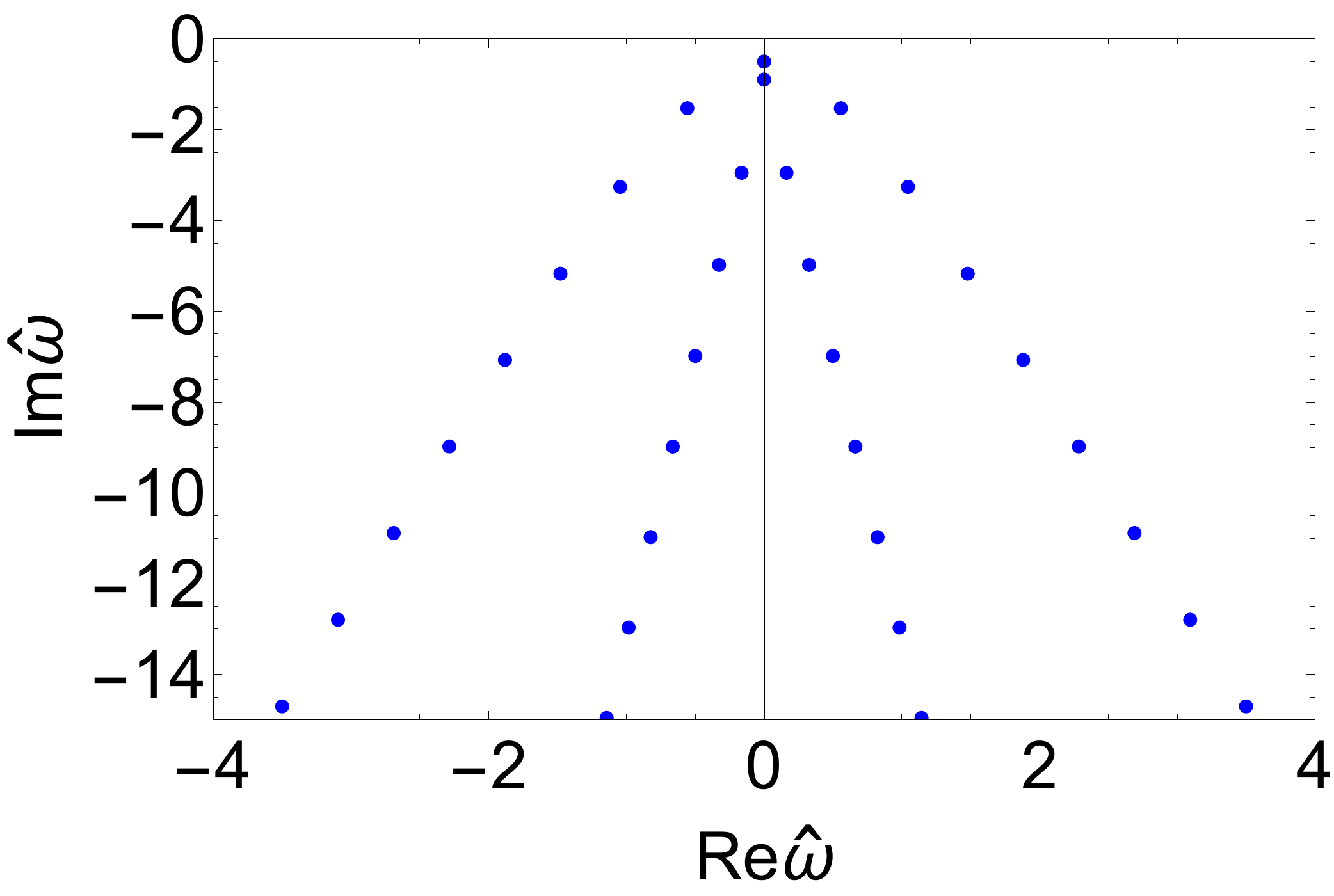}\ \hspace{0.1cm}
\includegraphics[scale=0.24]{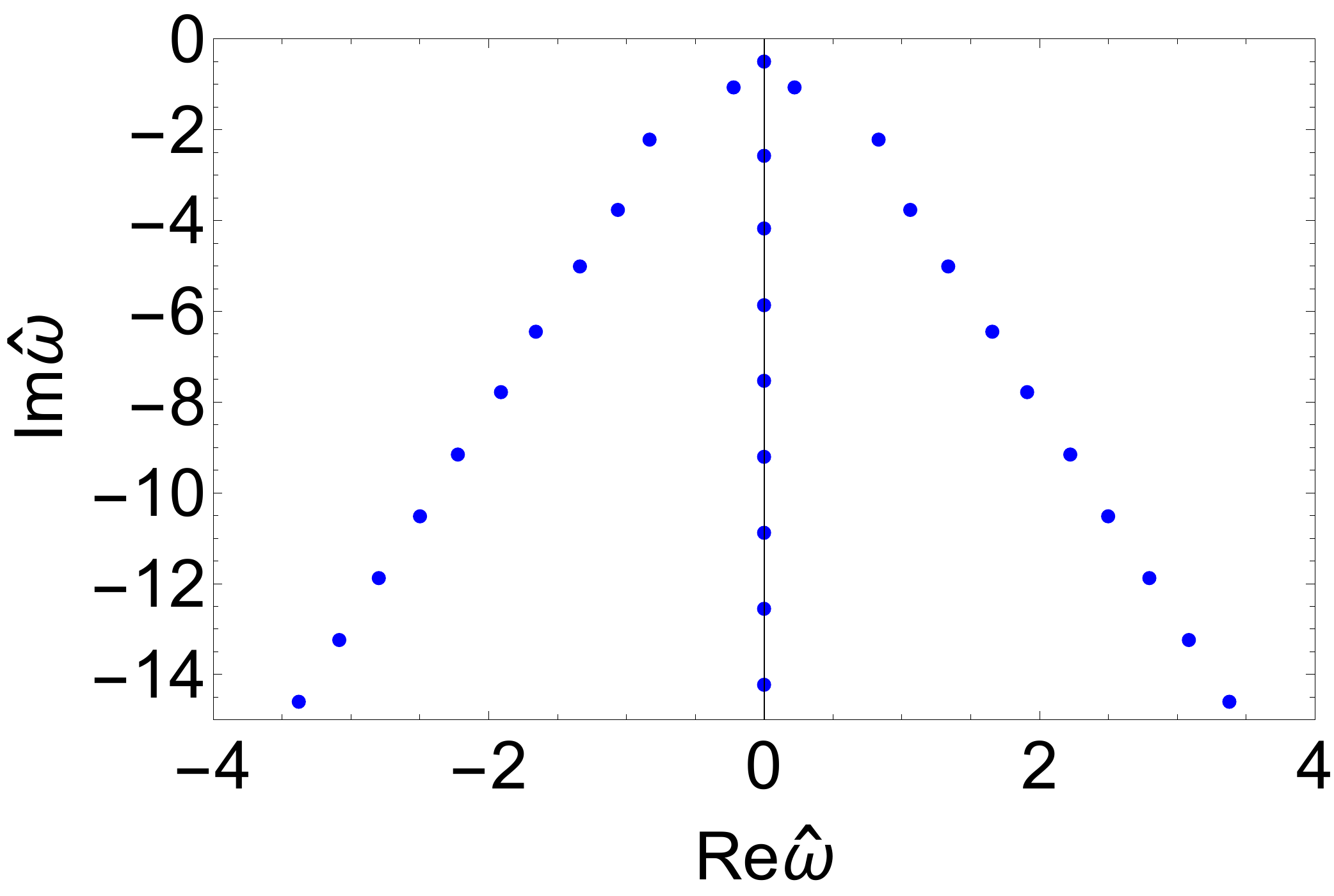}\ \hspace{0.1cm}
\includegraphics[scale=0.24]{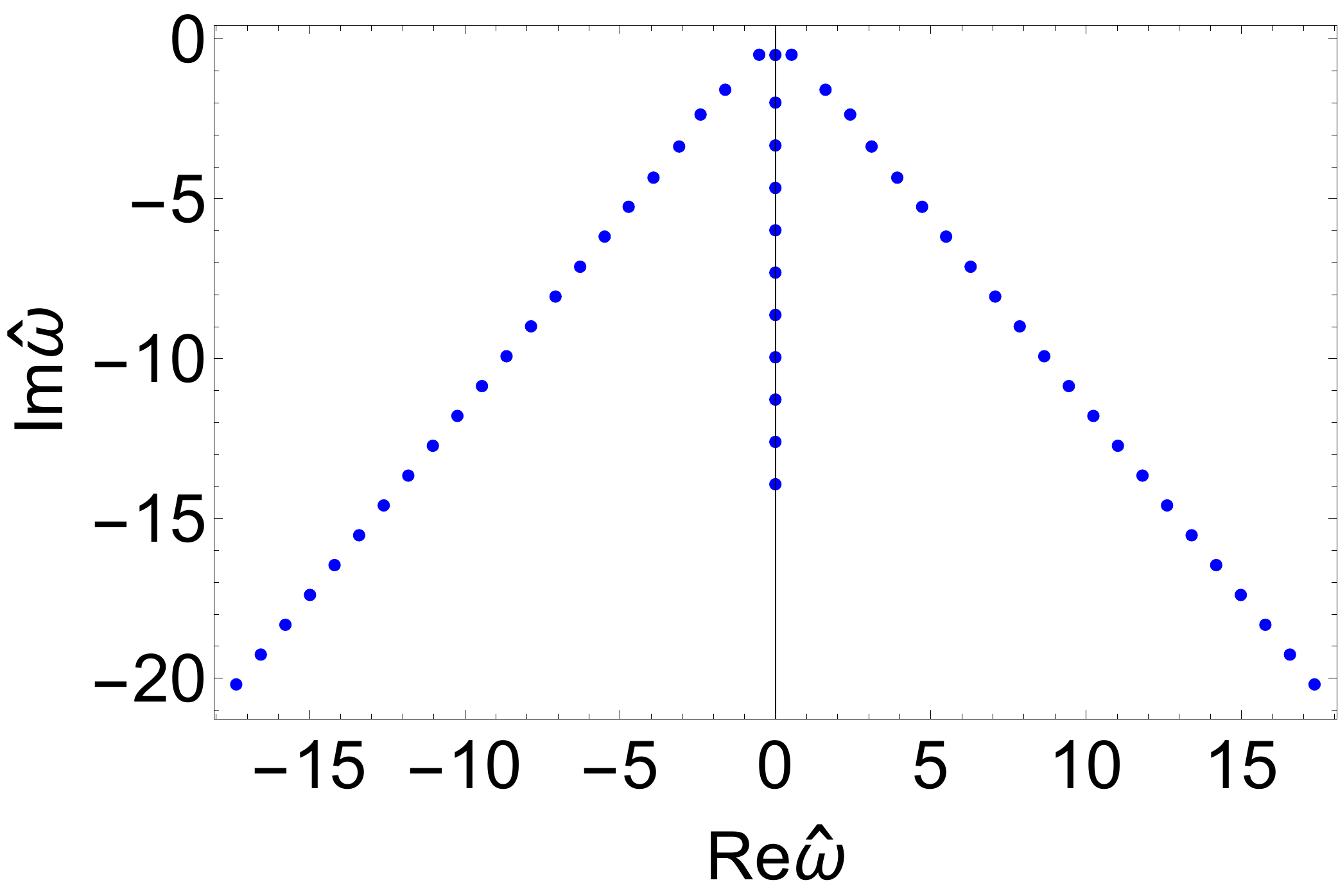}\ \\
\includegraphics[scale=0.24]{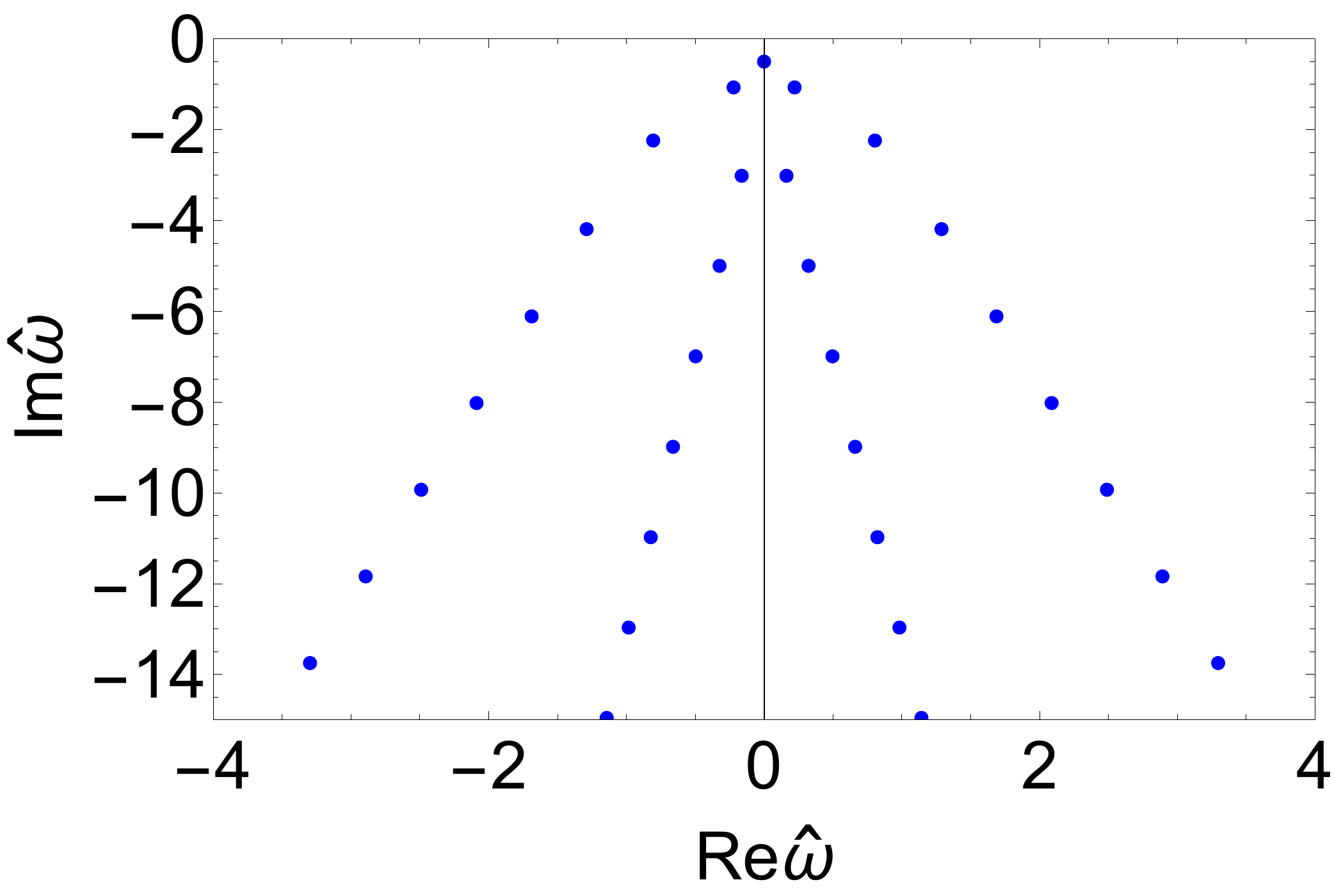}\ \hspace{0.1cm}
\includegraphics[scale=0.24]{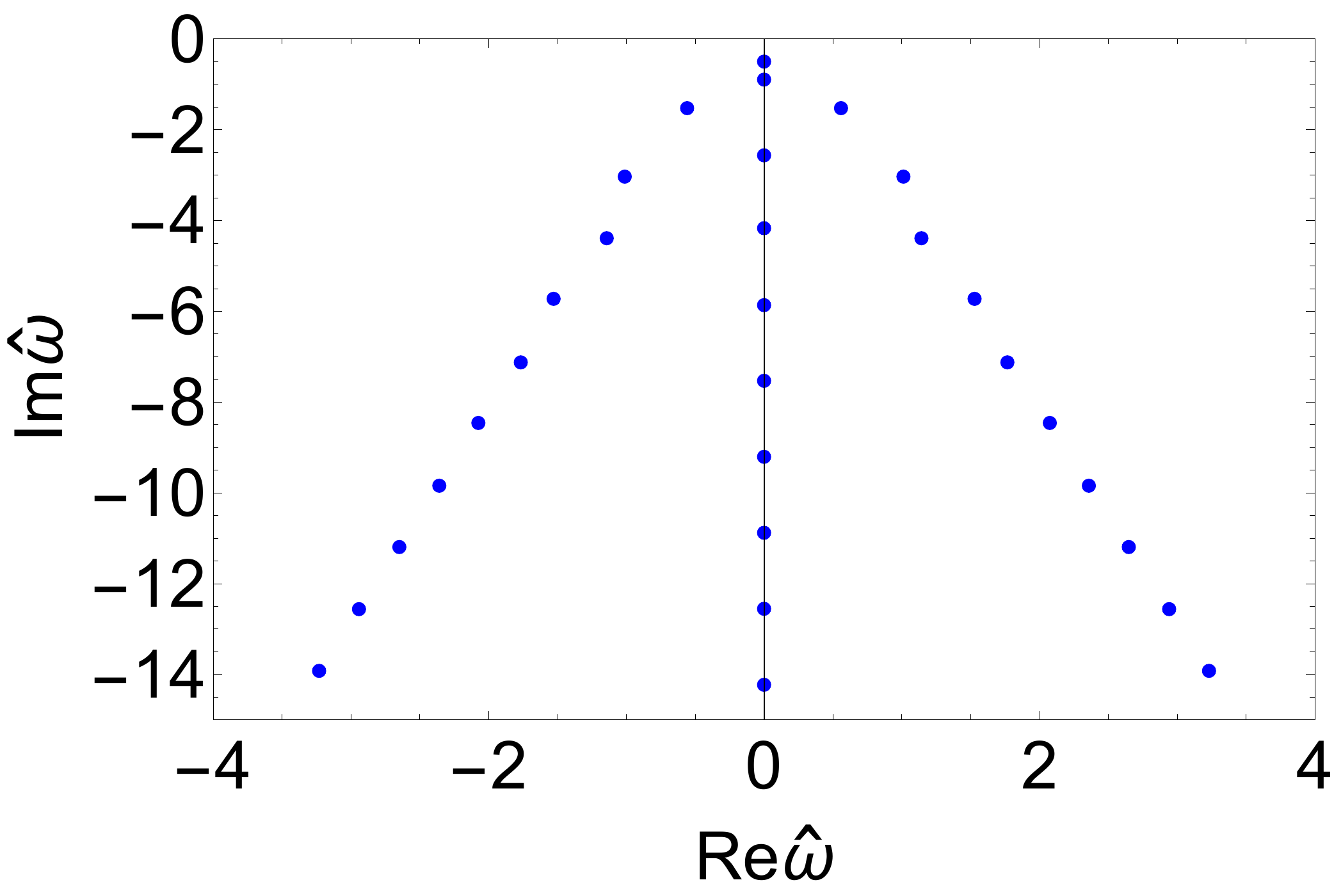}\ \hspace{0.1cm}
\includegraphics[scale=0.24]{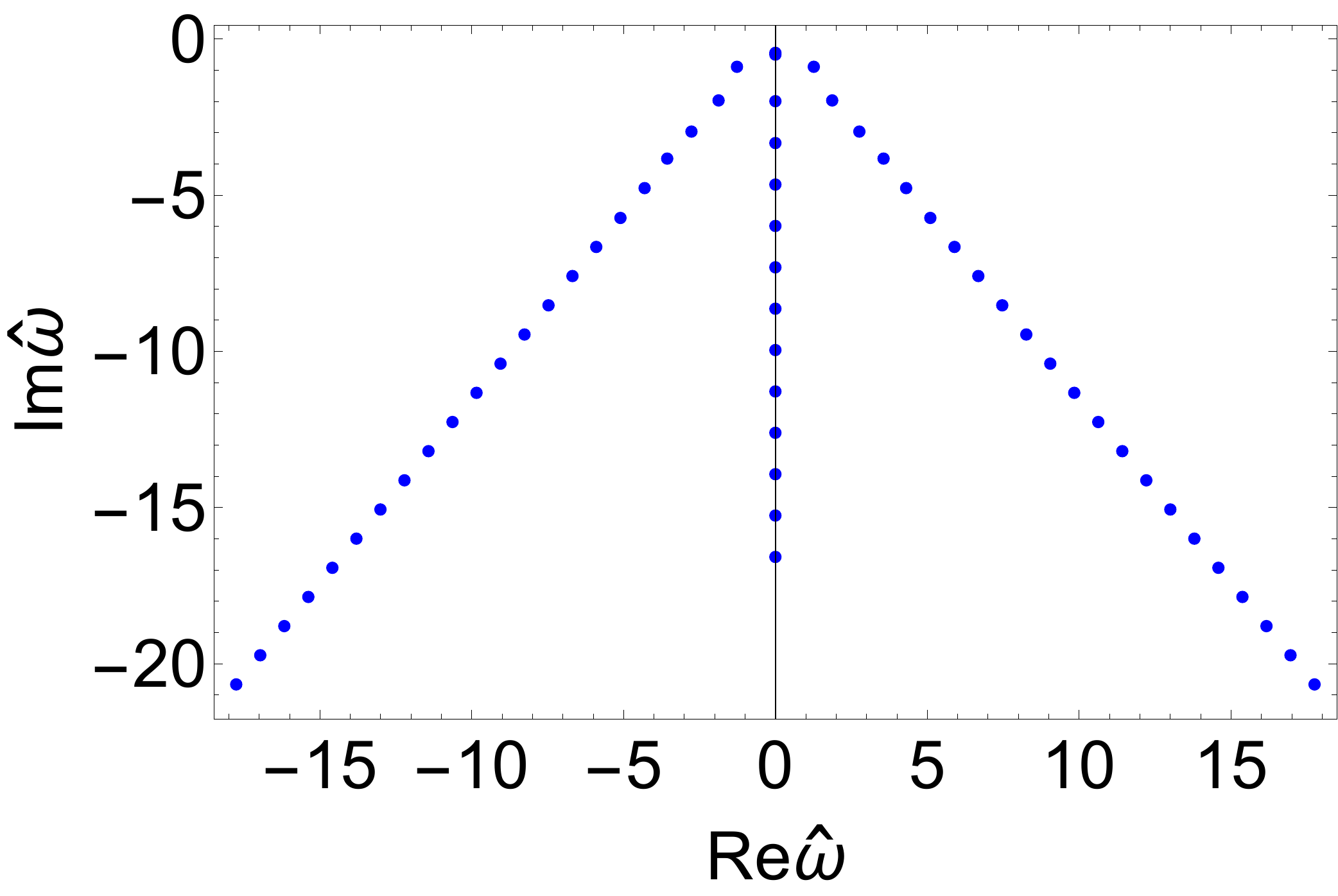}\ \\}
\caption{\label{fig_alpha_2osqrt3} QNMs (blue spots) of the gauge mode for $\hat{\alpha}=2/\sqrt{3}$.
The first, second and third columns are for $\gamma_1=0.02$, $-0.02$ and $-1$, respectively.
The panels above are for the gauge mode, while the ones below are the dual gauge mode.}
\end{figure}
\begin{figure}
\center{
\includegraphics[scale=0.24]{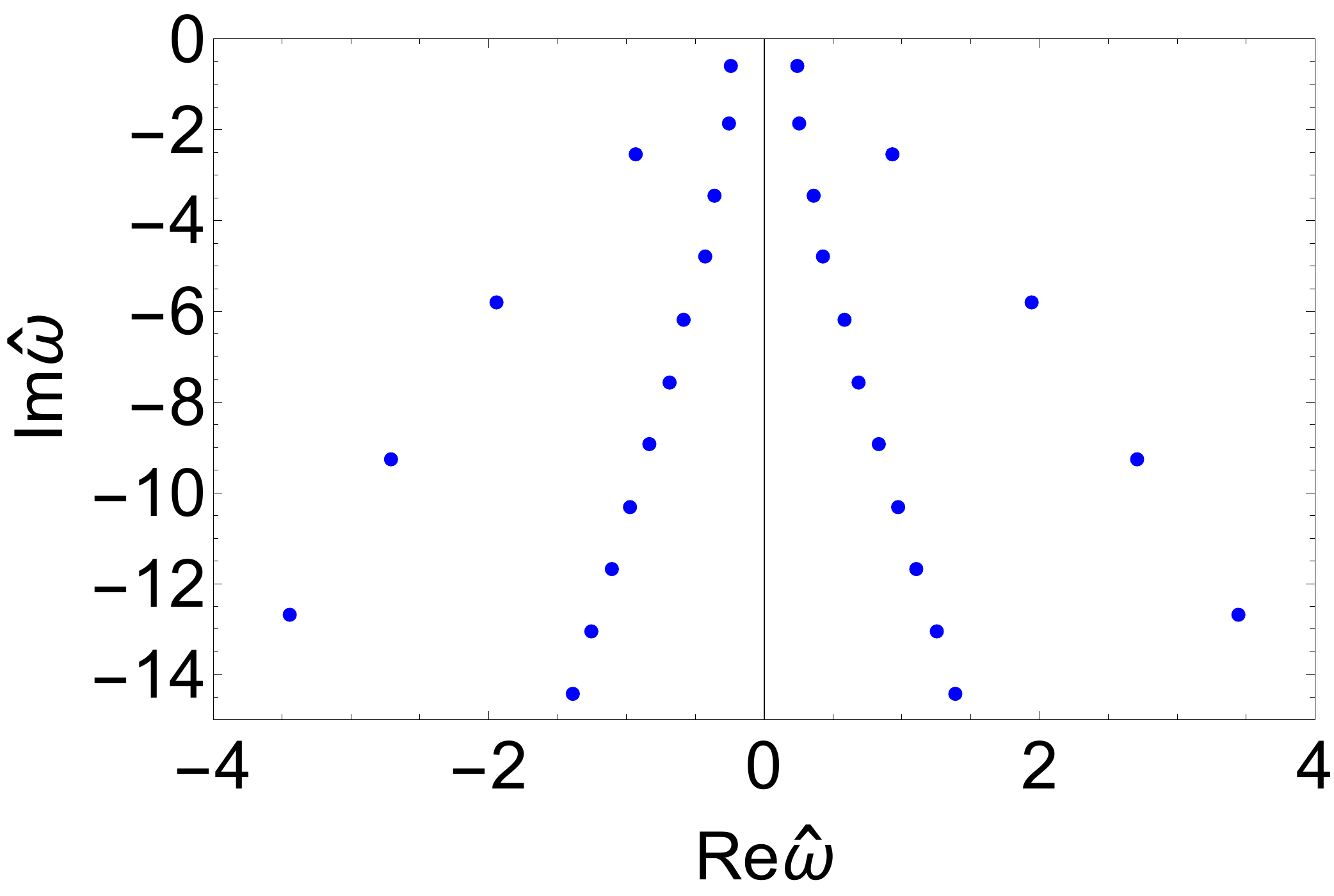}\ \hspace{0.1cm}
\includegraphics[scale=0.24]{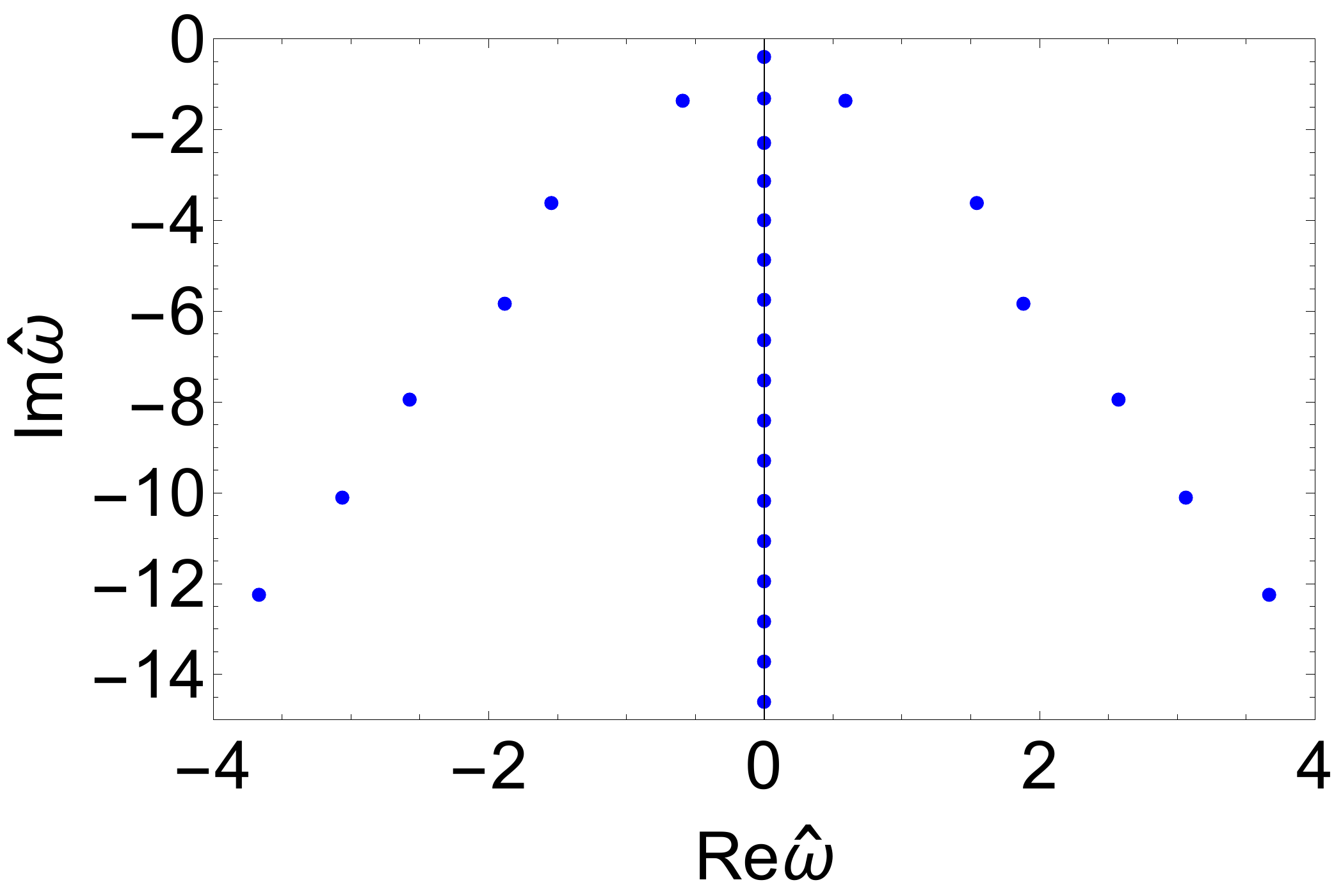}\ \hspace{0.1cm}
\includegraphics[scale=0.24]{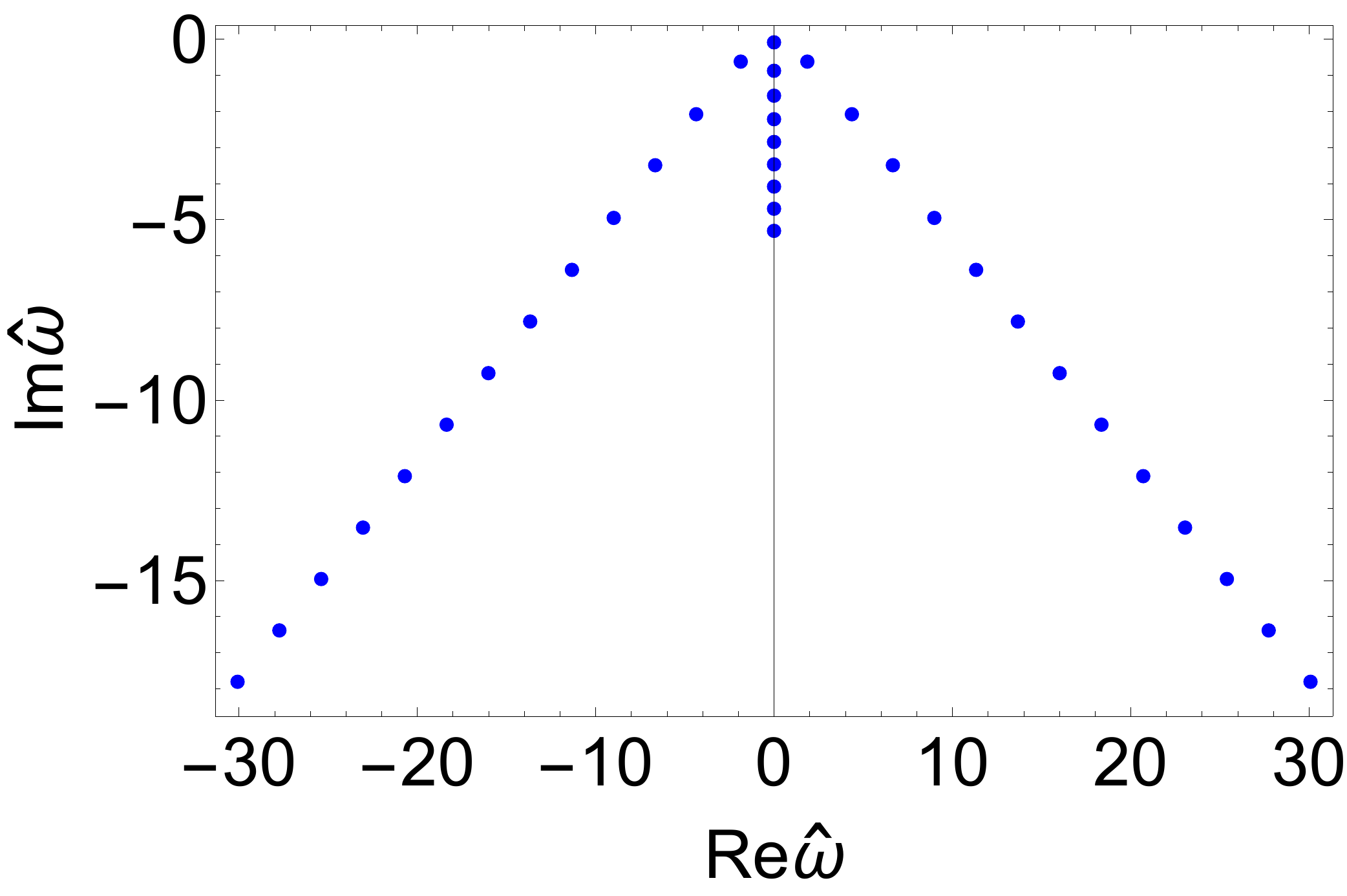}\ \\
\includegraphics[scale=0.24]{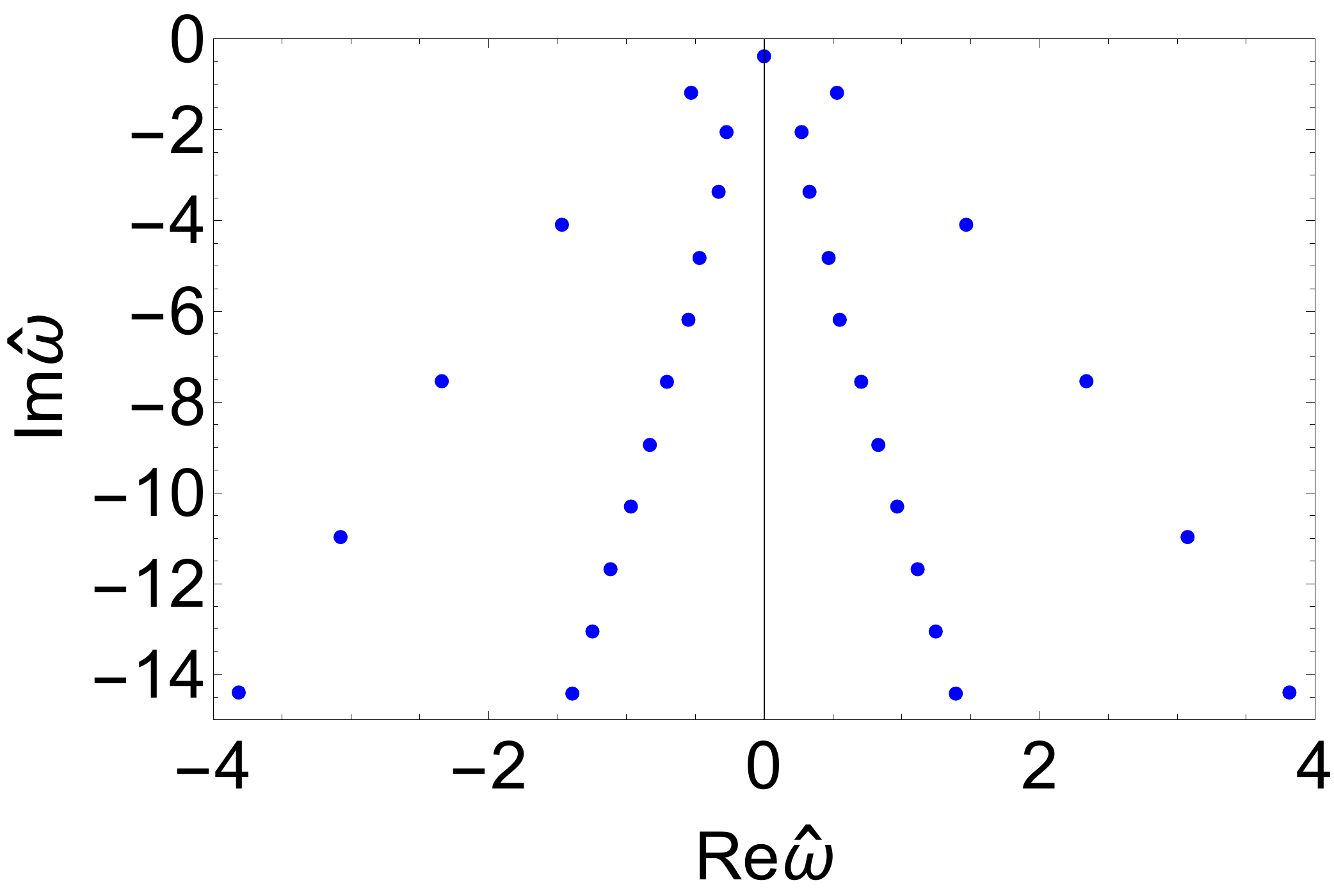}\ \hspace{0.1cm}
\includegraphics[scale=0.24]{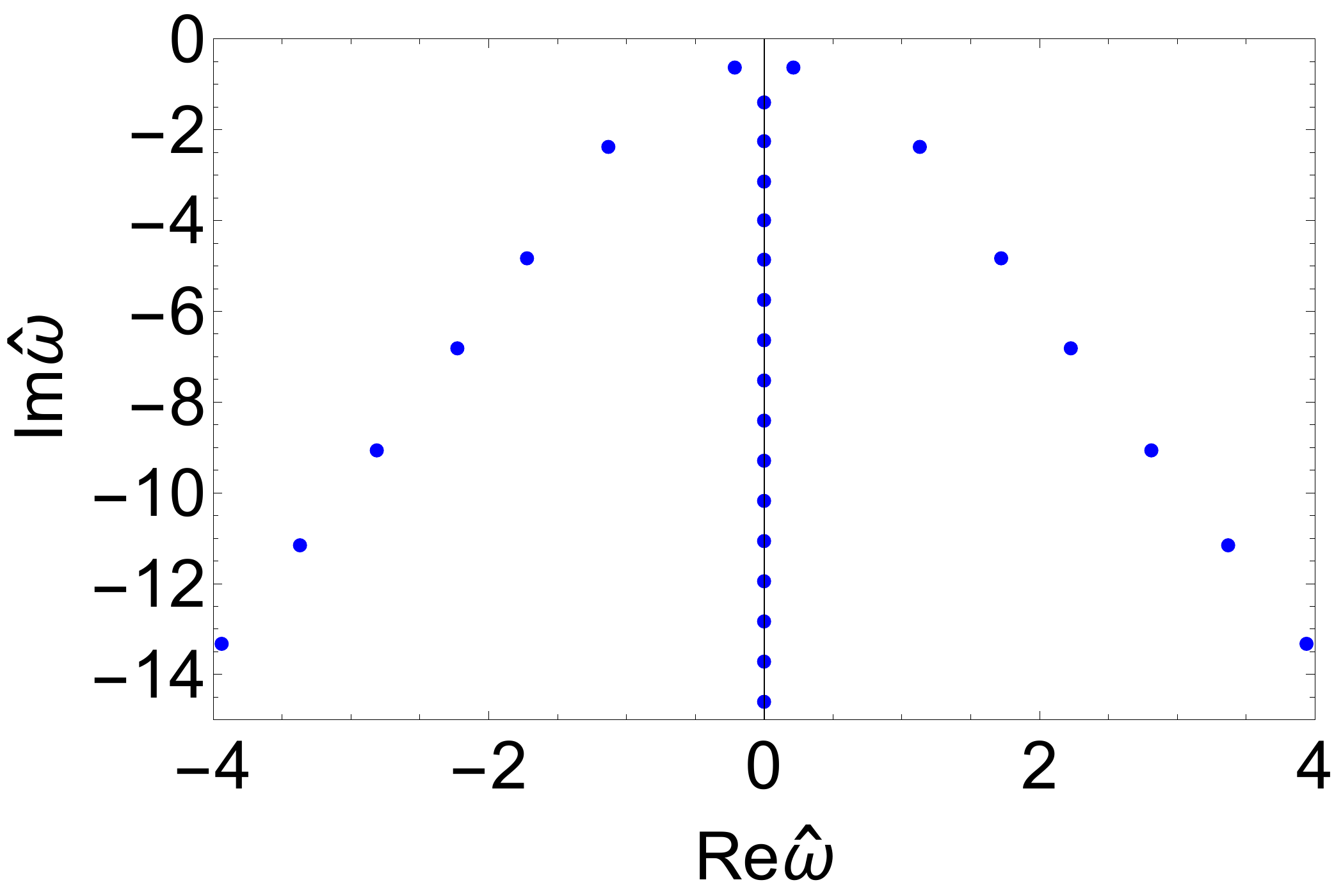}\ \hspace{0.1cm}
\includegraphics[scale=0.24]{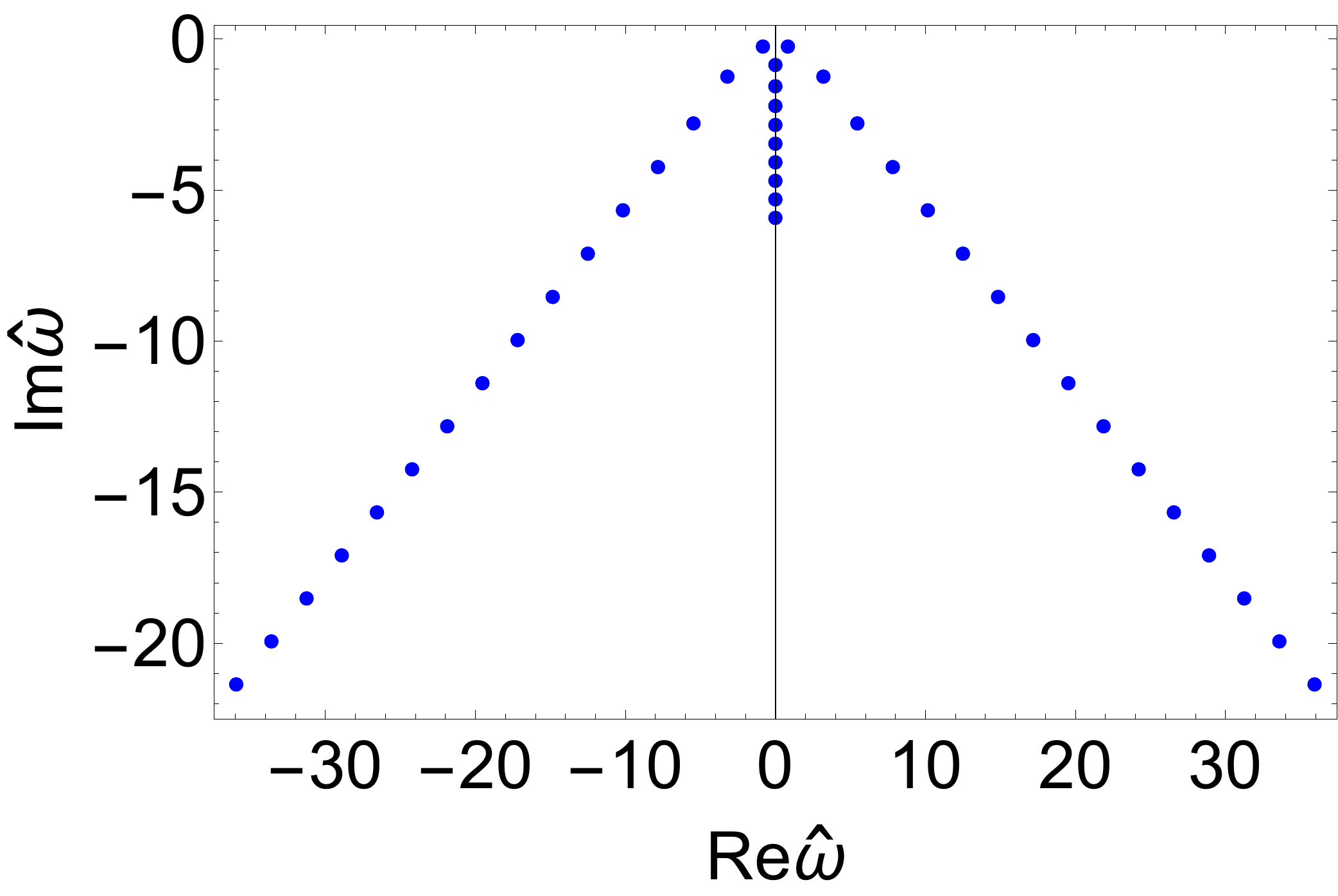}\ \\}
\caption{\label{fig_alpha_2} QNMs (blue spots) of the gauge mode for $\hat{\alpha}=2$.
The first, second and third columns are for $\gamma_1=0.02$, $-0.02$ and $-1$, respectively.
The panels above are for the gauge mode, while the ones below are the dual gauge mode.}
\end{figure}

Now we turn to the study of the effects of homogeneous disorder.
FIG.\ref{fig_alpha_0p5}, \ref{fig_alpha_2osqrt3} and \ref{fig_alpha_2} show the QNMs
with representative $\hat{\alpha}$ for $\gamma=0.02$, $-0.02$ and $-1$ (from left to right) in the complex frequency plane.
The panels above are the QNMs of the gauge mode, while the ones below are that of the dual gauge mode.
Rich information and insights on the pole structures are exhibited.
At the first sight, we find that all the poles are in the lower-half plane (LHP).
It indicates that the the gauge mode is stable. Further analysis on stability in terms of QNMs is presented in Appendix \ref{sec-stability}.

Next, we shall study the main properties of the QNMs when the homogeneous disorder is introduced,
in particular, the duality between the poles of $\texttt{Re}\sigma(\hat{\omega};\gamma_1)$
and the ones of $\texttt{Re}\sigma_{\ast}(\hat{\omega};-\gamma_1)$.
They are briefly summarized as what follows for $|\gamma_1|=0.02$.
\begin{itemize}
  \item For small $\hat{\alpha}$ ($\hat{\alpha}=1/2$ and $2/\sqrt{3}$), the approximate correspondence between the poles of $\texttt{Re}\sigma(\hat{\omega};\gamma_1)$
and the ones of $\texttt{Re}\sigma_{\ast}(\hat{\omega};-\gamma_1)$ recovers in low frequency region,
comparing with that without homogeneous disorder, for which the correspondence is violated.
But in the high frequency region, the case becomes different.
For $\hat{\alpha}=1/2$, some modes emerge at the imaginary frequency axis
for negative $\gamma_1$ as that without homogeneous disorder but not for positive $\gamma_1$.
For $\hat{\alpha}=2/\sqrt{3}$, there are still the modes locating in the imaginary frequency axis for negative $\gamma_1$
but not for positive $\gamma_1$, for which new branch cuts emerge.
These emerging modes in the high frequency region violate this correspondence.
\item For large $\hat{\alpha}$ ($\hat{\alpha}=2$), only for the dominate QNM, i.e., the ones closest to the real axis,
the correspondence between the poles of $\texttt{Re}\sigma(\hat{\omega};\gamma_1)$
and the ones of $\texttt{Re}\sigma_{\ast}(\hat{\omega};-\gamma_1)$ approximately holds.
As a whole, the pole structures for $\hat{\alpha}=2$ are similar with that for $\hat{\alpha}=2/\sqrt{3}$.
\end{itemize}
We also exhibit the QNMs for $\gamma_1=-1$ with homogeneous disorder (the third columns in FIG.\ref{fig_alpha_0p5}, \ref{fig_alpha_2osqrt3} and \ref{fig_alpha_2}).
When the homogeneous disorder is introduced, the new branch cuts attach to the imaginary frequency axis.
Such pole structures are interesting and are worthy of further study such that we can understand the physics of these pole structures.

The dominant poles reveal the late time dynamics of the system.
So, we shall study the dominant pole behaviors such that we have a well understanding on the effect of the homogeneous disorder.

\begin{figure}
\center{
\includegraphics[scale=0.25]{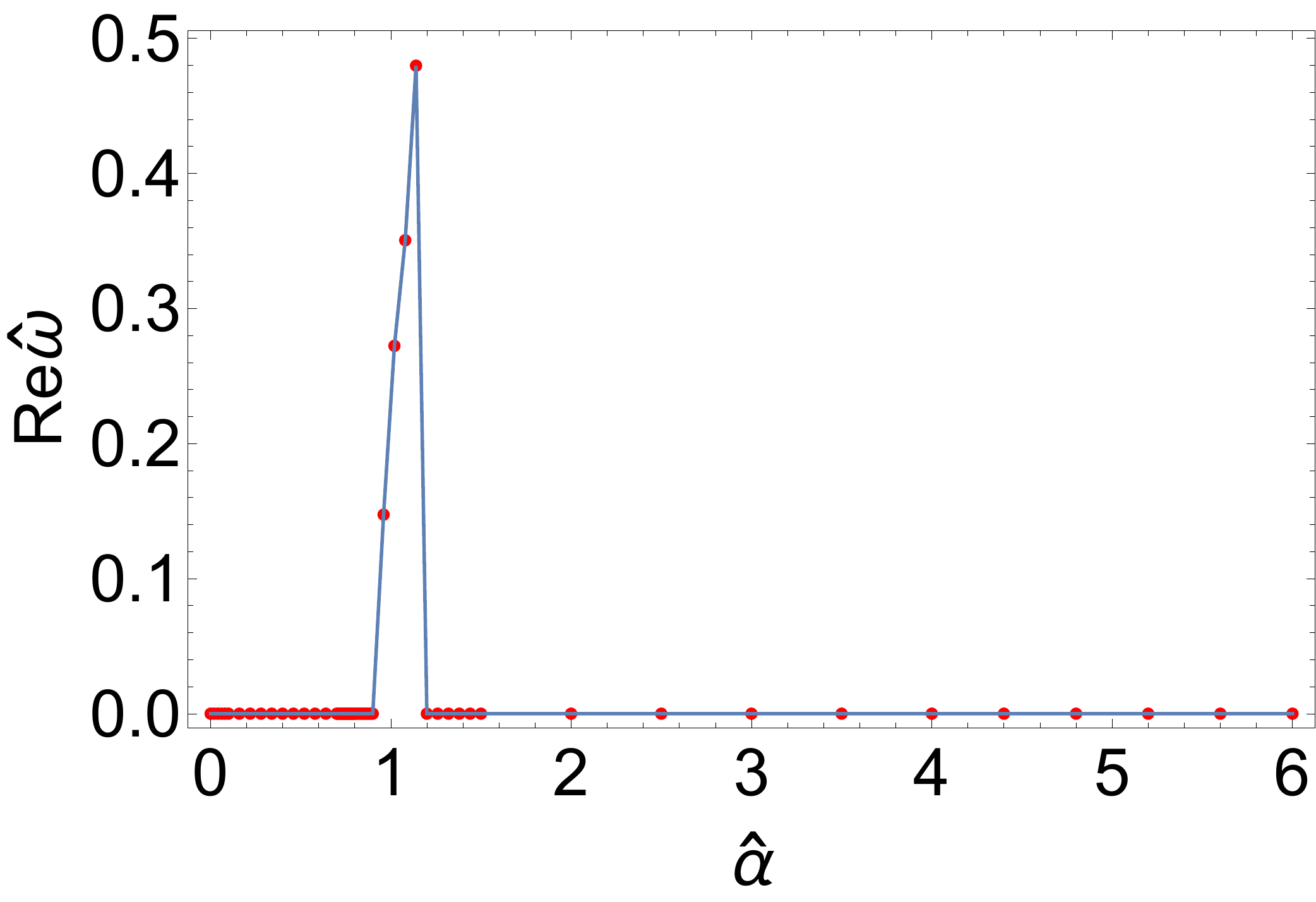}\ \hspace{0.4cm}
\includegraphics[scale=0.25]{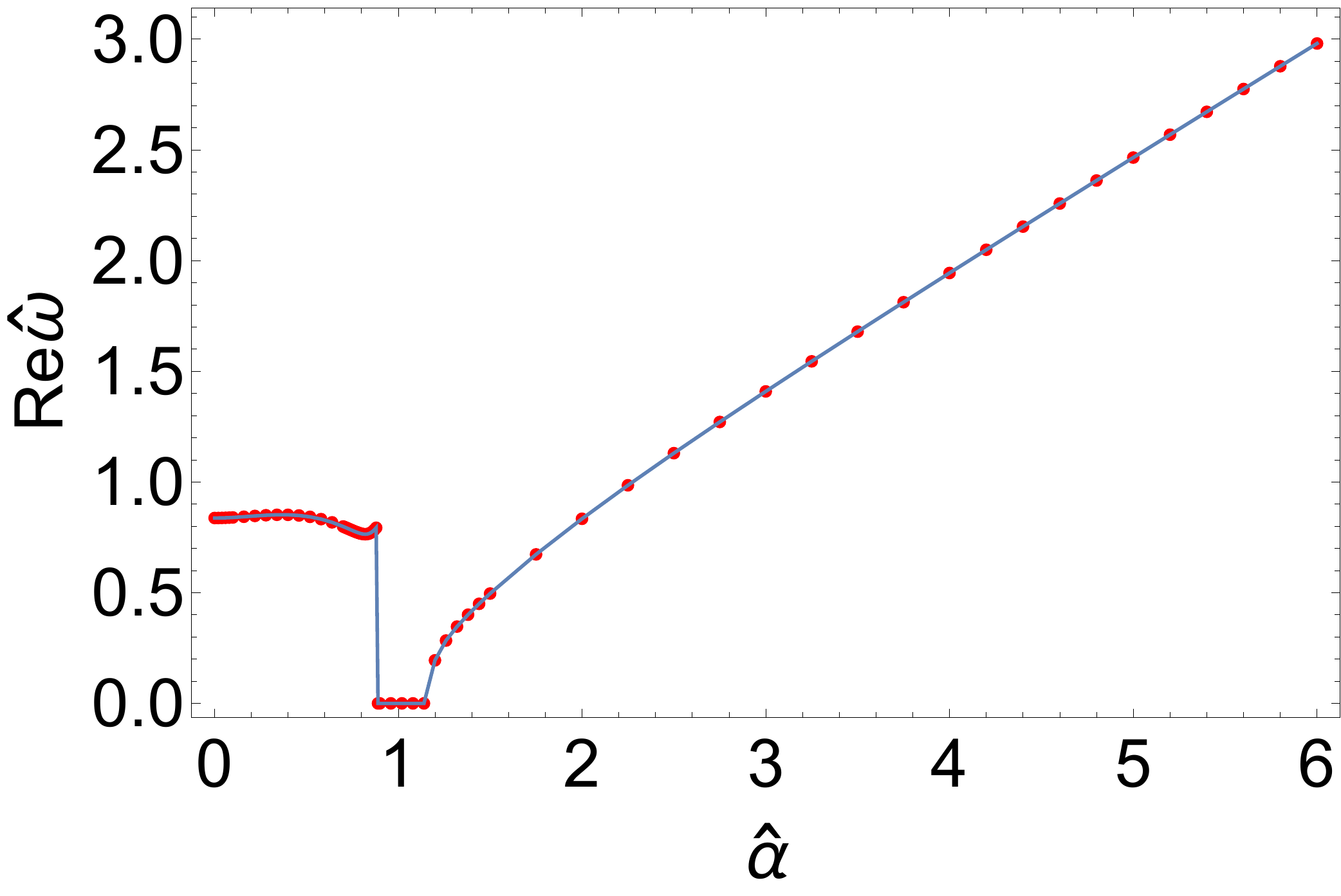}\ \\
\includegraphics[scale=0.25]{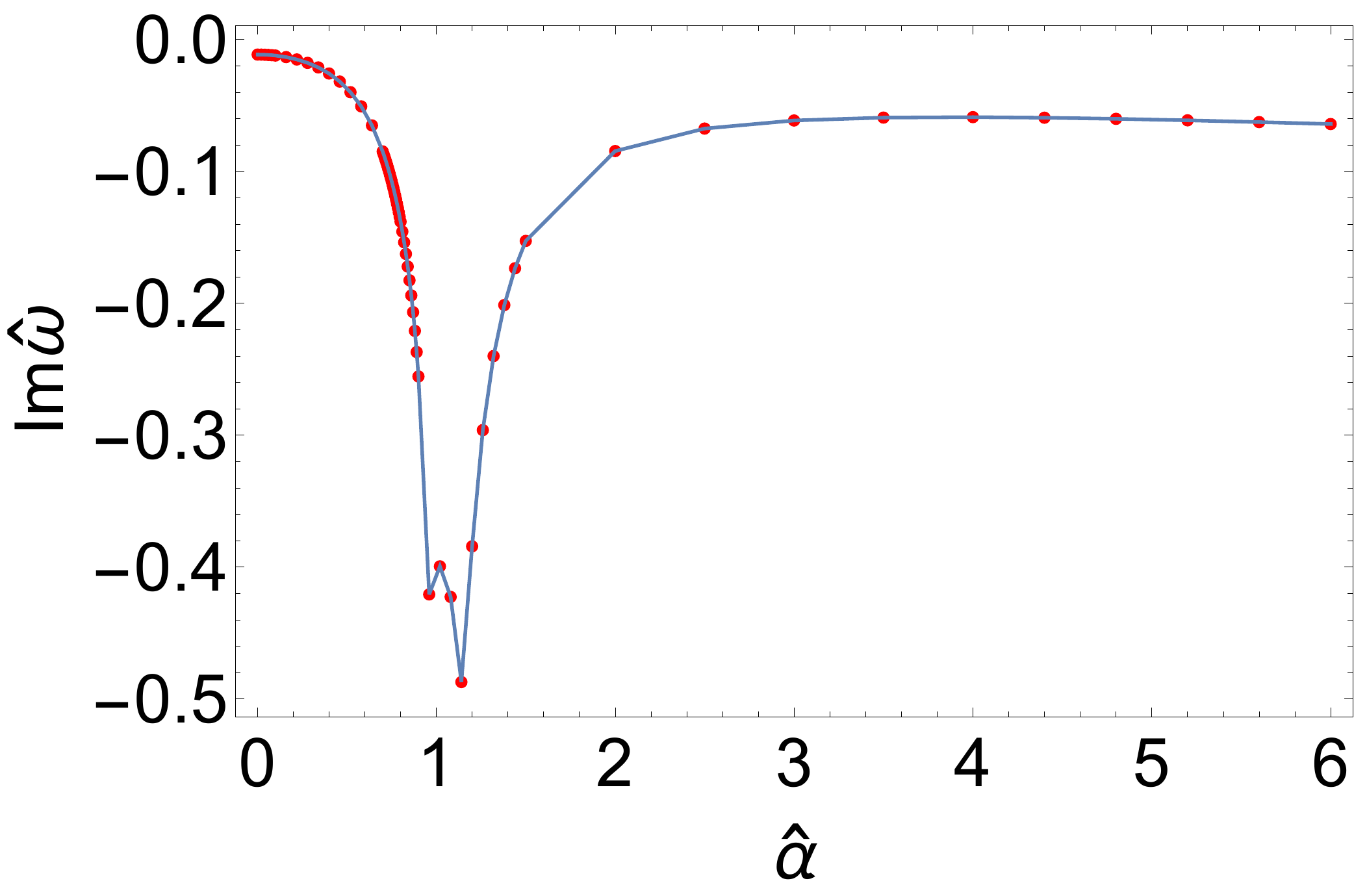}\ \hspace{0.4cm}
\includegraphics[scale=0.25]{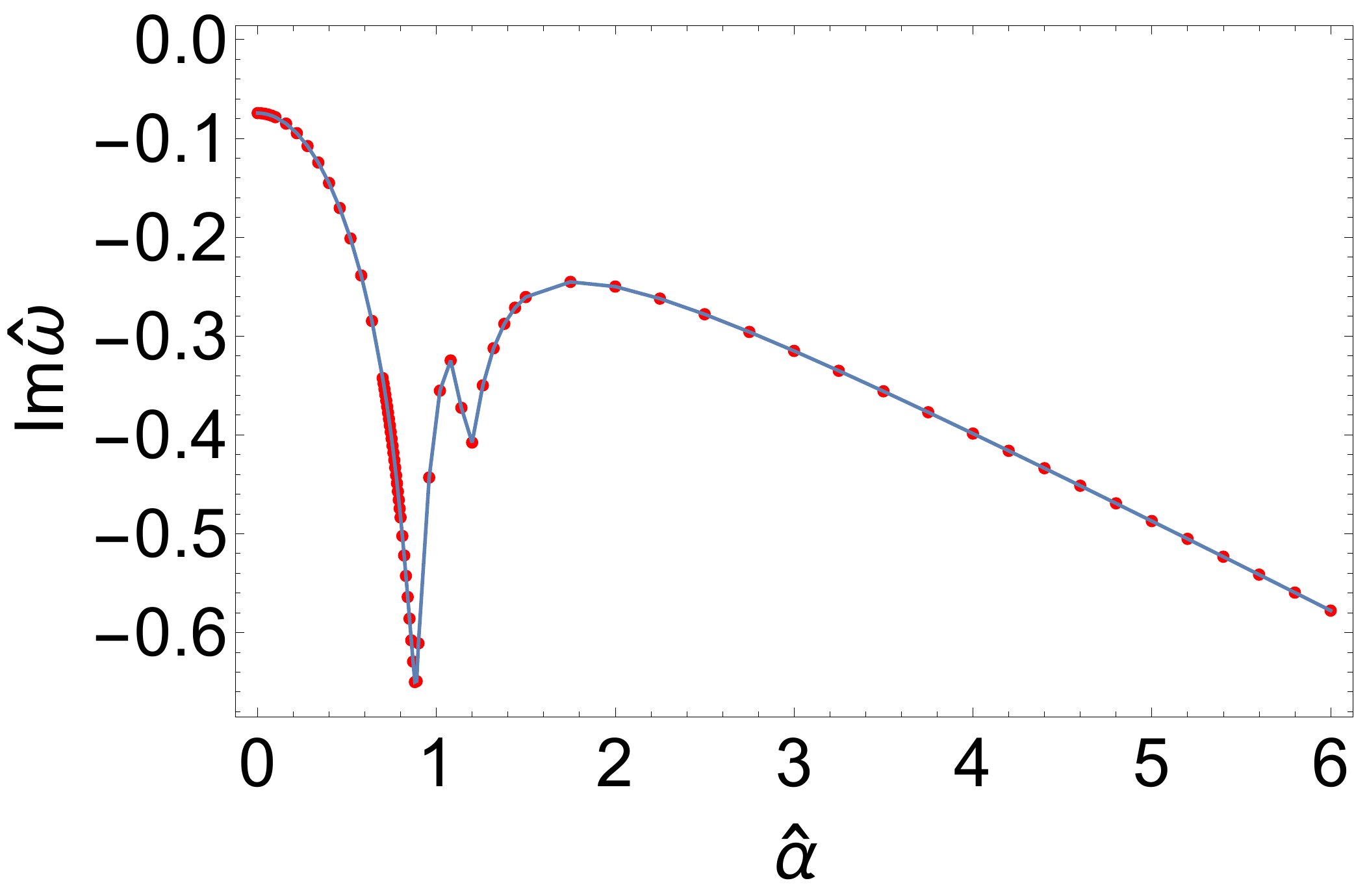}\ \\}
\caption{\label{fig_dqnm_gamma1_n1} Evolution of the dominant QNMs with $\hat{\alpha}$ for $\gamma_1=-1$.
The left column is the QNMs for the original theory, and the right one is that for its dual theory.}
\end{figure}

In our previous work \cite{Fu:2017oqa}, we have studied the conductivity at the low frequency in real axis
for $\gamma_1=-1$ with homogeneous disorder. For small $\hat{\alpha}$, it exhibits a sharp DS peak.
Though it is not the Drude peak, we can phenomenologically describe this peak by the Drude formula (see FIG.5 in \cite{Fu:2017oqa})
\fa
\sigma(\hat{\omega})=\frac{K}{\Gamma-i\hat{\omega}}\,,
\label{Drude}
\ffa
where $\Gamma$ is the relaxation rate, which relates the relaxation time $\tau$ as $\Gamma=1/\tau$.
This peak in the real axis corresponds to a purely imaginary mode in complex frequency panel (see the left column in FIG.\ref{fig_dqnm_gamma1_n1}).
We also locate the position of the purely imaginary mode, which is listed in Table \ref{table-Drude}.
We find that the relaxation rates fitted by the Drude formula \eqref{Drude} are in very agreement with that by locating the position of the purely imaginary mode.
\begin{widetext}
\begin{table}[ht]
\begin{center}
\begin{tabular}{|c|c|c|c|c|c|c|}
         \hline
~$\hat{\alpha}$~ &~$0$~&~$0.01$~&~$0.1$~&~$0.5$~
          \\
        \hline
~$\Gamma_1$~ & ~$0.0114$~&~$0.0115$~&~$0.0120$~&~$0.0284$~
          \\
        \hline
~$\Gamma_2$~ & ~$0.0114$~&~$0.0115$~&~$0.0122$~&~$0.0370$~
          \\
        \hline
\end{tabular}
\caption{\label{table-Drude} The relaxation rate $\Gamma_i$ of the holographic system with $\gamma_1=-1$ for different $\hat{\alpha}$.
$\Gamma_1$ is fitted by the Drude formula (\ref{Drude})
and $\Gamma_2$ is obtained by locating the position of the purely imaginary mode in complex frequency panel.
}
\end{center}
\end{table}
\end{widetext}

More detailed evolution of the dominant QNMs with $\hat{\alpha}$ for $\gamma_1=-1$ is presented in FIG.\ref{fig_dqnm_gamma1_n1}.
Most of the dominant QNMs for the original theory, except for that approximately in the region of $\hat{\alpha}\in(1,2/\sqrt{3})$,
are the purely imaginary modes.
Correspondingly, the dominant QNMs for the dual theory, except for that approximately in the region of $\hat{\alpha}\in(1,2/\sqrt{3})$,
are off axis.
\begin{figure}
\center{
\includegraphics[scale=0.25]{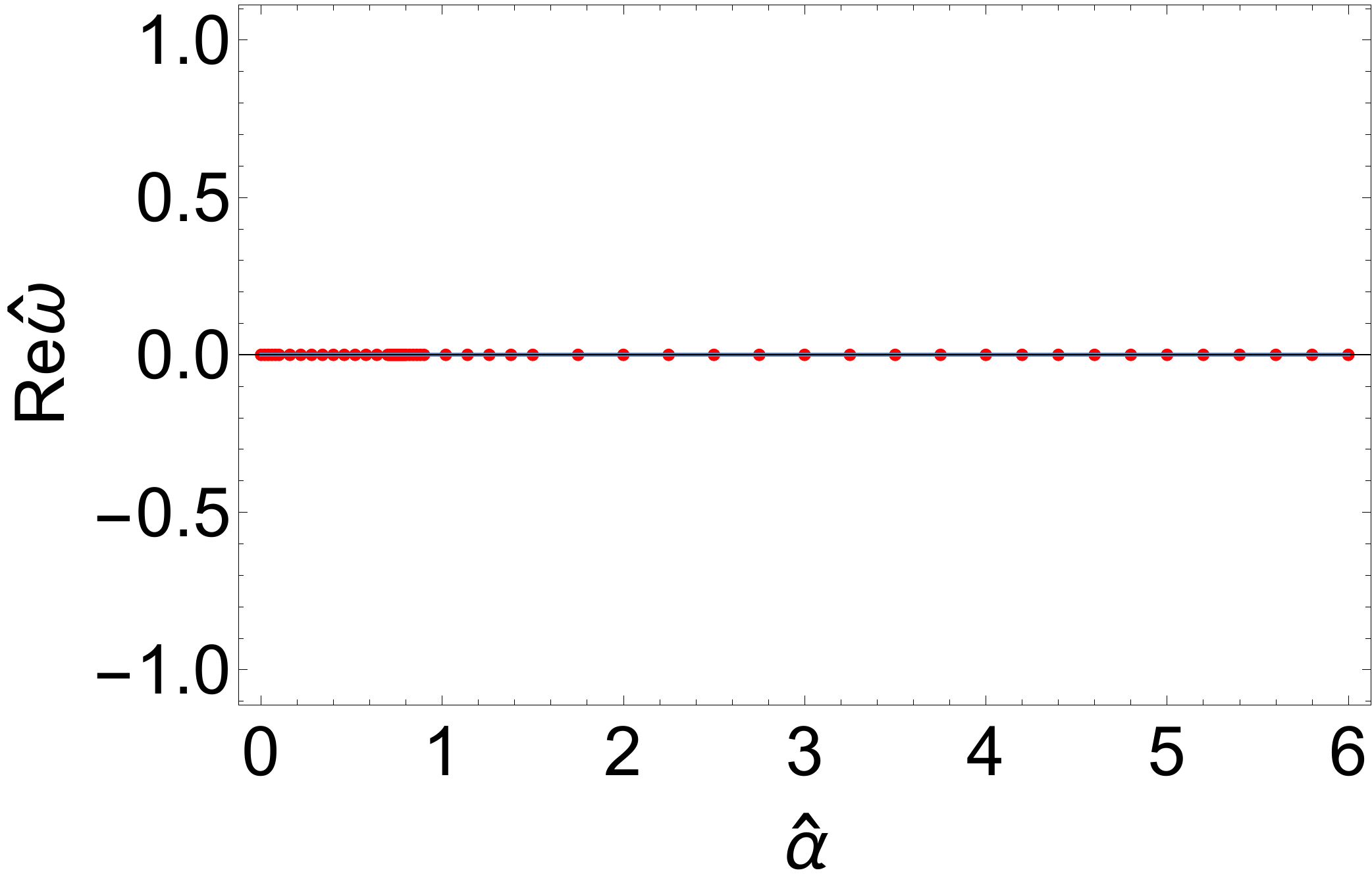}\ \hspace{0.4cm}
\includegraphics[scale=0.25]{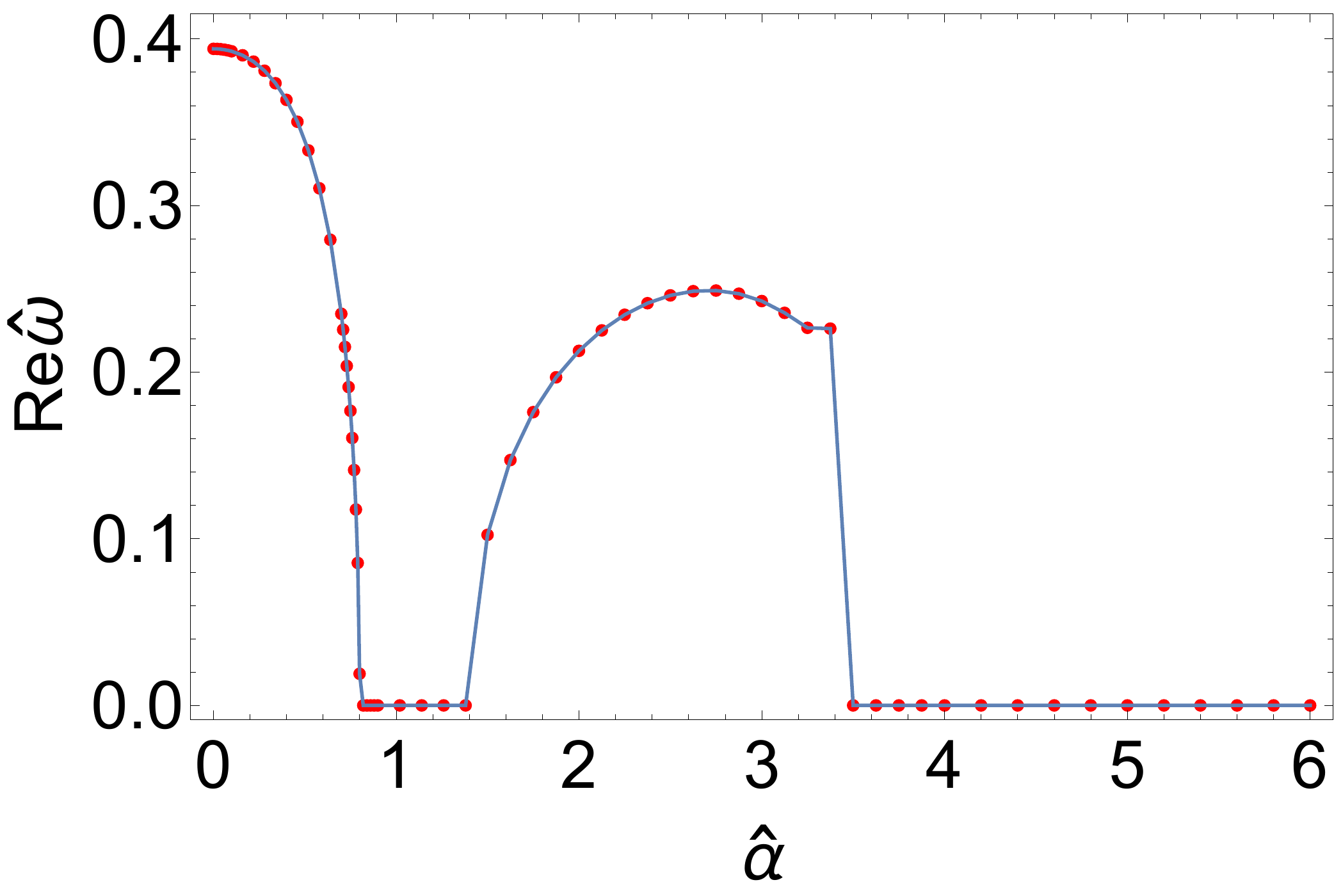}\ \\
\includegraphics[scale=0.25]{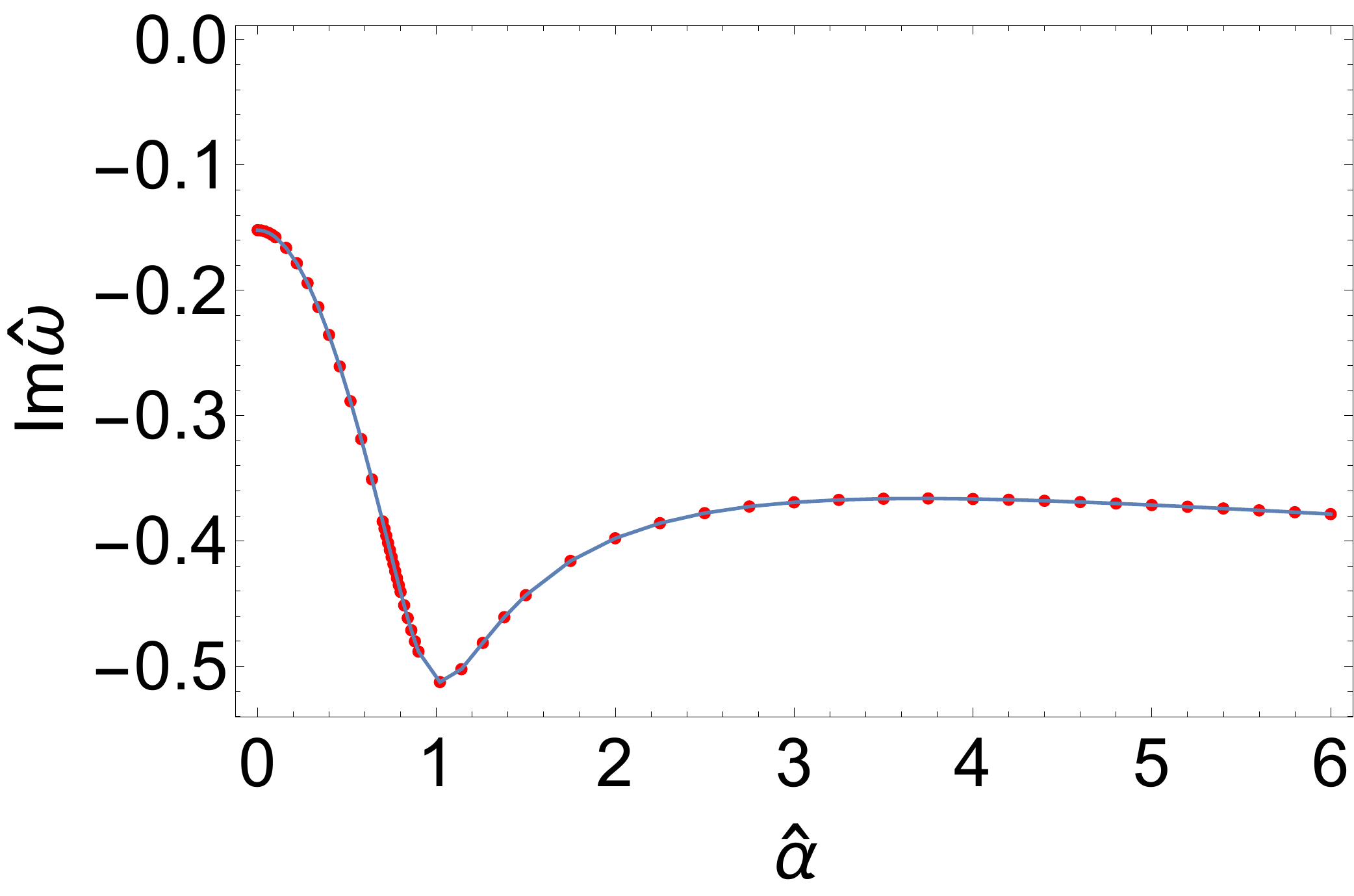}\ \hspace{0.4cm}
\includegraphics[scale=0.25]{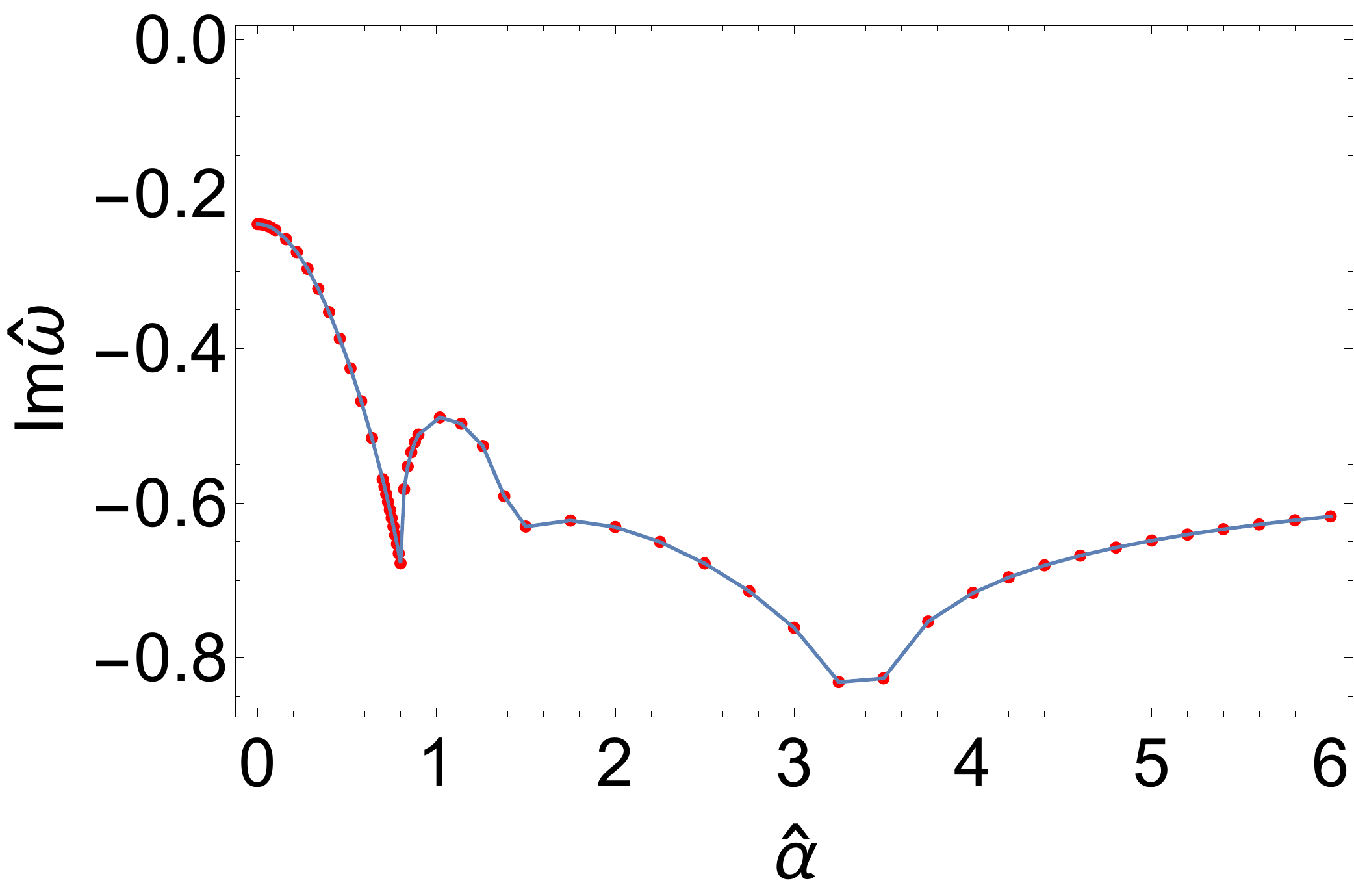}
\includegraphics[scale=0.25]{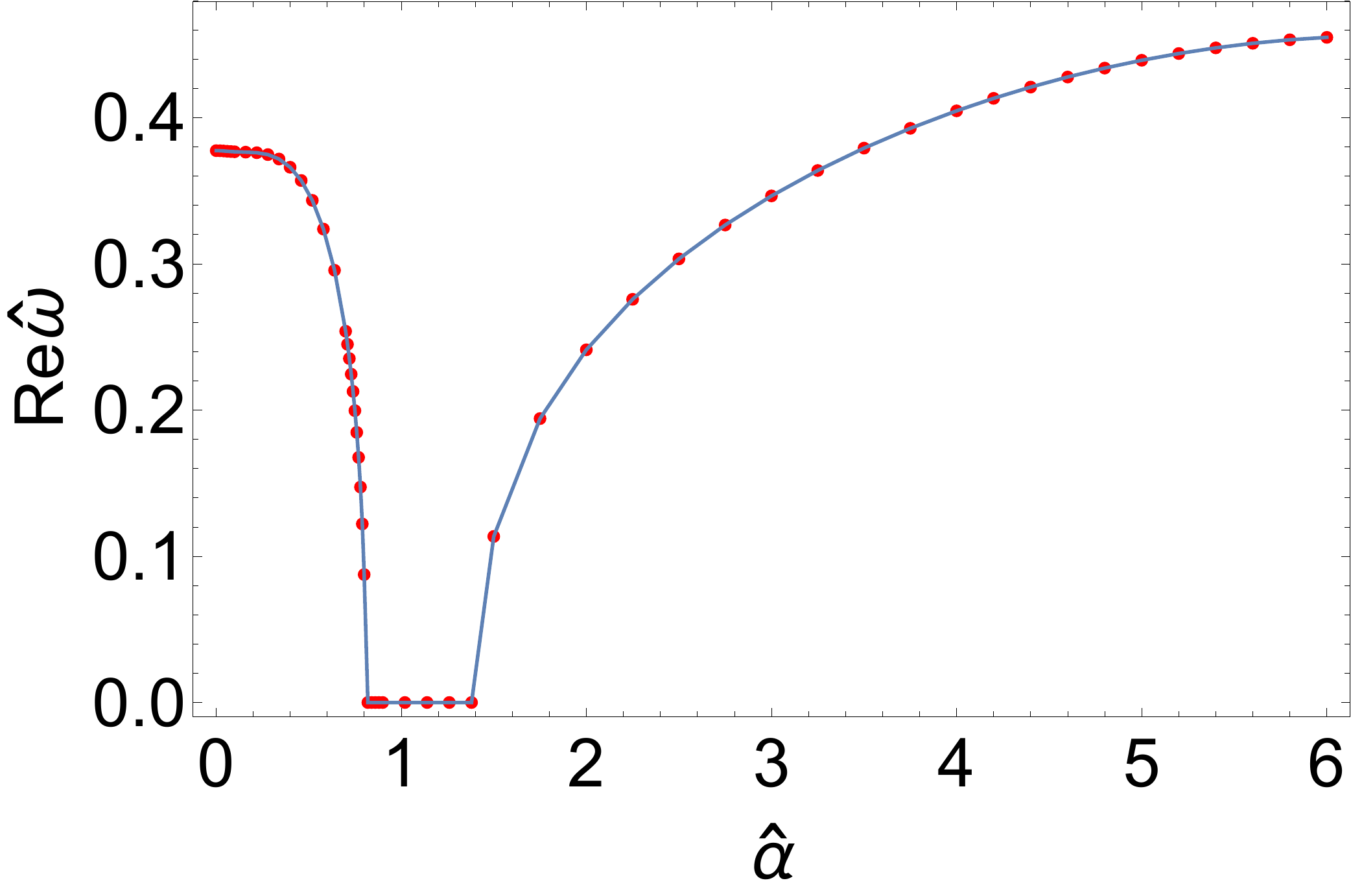}\ \hspace{0.4cm}
\includegraphics[scale=0.25]{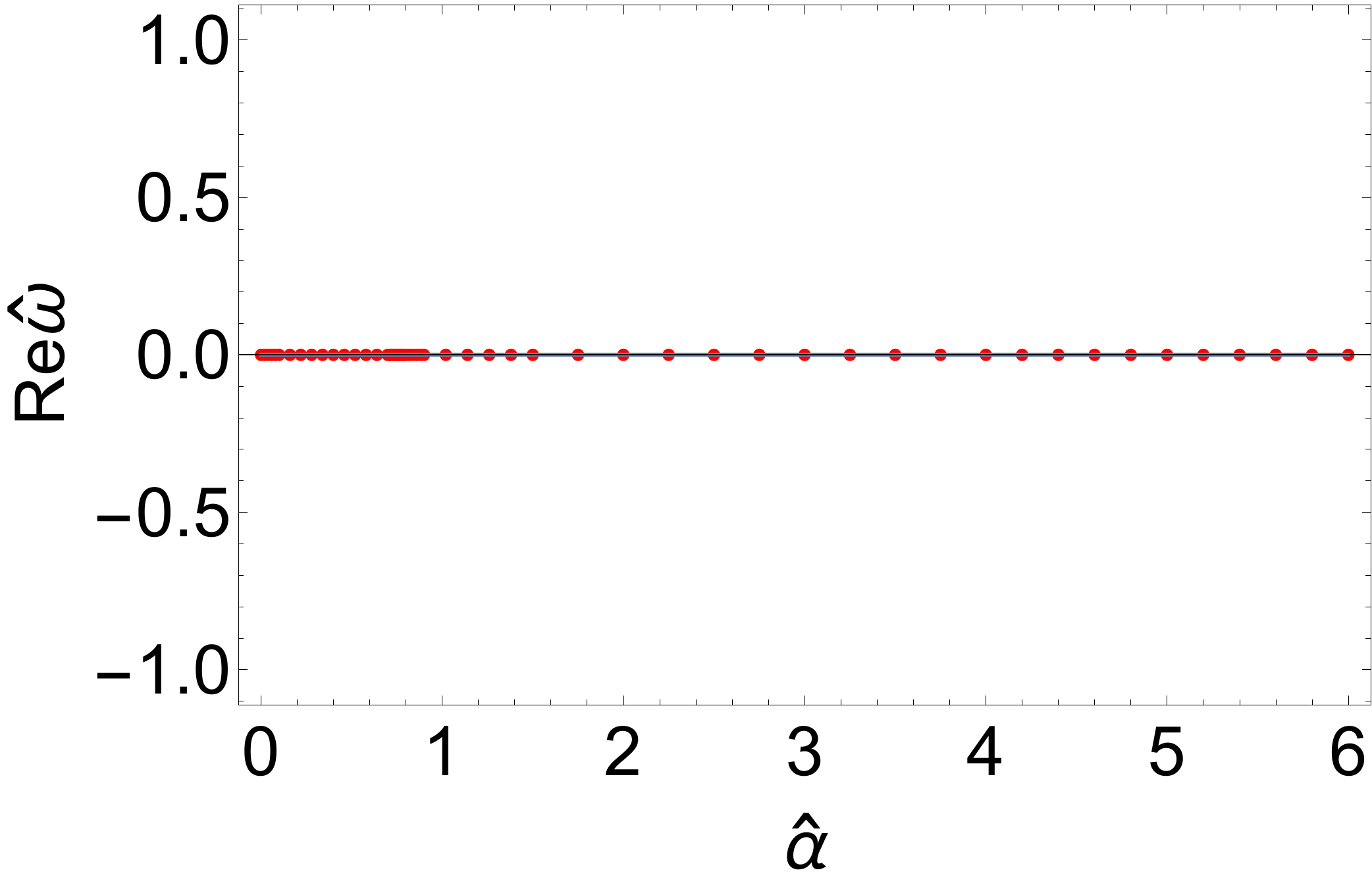}\ \\
\includegraphics[scale=0.25]{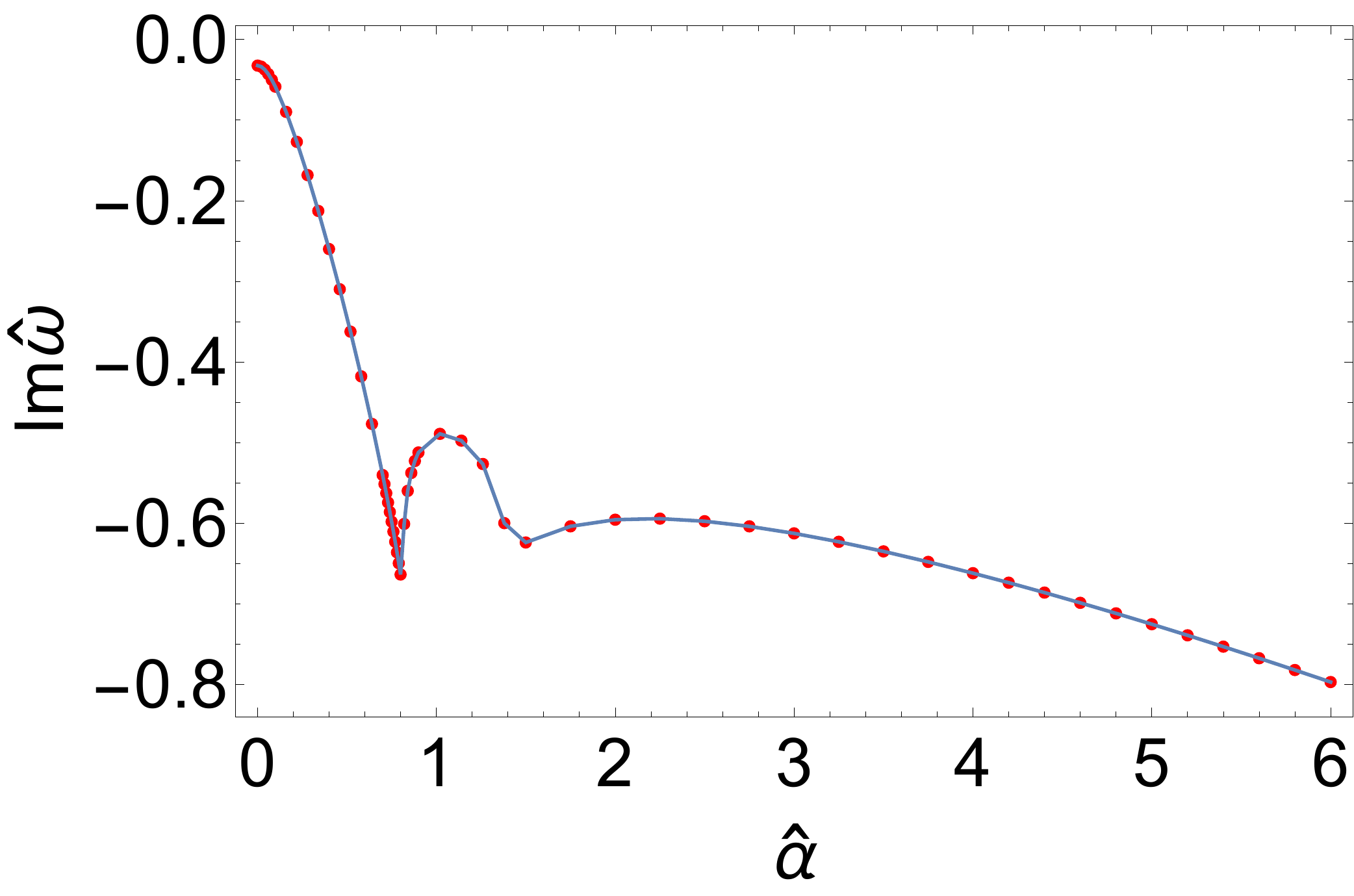}\ \hspace{0.4cm}
\includegraphics[scale=0.25]{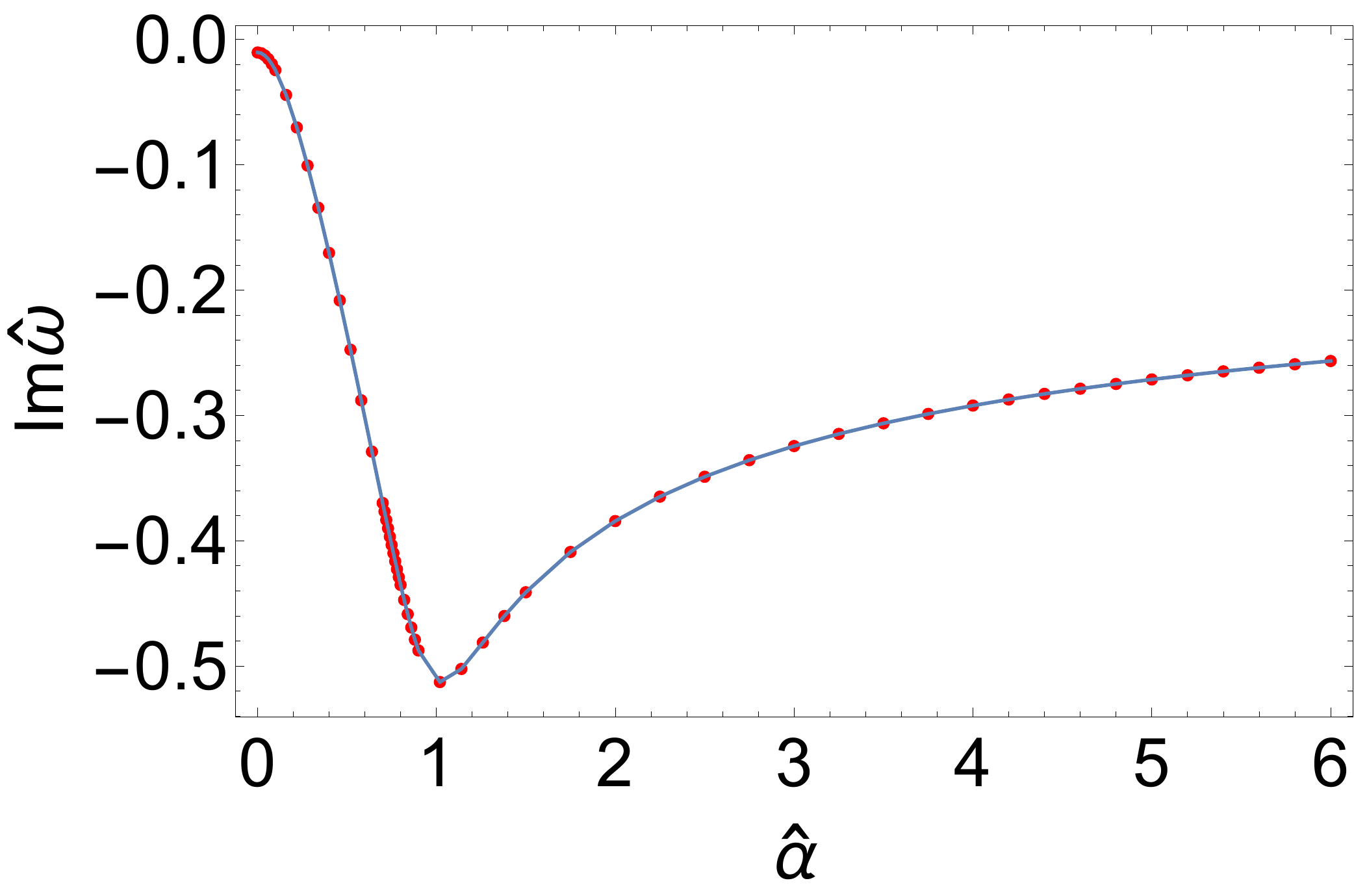}\ \\
\ \\
}
\caption{\label{fig_dqnm} Evolution of the dominant QNMs with $\hat{\alpha}$ for $|\gamma_1|=0.02$ (the first two rows are
for $\gamma_1=-0.02$ and the last two rows for $\gamma_1=0.02$). The left columns are the QNMs
for original theory, and the right ones are that for its dual theory.}
\end{figure}

Further, we study the evolution of the dominant QNMs with $\hat{\alpha}$ for $|\gamma_1|=0.02$, which is shown in FIG.\ref{fig_dqnm}.
We describe the properties as what follows.
\begin{itemize}
  \item For $\gamma_1=-0.02$, all the dominant QNMs for the original theory are purely imaginary modes.
  With the increase of $\hat{\alpha}$, these modes firstly migrate downwards,
  and then migrate upwards, finally approaches to a certain value.
  The evolution of the dominant QNMs for the dual theory with $\gamma_1=0.02$ is similar with that for the original theory with $\gamma_1=-0.02$.
  Such evolution is also similar with that for $4$ derivative theory studied in \cite{Wu:2018vlj}.
  As a whole, there is a qualitative correspondence between the evolution of the poles for the original theory for $\gamma_1=-0.02$
  and that for its dual theory for $\gamma_1=0.02$. But we would like to point out that as discussed above, for small $\hat{\alpha}$,
  the poles for the dual theory are closer to the real frequency axis
  than that for the original theory.
  In contrast, the correspondence hold more well for large $\hat{\alpha}$.
  \item For $\gamma_1=0.02$, the dominant poles for the original theory are off-axis for small $\hat{\alpha}$.
  With the increase of $\hat{\alpha}$, the poles migrate downwards and is closer to the real axis.
  When $\hat{\alpha}$ reaches the value of $\hat{\alpha}\simeq 0.8$, the poles merge into one purely imaginary pole.
  And then, when $\hat{\alpha}$ is beyond $\hat{\alpha}\simeq 1.4$, the purely imaginary pole splits into two off-axis modes.
  While for $\gamma_1=-0.02$, the evolution of the dominant poles for the dual theory is similar with that for the original theory with $\gamma_1=0.02$.
  But when $\hat{\alpha}>3.5$, the pole for the dual theory with $\gamma_1=-0.02$ becomes a purely imaginary one.
\end{itemize}

\section{Conclusion and discussion}\label{sec-con}

In this paper, we extend our previous work \cite{Fu:2017oqa},
which studied the optical conductivity from $6$ derivative theory on top of EA-AdS geometry,
to the holographic response of its EM dual theory. In particular, we explore thoroughly the EM duality.
Also we study the QNMs and the EM duality in complex frequency panel.

In absence of the homogeneous disorder, with the change of the sign of $\gamma_1$,
the particle-vortex duality only holds for small $|\gamma_1|$.
With the increase of $|\gamma_1|$, this duality violates.
When the homogeneous disorder is introduced, as found in $4$ derivative theory,
we find that for the specific value of $\hat{\alpha}=2/\sqrt{3}$, the optical conductivity of
the original theory is almost the same as that of the dual one when the sign of $\gamma_1$ changes.
For other value of $\hat{\alpha}$, the particle-vortex also approximately holds with the change of the sign of $\gamma_1$.
Therefore, we can conclude that the homogeneous disorder make the pseudogap-like of the low frequency conductivity of the original theory
becomes small dip, while suppresses the sharp peak of the EM dual theory
such that we have an approximate particle-vortex duality with the change of the sign of $\gamma_1$.

The properties of the QNMs are also analyzed.
In absence of the homogeneous disorder,
the qualitative correspondence between the poles of $\texttt{Re}\sigma(\hat{\omega};\gamma_1)$
and the ones of $\texttt{Re}\sigma_{\ast}(\hat{\omega};-\gamma_1)$ holds well at low frequency only for small $\gamma_1$.
When $\gamma_1$ becomes large, this correspondence is also violated even at the low frequency region.
For $\gamma_1=-1$, new branch cuts of QNMs are observed.
When the homogeneous disorder is introduced and its strength is small,
the approximate correspondence between the poles of $\texttt{Re}\sigma(\hat{\omega};\gamma_1)$
and the ones of $\texttt{Re}\sigma_{\ast}(\hat{\omega};-\gamma_1)$ recovers in low frequency region even for $|\gamma_1|=0.02$.
But in the high frequency region, this correspondence is violated.
For large $\hat{\alpha}$, this correspondence is strongly violated and
only holds for the dominate QNMs.
For $\gamma_1=-1$, the homogeneous disorder drives the off-axis new branch cuts for $\hat{\alpha}=0$ to the purely imaginary modes.

The evolution of the dominant QNMs with $\hat{\alpha}$ are also explored.
We find that all the dominant QNMs for the original theory with $\gamma_1=-0.02$ and that for the dual theory with $\gamma_1=0.02$ are purely imaginary modes.
Their evolutions are also similar.
As a whole, there is a qualitative correspondence between them.
However, for the evolution of the dominant QNMs for the original theory with $\gamma_1=0.02$ and that for the dual theory with $\gamma_1=-0.02$,
the case is somewhat different. The main difference is that the modes are not again the purely imaginary ones except for some specific $\hat{\alpha}$.

There are lots of open questions worthy of further study.
\begin{itemize}
  \item We would like to carry out an analytical study on the complex frequency conductivity by matching method developed in \cite{Faulkner:2009wj}.
  Also we can analytically work out the QNMs by WKB method as \cite{WitczakKrempa:2012gn,Witczak-Krempa:2013aea}.
  These analytical analysis can surely provide more physical insight and understanding into our present observation.
  \item At present, the perturbative black brane solution to the first order of the coupling parameter from $4$ derivative theory has been worked out in \cite{Ling:2016dck,Li:2017nxh,Wu:2018pig,Mahapatra:2016dae,Dey:2015ytd,Dey:2015poa,Mokhtari:2017vyz,Wu:2018cge} and the related exploration,
      including holographic metal-insulator transition, holographic entanglement, holographic thermalization etc., at finite charge density on top of the perturbative background with HD correctons.
  In future, it is also interesting to further study the QNMs, including scalar, vector and tensor modes, from HD theory at finite charge density\footnote{The QNMs of massless scalar field over the perturbative black hole with spherical symmetry horizon have been studied in \cite{Mahapatra:2016dae}.}.
  \item The dispersing QNMs with finite momentum deserve further studying.
  It surely reveals richer physics of the system.
  \item The superconducting phase from HD theory has been widely explored in \cite{Wu:2010vr,Ma:2011zze,Wu:2017xki,Ling:2016lis,Momeni:2011ca,Momeni:2012ab,Zhao:2012kp,Momeni:2013fma,Momeni:2014efa,Zhang:2015eea}
and references therein. It is also interesting to study the QNMs in the superconducting phase of these models such that we can get richer insight and understanding on the HD theory.
\end{itemize}

\begin{acknowledgments}

This work is supported by the Natural Science Foundation of China under
Grant No.11775036.

\end{acknowledgments}

\begin{appendix}

\section{Brief analysis on the stability}\label{sec-stability}

\begin{figure}
\center{
\includegraphics[scale=0.4]{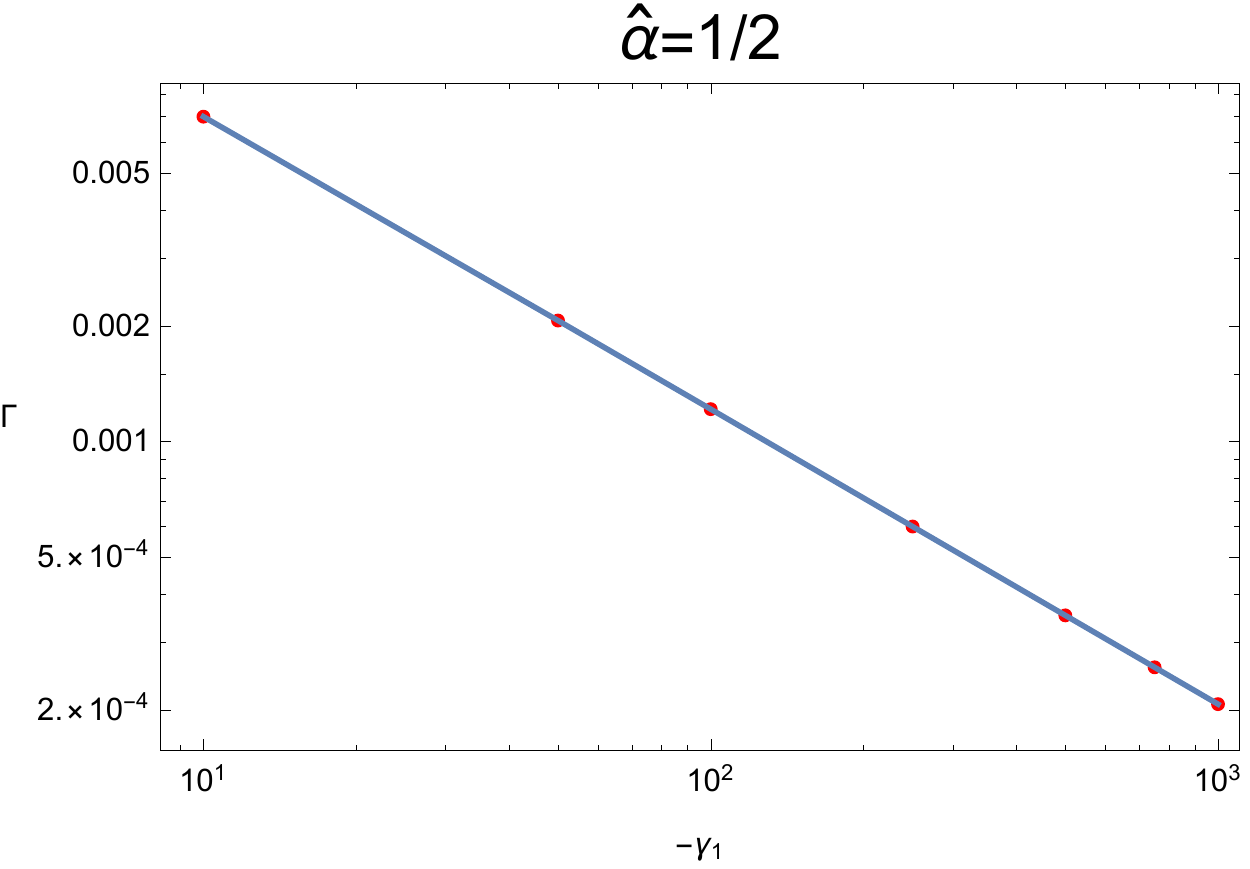}\ \hspace{0.5cm}
\includegraphics[scale=0.4]{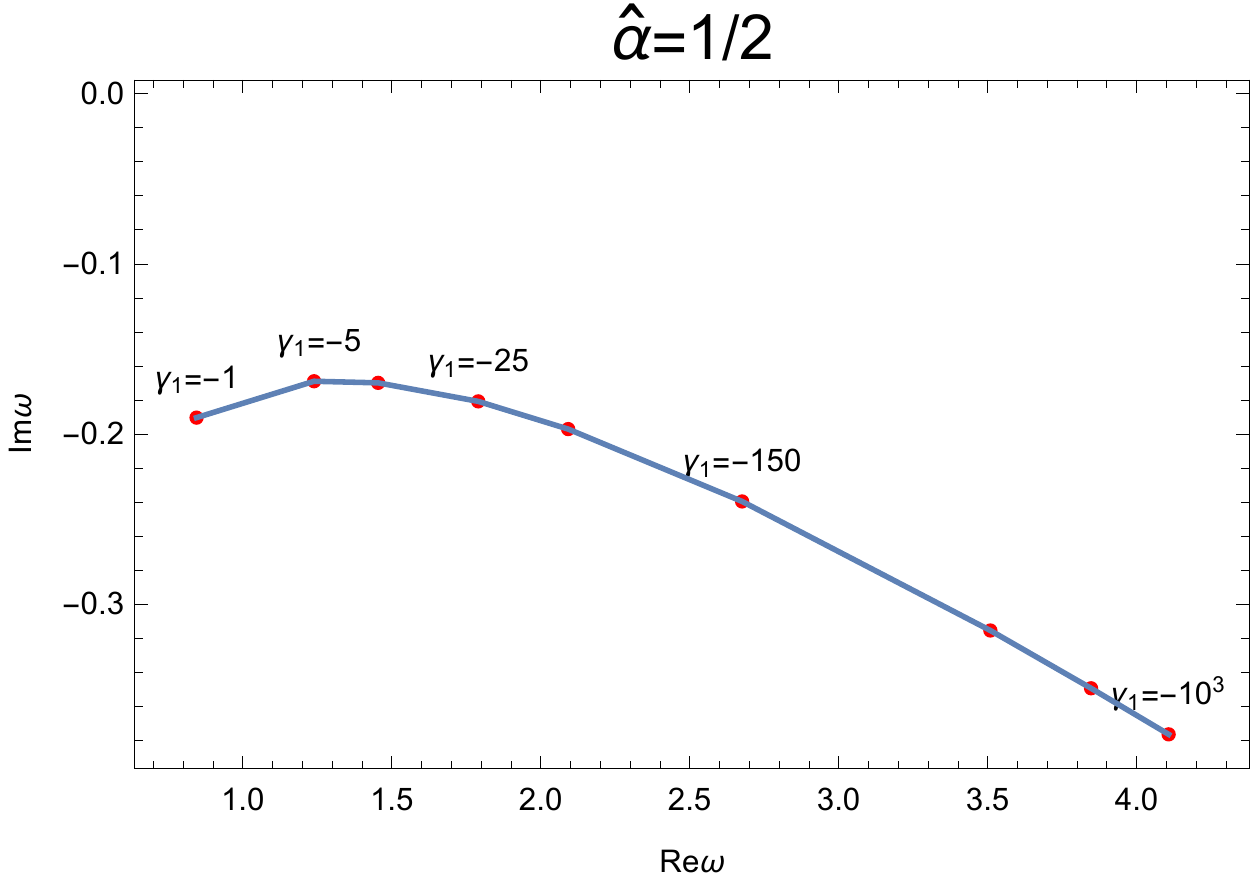}\ \\
\includegraphics[scale=0.4]{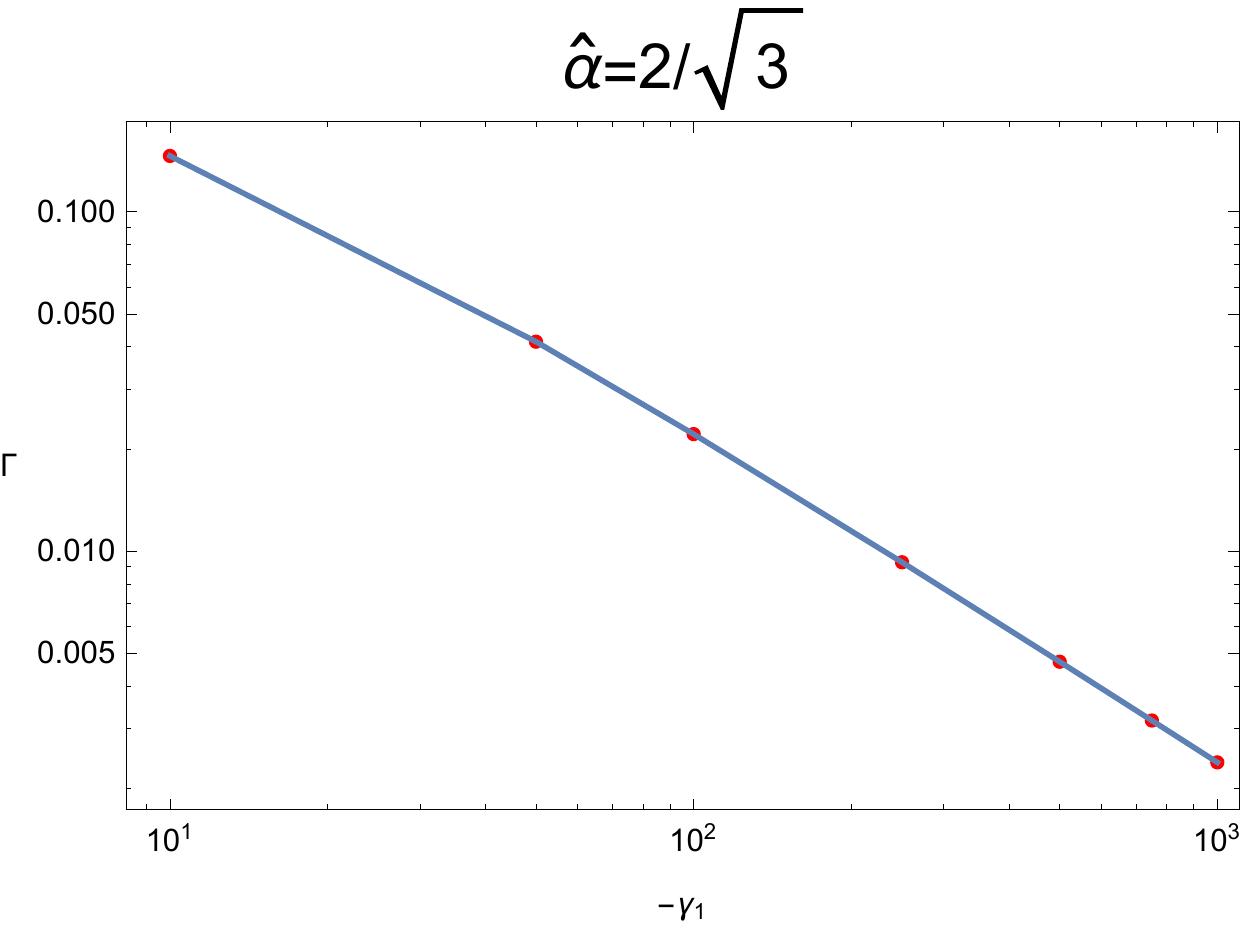}\ \hspace{0.1cm}
\includegraphics[scale=0.4]{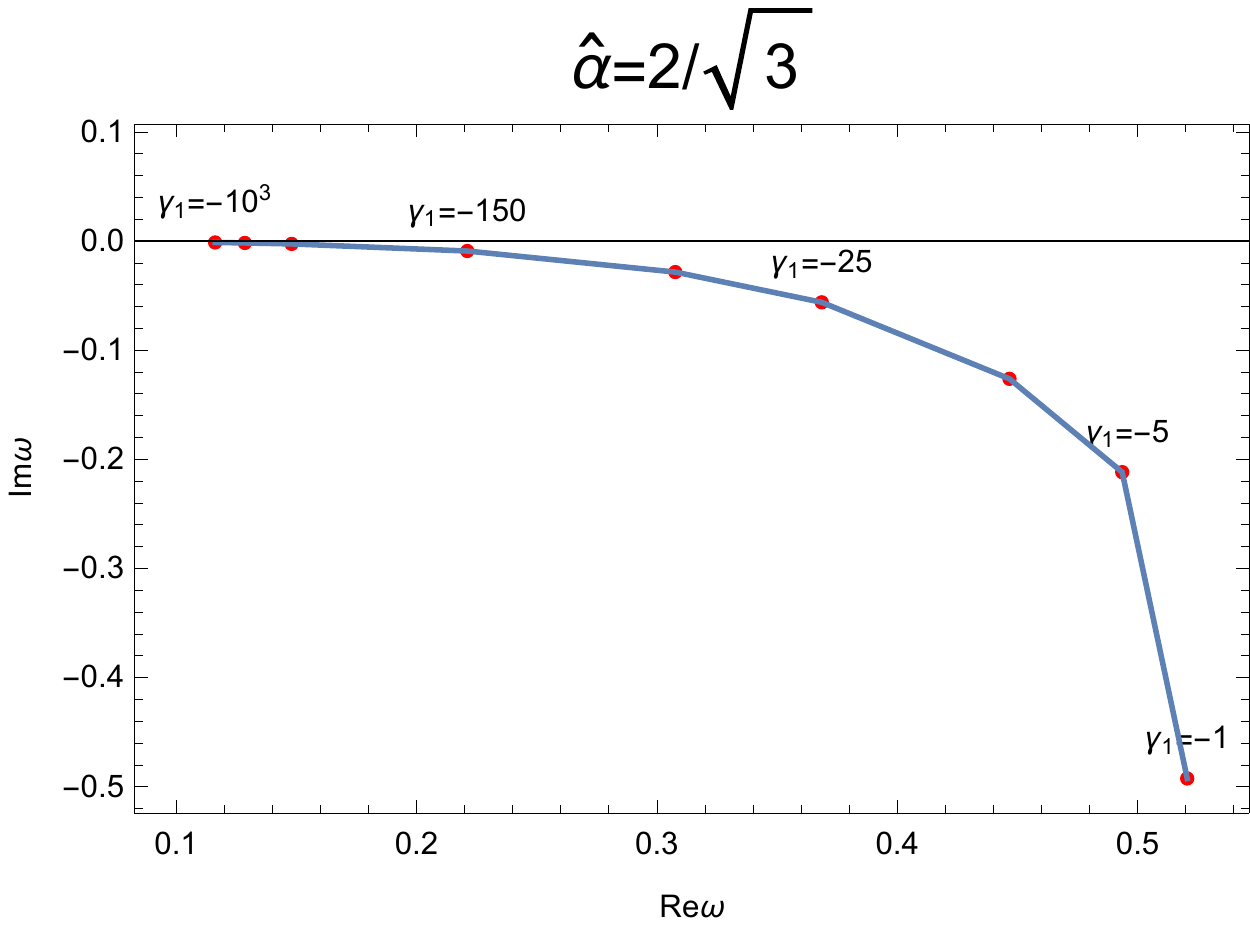}\ \\
\includegraphics[scale=0.4]{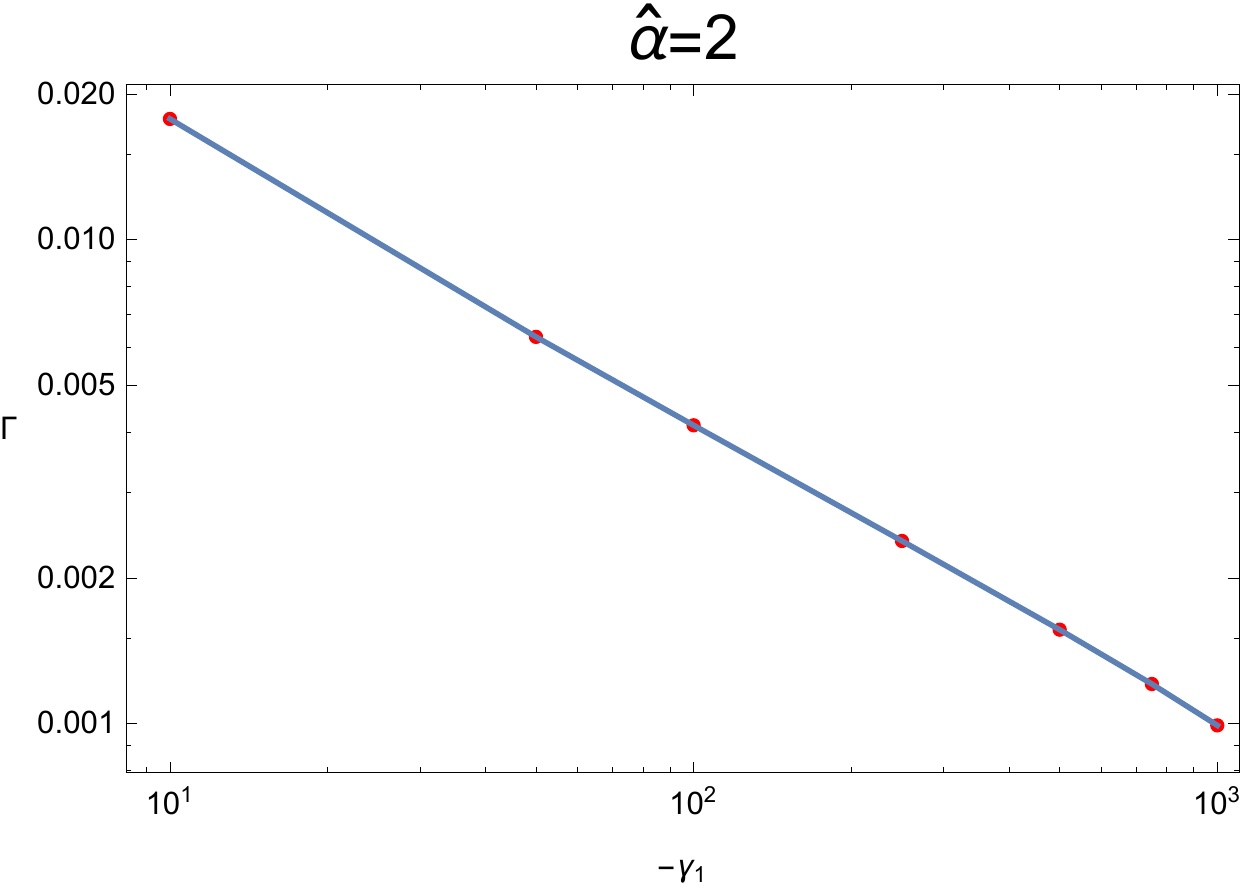}\ \hspace{0.1cm}
\includegraphics[scale=0.4]{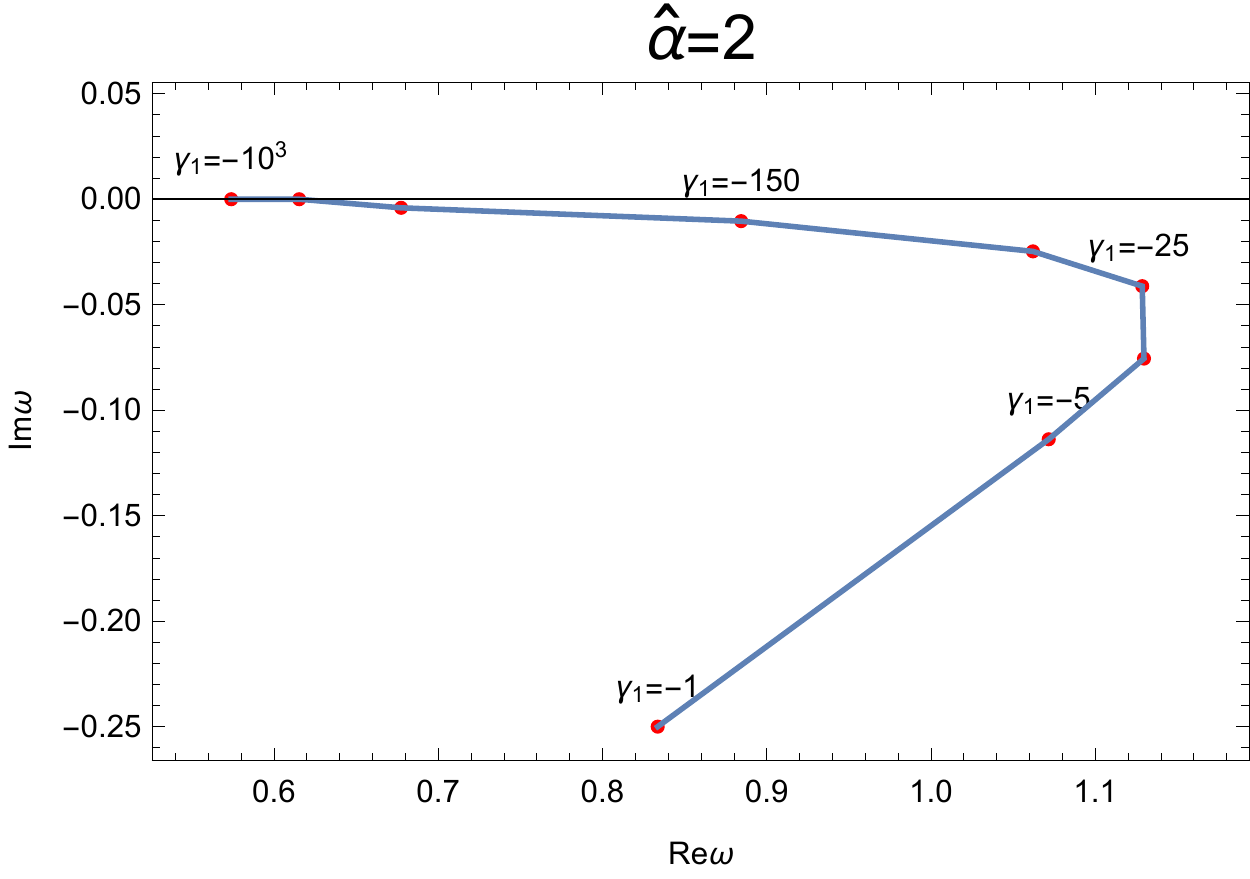}\ \\
\includegraphics[scale=0.4]{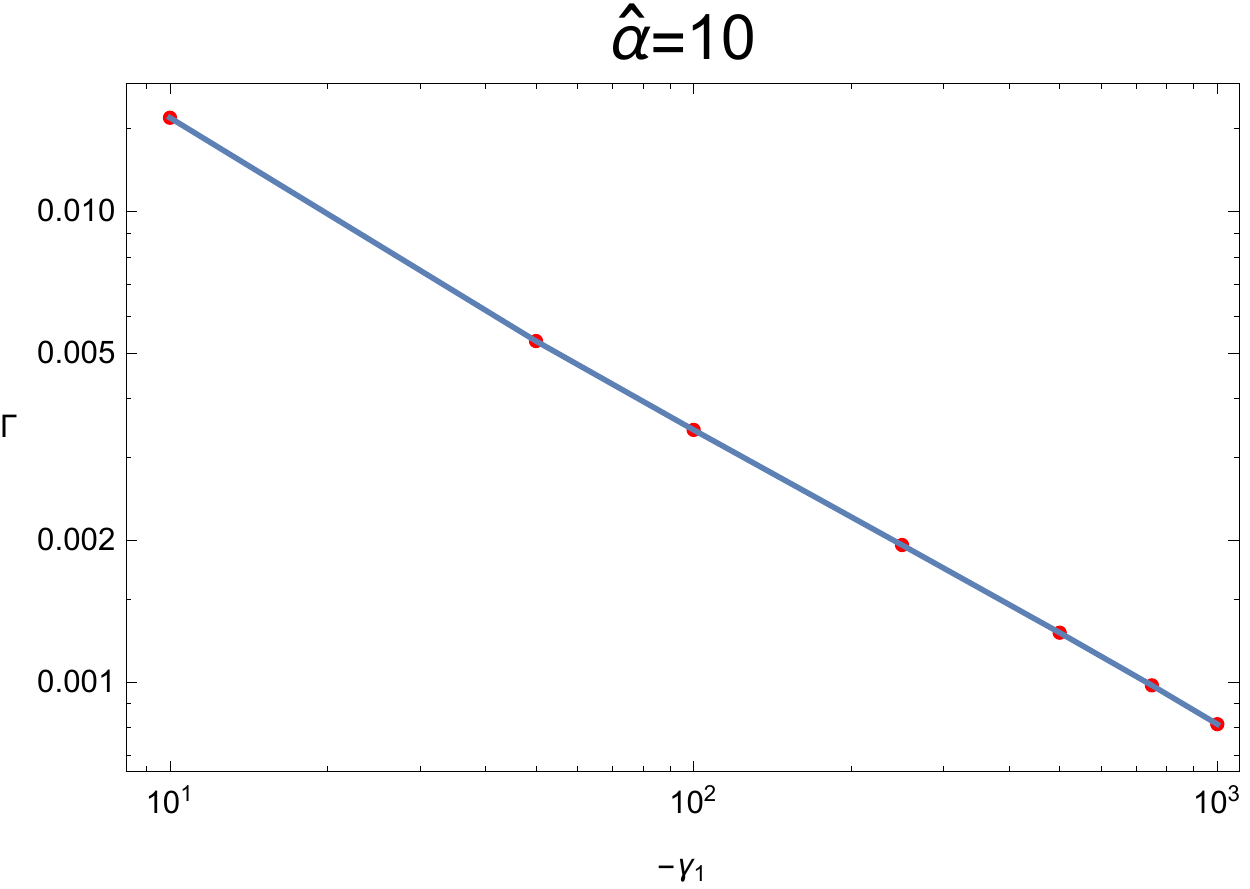}\ \hspace{0.1cm}
\includegraphics[scale=0.4]{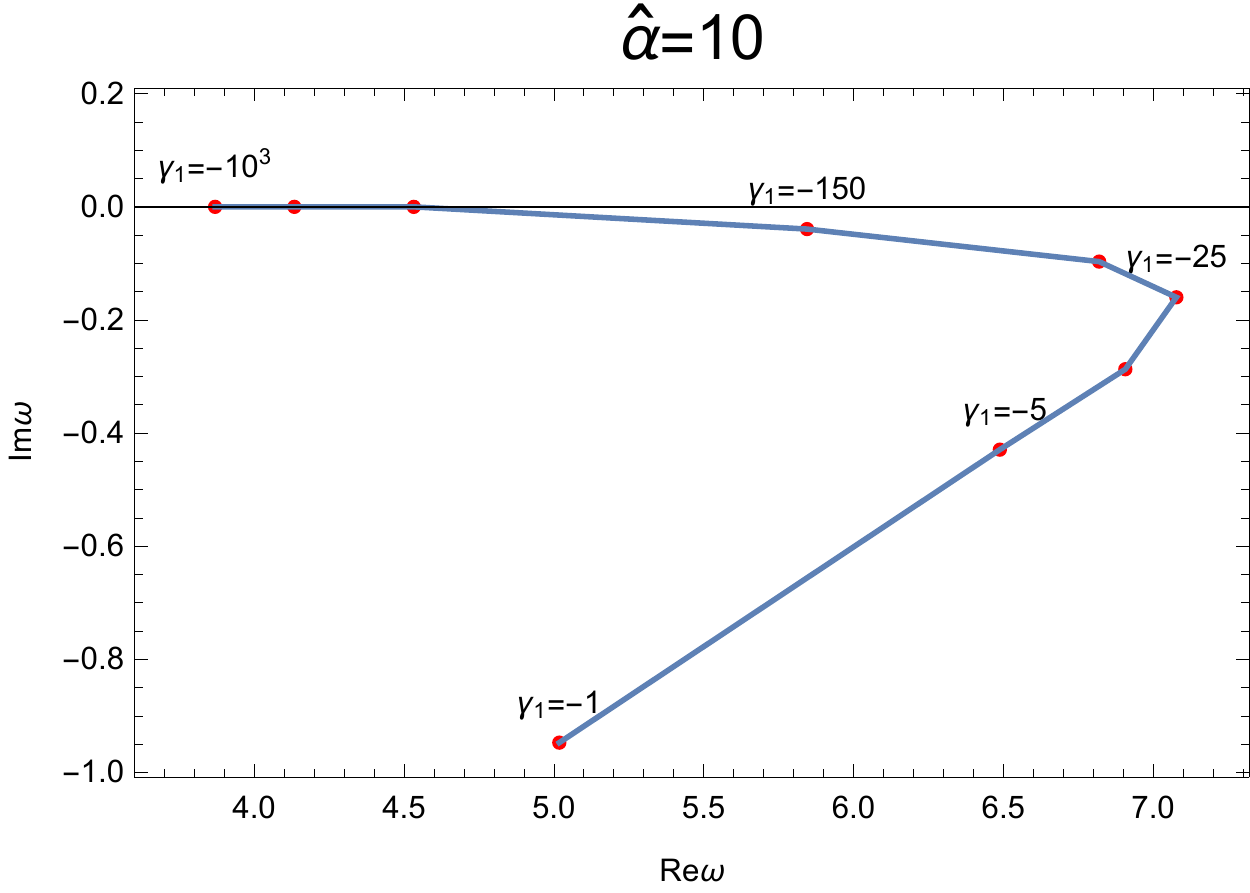}\ \\}
\caption{\label{fig_QNMs_bound} Left plot: Location of the dominate QNM for different $\hat{\alpha}$.
Red dots is the numerical data and the blue line is the power-law fit of the form $a/|\gamma_1|^b$.
Right plot: Location of the next QNM closet to the real axis, a zero, for different $\hat{\alpha}$.}
\end{figure}

In \cite{Fu:2017oqa}, by analyzing the instabilities of gauge mode and the causality in CFT, in addition to the constraint from $\texttt{Re}\sigma(\omega)\geq 0$,
we confirm that the constraint $\gamma_1\leq 1/48$, which has been obtained over SS-AdS geometry in \cite{Witczak-Krempa:2013aea}, also holds over EA-AdS geometry.
In this section, we further examine this constraint by studying the QNMs.
Indeed, we find that when $\gamma_1\leq 1/48$, all the QNMs are in the LHP.

First, it has been shown in FIG.\ref{fig_dqnm_gamma1_n1} and \ref{fig_dqnm}
that for $\gamma_1=-1$, $|\gamma_1|=0.02$ and $\hat{\alpha}\in[0,6]$,
all the poles and zeros, which are the poles of the dual theory, are in the LHP.
In addition, we also see that when $\hat{\alpha}\rightarrow\infty$,
the imaginary modes approach to a constant or even continue to migrate downwards.
Therefore, we can conclude that for the selected $\gamma_1$,
the modes are stable for all $\hat{\alpha}$.
\begin{table}[!h]
\begin{tabular}{|c|c|c|c|c|}
     \hline
           $~\hat{\alpha}~$&$~~1/2~~$&$~~~~2/\sqrt{3}~~~~$&$~~~~2~~~~$&$~~~~10~~~~$\\
     \hline
  $a$&$0.040$&$0.957$&$0.076$&$0.072$\\
     \hline
  $b$&$0.761$&$0.816$&$0.632$&$0.662$\\
     \hline
\end{tabular}
\caption{\label{Table ab} The coefficients of the power-law fit of the form $a/|\gamma_1|^b$ for the dominant poles for different $\hat{\alpha}$.}
\end{table}

And then, for represent $\hat{\alpha}$, we show the poles and zeros for larger region of $\gamma_1\leq 1/48$.
As that on top of SS-AdS geometry in \cite{Witczak-Krempa:2013aea}, the dominant poles from EA-AdS geometry are purely imaginary modes
and also asymptotically approaches the real $\hat{\omega}$-axis as $\gamma_1\rightarrow-\infty$.
In addition, the data can be well fitted by a power-law formula as $a/|\gamma_1|^b$.
The coefficients $a$ and $b$ are listed in Table \ref{Table ab}.
We can see that at least over the $3$ decades we have studied, the fit is very well as shown in FIG.\ref{fig_QNMs_bound}.
We would like to point out that this result is consistent with that the peak of the conductivity at low frequency
becomes sharper with the decrease of $\gamma_1$ and finally approaches a delta function as $\gamma_1\rightarrow-\infty$.
For the zeros, we also see that they all locate at the LHP at least over the $3$ decades we have studied here (see right plots in FIG.\ref{fig_QNMs_bound}).
Therefore, by the analysis of QNMs, we again confirm that the gauge mode is stable over EA-AdS geometry when the coupling parameter is confined to $\gamma_1\leq 1/48$.

\end{appendix}

\end{document}